\documentclass[letterpaper,11pt,fleqn]{article}
\usepackage{jheppub}

\usepackage{graphicx}
\usepackage{bm,amsmath,amssymb}
\usepackage[mathscr]{eucal}

\long\def\comment#1{ }
\newcommand{\eqn}[1]{Eq.~(\ref{#1})}
\newcommand{\beq}{\begin{equation}}
\newcommand{\eeq}{\end{equation}}
\newcommand{\nn}{\nonumber\\}
\newcommand{\dif}{{\rm d}}
\newcommand{\rmd}{{\rm d}}
\newcommand{\rme}{{\rm e}}
\newcommand{\rmi}{{\rm i}}
\newcommand{\rmP}{{\rm P}}
\newcommand{\rmtr}{{\rm tr}}
\newcommand{\rmTr}{{\rm Tr}}
\newcommand{\del}{\partial}
\newcommand{\lan}{\langle}
\newcommand{\ran}{\rangle}

\newcommand{\order}[1]{\mcal{O}{(#1)}}
\newcommand{\mcal}{\mathcal}

\newcommand{\bk}{\bm{k}}

\newcommand{\bp}{\bm{p}}
\newcommand{\bx}{\bm{x}}
\newcommand{\by}{\bm{y}}
\newcommand{\bu}{\bm{u}}

\newcommand{\bz}{\bm{z}}

\newcommand{\br}{\bm{r}}
\newcommand{\bR}{\bm{R}}

\newcommand{\atpi}{\frac{\bar{\alpha}}{2 \pi}}
\newcommand{\abar}{\bar{\alpha}}

\newcommand{\lap}{\nabla_{\perp}^2}

\title{\Large The non--linear evolution of jet quenching}

\author{Edmond Iancu}

\affiliation{Institut de Physique Th\'{e}orique de Saclay,
F-91191 Gif-sur-Yvette, France}

\emailAdd{edmond.iancu@cea.fr}

\abstract{We construct a generalization of the JIMWLK Hamiltonian, going beyond the eikonal approximation, which governs the high-energy evolution of the scattering between a dilute projectile and a dense target with an arbitrary longitudinal extent (a nucleus, or a slice of quark--gluon plasma). Different physical regimes refer to the ratio $L/\tau$ between the longitudinal size $L$ of the target and the lifetime $\tau$ of the gluon fluctuations. When $L/\tau \ll 1$, meaning that the target can be effectively treated as a shockwave, we recover the JIMWLK Hamiltonian, as expected. When $L/\tau \gg 1$, meaning that the fluctuations live inside the target, the new Hamiltonian governs phenomena like the transverse momentum broadening and the radiative energy loss, which accompany the propagation of an energetic parton through a dense QCD medium. Using this Hamiltonian, we derive a non--linear equation for the dipole amplitude (a generalization of the BK equation), which describes the high--energy evolution of jet quenching. As compared to the original BK--JIMWLK evolution, the new evolution is remarkably different: the plasma saturation momentum evolves much faster with increasing energy (or decreasing Bjorken's $x$) than the corresponding scale for a shockwave (nucleus). This widely opens the transverse phase-space for the evolution and implies the existence of large radiative corrections, enhanced by the double logarithm $\ln^2(LT)$, with $T$ the temperature of the medium. This confirms and explains from a physical perspective a recent result by Liou, Mueller, and Wu (arXiv:1304.7677). The dominant,
double--logarithmic,  corrections are smooth enough to be absorbed into a renormalization of the jet quenching parameter $\hat q$. This renormalization is controlled by a linear equation supplemented with a saturation boundary, which emerges via controlled approximations from the generalized BK equation alluded to above.
}

\begin{document}
\maketitle


\section{Introduction}

The concept of jet quenching globally denotes the modifications in the properties of a
`hard probe' (an energetic parton, or the jet generated by its evolution) which occur when this
`jet'  propagates through the dense QCD matter (`quark--gluon plasma') created 
in the intermediate stages of a ultrarelativistic nucleus--nucleus collision \cite{Mehtar-Tani:2013pia,Majumder:2010qh,d'Enterria:2009am,CasalderreySolana:2007zz,Kovner:2003zj}.
This encompasses several related phenomena like the transverse momentum broadening,
the (radiative) energy loss, or the jet fragmentation via medium--induced gluon branching,
and also the associated observables, like the nuclear modification factor, or the di--jet asymmetry. 
A common denominator of these phenomena is that, within most of their theoretical descriptions
to date, they depend upon the medium properties via a single parameter : a transport coefficient
known as the `jet quenching parameter' $\hat q$.  This explains the importance of this quantity $\hat q$  
for both theory and phenomenology, and motivates the recent attempts to obtain better estimates for it
from first principles, at least
in special cases \cite{Arnold:2008vd,CaronHuot:2008ni,Majumder:2012sh,
Liou:2013qya,Laine:2013lia,Panero:2013pla}.

Roughly speaking, the jet quenching parameter measures 
the dispersion in transverse momentum\footnote{By `transverse' we mean the two
dimensional plane $\bx=(x^1,x^2)$ orthogonal to the parton direction of motion, conventionally
chosen along $x^3$.} accumulated by a fast parton after 
crossing the medium over a distance $L$: $\langle p_\perp^2\rangle \simeq\hat q L$. 
At weak coupling, the dominant mechanism
responsible for this dispersion is multiple scattering off the medium constituents. At leading
order in $\alpha_s$, $\hat q$ can be computed as the second moment of the `collision kernel' (the differential 
cross--section for elastic scattering in the medium; see Sect.~\ref{sec:tree} for details).
Beyond leading order, one needs a non--perturbative definition for the
transverse momentum broadening. The one that we shall adopt here and which is often used in the
literature involves the `color dipole', a light--like Wilson loop in the color representation of the fast
parton. Physically, this Wilson loop describes the $S$--matrix for a small `color dipole' (a quark--antiquark
pair, or a set of two gluons, in a color singlet state) which propagates through the medium.
Via unitarity, the Fourier transform of this $S$--matrix determines the cross--section $\rmd N/\rmd^2\bp$ 
for transverse momentum broadening \cite{Mueller:2001fv,Mueller:2012bn}.  At tree--level, these
definitions imply $\langle p_\perp^2\rangle^{(0)} \simeq\hat q^{(0)}(L) L$, with $\hat q^{(0)}(L)$
logarithmically dependent upon the medium size $L$. This dependence
enters via the resolution of the scattering process: the transverse momenta transferred by 
the medium can be as large as the `saturation momentum' $Q_s^2\equiv \hat q L$ 
 (see Sect.~\ref{sec:tree}). Beyond leading order, it is
a priori unclear whether the notion of  `jet quenching parameter' (as a quasi--local 
transport coefficient) is still useful, or even well--defined. Our criterion in that sense will be to check
whether a formula like $\langle p_\perp^2\rangle \simeq\hat q(L) L$ does still hold, with $\hat q(L)$
a reasonably slowly--varying function. But even when this appears to be the case, we shall see that
the $L$--dependence of $\hat q$ is generally enhanced by the radiative corrections, 
due to the intrinsic non--locality of the quantum fluctuations. 

\comment{For what follows, it is useful to keep in mind that there are two main
sources of  $L$--dependence within $\hat q$~: the transverse resolution of the scattering process
--- the transverse momenta transferred by the medium can be as large as the `saturation
momentum' $Q_s^2\equiv \hat q L$ ---, and the (transverse and longitudinal) kinematics of the
fluctuations --- their transverse momentum can be as large as $Q_s$, and 
their lifetime as large as $L$.}

So far, two different classes of next--to--leading order corrections, which correspond to very
different kinematical regimes, have been computed at weak coupling \cite{CaronHuot:2008ni,Liou:2013qya}.
In Ref.~\cite{CaronHuot:2008ni}, Caron-Huot considered a
medium which is a weakly--coupled quark--gluon plasma (QGP) with temperature $T$
and computed the corrections of 
$\order{g}$ to the `collision kernel', as  generated by the soft, highly--populated (and hence semi--classical), 
thermal modes, with energies and momenta $\lesssim gT$. (The corresponding leading--order value
has been computed by Arnold and Xiao \cite{Arnold:2008vd}.) These corrections do not modify
the logarithmic dependence of $\hat q$ upon the medium size $L$, which is rather introduced 
by the hardest collisions, with transferred momenta $k_\perp\sim Q_s$. (We implicitly assume that
$Q_s\gg T$.)

By contrast, in Ref.~\cite{Liou:2013qya}, Liou, Mueller, and Wu have studied the relatively hard 
and nearly on--shell gluon fluctuations, with large transverse momenta $p_\perp\gg T$ and even larger
longitudinal momenta, $p^3\simeq p^0\gg p_\perp$ (in the plasma rest frame).
Such fluctuations, which are most naturally viewed as bremsstrahlung by the projectile,
are not sensitive to the detailed properties of the medium. They depend upon the latter
only via the tree--level value $\hat q^{(0)}$ of the jet quenching parameter and via two basic
scales --- the longitudinal extent $L$ of the medium and the wavelength $\lambda$
of its typical constituents (with $\lambda=1/T$ for the QGP)
--- which constrain the phase--space for bremsstrahlung. Ref.~\cite{Liou:2013qya} found large
one--loop corrections\footnote{See also Ref.~\cite{Wu:2011kc} for a similar but earlier observation, which 
has motivated the more elaborate analysis in Ref.~\cite{Liou:2013qya}.} 
to $\langle p_\perp^2\rangle$, of relative order 
$\alpha_s N_c\ln^2(L/\lambda)$, where the double logarithm comes from the 
phase--space: one logarithm is generated by integrating over the 
lifetime $\tau\sim p^0/p_\perp^2$ of the fluctuations, over the range $\lambda\ll \tau\ll L$, and
the other one comes from the respective transverse momenta,  within the interval $\hat q \tau
\ll p_\perp^2\ll Q_s^2$ (for a given value of $\tau$). The lower limit $\hat q\tau$ on $p_\perp^2$ 
refers to multiple scattering: the condition $p_\perp^2\gg \hat q\tau$ means that the relevant 
fluctuations are hard enough to suffer only one scattering during their lifetime.

Note that this `single scattering' property refers to a color {\em dipole}, and not to
a charged parton. That is, the large radiative corrections identified in Ref.~\cite{Liou:2013qya}
should not be viewed as a renormalization of the collision kernel above mentioned, which
represents the differential cross--section for individual scatterings in the plasma, but rather
as a change in the {\em transport} cross--section relevant for transverse momentum broadening,
which is controlled by the relatively rare collisions involving a sufficiently large momentum 
transfer (which can be as large as $Q_s$).

For what follows, it is important to notice that the medium size $L$ sets the upper
limit on the lifetime $\tau$ of the fluctuations, and hence on their energy $p^0$. Accordingly, 
when increasing $L$, one opens the phase--space for fluctuations which are more and more
energetic. Such fluctuations can then evolve towards lower energies, via soft gluon emissions.
This evolution is represented by Feynman graphs of higher--loop order  (gluon 
cascades which are strongly ordered in energy), which are enhanced by the phase--space:
the powers of $\abar\equiv\alpha_sN_c/\pi$ associated with soft gluon emissions
can be accompanied by either double, or at least single, logarithms of $L/\lambda$, depending upon
the kinematics of the emissions.
Ref.~\cite{Liou:2013qya} not only computed the first step in this evolution, for both the 
double--logarithmic and the single--logarithmic corrections, 
but also provided a simple recipe for resuming the
corrections enhanced by double--logarithms to all orders.

Yet, already the one--loop calculation of the single--logarithmic corrections 
in Ref.~\cite{Liou:2013qya} has met with serious
difficulties, which reflect the lack of a systematic theoretical framework for addressing this 
complicated, non--linear, high-energy evolution. Namely, in order to compute the effects of
order $\abar\ln(L/\lambda)$, one had to estimate the effects of multiple scattering beyond the 
eikonal approximation and also to heuristically include the `virtual' corrections, 
which were truly missed by that analysis, but were essential for that purpose.  Vice--versa,
the only reason why the double--logarithmic corrections appear to be comparatively simple,
is because they are neither sensitive to multiple scattering (except for the restriction on their
phase--space), nor to the effects of the `virtual' corrections (which are generally important to ensure
probability conservation\footnote{The `virtual' terms express the reduction in 
the probability that the evolving system remain in its original state, as it was prior to the evolution.
They become negligible to double--logarithmic accuracy because,
in that limit, the scattering of the original projectile is much weaker than that of the evolved
system including additional gluons.}, 
but become irrelevant at double--logarithmic accuracy).
Remarkably, it appears that the subset of radiative corrections which are enhanced by 
powers of $\ln^2(L/\lambda)$ forms an `island' of effectively linear evolution,
which besides being structurally simple, it also plays a major physical role, in that it gives the 
dominant contributions in the limit of a large medium  $L\gg\lambda$.

By itself, the prominence of a double--logarithmic approximation in the context
of pQCD evolution  is not new --- other familiar examples include the 
fragmentation of a virtual jet in the vacuum \cite{Dokshitzer:1991wu}, or the 
evolution of the parton distribution in the `double--leading--log--approximation' 
(a common limit of the DGLAP and BFKL equations \cite{Kovchegov:2012mbw}).
What is a bit surprising though, is the importance of such an approximation in the context
of a {\em non--linear} evolution. All the other examples listed above refer to linear processes.
And in the only other example of a non--linear pQCD evolution at our disposal
--- the BK--JIMWLK evolution of the gluon distribution in a large nucleus
(or of particle production in proton--nucleus collisions)  
\cite{Balitsky:1995ub,Kovchegov:1999yj,JalilianMarian:1997jx,JalilianMarian:1997gr,JalilianMarian:1997dw,
Kovner:2000pt,Weigert:2000gi,Iancu:2000hn,Iancu:2001ad,Iancu:2001md,Ferreiro:2001qy} ---,
it is well known that the `double--logarithmic
approximation' (DLA) is not a good approximation at high energy. 

For instance, the high--energy scattering between a small dipole and a dense nucleus
is described by the non--linear BK equation or, at least (in the single scattering
regime), by a linear approximation to it that can be described as `the BFKL equation supplemented with
a saturation boundary' \cite{Iancu:2002tr,Mueller:2002zm,Munier:2003vc}. 
The `saturation boundary' expresses
the reduction in the phase--space for linear evolution introduced by multiple scattering (or, equivalently,
by gluon saturation in the nucleus). Both the saturation effects and the `virtual' BFKL corrections
are essential for that dynamics and together they lead to a dramatic change in the behavior of the 
scattering amplitude for small dipole sizes $r\lesssim 1/Q_s$: 
they introduce a non--perturbative anomalous dimension (the amplitude behaves like $r^{2\gamma_s}$
with $\gamma_s\simeq 0.63$, instead of the tree--level result $\propto r^2$).
If a similar change was to occur in the context of transverse momentum 
broadening, it would have important consequences for the phenomenology : 
such corrections could not be simply absorbed into a redefinition of $\hat q$, unlike 
the comparatively smooth corrections introduced by the DLA (see below).  
We thus see that the prominence of the DLA in the problem of jet quenching is a remarkable
simplification, which is unusual in the context of the non--linear evolution and is
important  for the phenomenology.

Such considerations invite us to a deeper understanding 
of the high--energy evolution of jet quenching from first principles.
It is our main purpose in this paper to provide a general framework in that sense
--- that is, a theory for the non--linear 
evolution of jet quenching to leading order in perturbative QCD at high energy 
--- and then use this framework to address some of the questions aforementioned. In 
particular,  we shall try to clarify issues like the comparison with the BK--JIMWLK evolution, 
the origin and calculation of the virtual corrections (which is particularly tricky for an extended
target), the physics of gluon evolution and saturation in the plasma, 
the emergence of the double--logarithmic approximation (including
the precise phase--space), the possibility to include 
the radiative corrections into a renormalization of the jet quenching parameter, and the consequences
of such a renormalization for the related problem of the radiative energy loss.

In developing the general formalism below, it will be convenient to assume that the projectile
enters the medium from the outside and that it was on--shell prior to the collision. This
guarantees that the quantum fluctuations which matter for the evolution of the $S$--matrix
are generated exclusively via interactions in the target\footnote{If the projectile is produced by a hard
process occurring inside the medium or at some finite distance from it, then there is additional
radiation, associated with the initial virtuality, that would mix with the evolution that we are
here interested in (see e.g. the discussion in \cite{Kang:2013raa}) By choosing an on--shell
projectile, we avoid this mixing.}. Then the main difference
between the problem of jet quenching and the BK--JIMWLK formalism for $pA$ collisions refers
to the longitudinal extent $L$ of the target and, more precisely, to the ratio between $L$ and the
lifetime $\tau$ of the typical gluon fluctuations. In $pA$ collisions, the center--of--mass energy
is so high that the nuclear target looks effectively like a shockwave ($L\ll\tau$), due to Lorentz contraction. 
Then  the multiple  scattering can be treated in the strict eikonal approximation, which assumes that 
the transverse coordinates of the projectile partons are not affected by the interactions. 
By contrast, in the context of jet quenching, the energies are much lower and the fluctuations
live fully inside the medium ($L\gtrsim \tau$), so the effects of the multiple scattering can 
accumulate during their whole lifetime. Then the strict eikonal approximation is not applicable anymore,
although the individual scatterings are still soft: one cannot ignore the transverse motion 
of the fluctuations during their lifetime.

These considerations also show that the two problems aforementioned 
($pA$ collisions and jet quenching) can be viewed as limiting situations
of a common set--up: the high--energy scattering between a dilute projectile and a dense target
{\em with an arbitrary longitudinal extent}. This is the first problem that we shall address and solve
in this paper. Specifically, in Sect.~\ref{sec:H} and Appendix~\ref{sec:deriv},
we shall construct an effective Hamiltonian which, when acting on the 
$S$--matrix of the projectile (a gauge--invariant product of Wilson lines), generates one additional
soft gluon emission in the background of a strong color field representing the medium. (The medium
correlations are reproduced by averaging over this background field, in the spirit of the color glass
condensate \cite{Iancu:2002xk,Gelis:2010nm}.) This Hamiltonian may be viewed as a generalization
of the JIMWLK Hamiltonian beyond the strict eikonal approximation. It looks compact and simple, but
it is less explicit than the JIMWLK Hamiltonian, in the sense that the integrals over the 
emission times cannot be performed in general (i.e. for an arbitrary target). Accordingly, the general
Hamiltonian is non--local both in the transverse coordinates and in the light--cone (LC) times.
The formal manipulations with this Hamiltonian
are complicated by potential (infrared and ultraviolet) divergences which require
prescriptions at the intermediate steps and cancel only in the final results.
In Sect.~\ref{sec:prob}, we 
demonstrate a general mechanism ensuring such cancellations --- this involves a particular
`sum--rule' for the gluon propagator in the LC gauge, \eqn{ident} --- and clarify its connexion
to probability conservation. In particular, we show that the `virtual' corrections can be alternatively 
implemented as a local `counter--term', which is particularly convenient when the target is an 
extended medium.


As a first test of the new Hamiltonian and of our ability to use it for explicit calculations,
we consider  in  Sect.~\ref{sec:JIM} the example of a shockwave target ($L\ll\tau$). Then the
integrals over the emission times can be explicitly performed (the adiabatic prescription for
regulating the large time behavior turns out to be important for that purpose) and, as a result, 
we recover the JIMWLK 
Hamiltonian \cite{JalilianMarian:1997jx,JalilianMarian:1997gr,JalilianMarian:1997dw,
Kovner:2000pt,Weigert:2000gi,Iancu:2000hn,Iancu:2001ad,Iancu:2001md,Ferreiro:2001qy}, as expected.
We also show that the `counter--term' alluded to above generates the `virtual' piece in the BK
equation, as it should.

Starting with Sect.~\ref{sec:JET}, we turn to the case of an extended target ($L\gg\tau$), 
as appropriate for the physics of jet quenching and related phenomena. The general equations
generated by the evolution Hamiltonian in that case are extremely complicated (see the
discussion in Sect.~\ref{sec:dipole}): they are {\em non--local
in LC time} (because gluon emissions can occur anywhere inside the medium and they can have
any lifetime $\tau$) and also {\em functional} (the transverse trajectories of the gluons are random, due
to the quantum diffusion, and they are distributed according to a path--integral). An useful approximation
is to assume that the medium correlations are Gaussian and local in LC time. (A similar mean field
approximation proved to be successful in the case of the BK--JIMWLK equations
\cite{Kovner:2001vi,Iancu:2002xk,Iancu:2002aq,Blaizot:2004wv,Kovchegov:2008mk,Dominguez:2011wm,Iancu:2011ns,Iancu:2011nj,Dumitru:2011vk}.) Under this assumption,
the equation obeyed by the dipole $S$--matrix takes the form shown in \eqn{DeltaHeq},
which is recognized as a functional 
generalization of the BK equation. The solution to this equation resums all the corrections
enhanced by at least one power of the large logarithm $\ln(L/\lambda)$. It remains as an open question
whether such a functional equation can be solved via numerical methods. Our main point
though is that, for the present purposes --- i.e. for a study of the leading--order evolution of the
jet quenching in the limit $L\gg\lambda$ ---, one can drastically simplify this equation and even obtain
analytic results. 

This is based on the following two observations. 
First, in order to compute the transverse momentum broadening $\langle p_\perp^2\rangle$,
one needs the dipole $S$--matrix $\mathcal{S}(r)$ for dipole
sizes in the vicinity of the {\em plasma saturation line}~; that is, when increasing the medium size
$L$, one must simultaneously decrease the typical dipole size, according to
$r\sim 1/Q_s(L)$, with $Q_s^2(L)=\hat q L$. Second, the dominant radiative corrections in the
interesting regime at  $L/\lambda\gg 1$ and $r\sim 1/Q_s(L)$ are those which are enhanced by a 
{\em double} logarithm $\ln^2(L/\lambda)$. They can be resumed to all orders 
by solving a simplified, {\em linear}, equation, namely \eqn{DLAint}, which emerges from
 the generalized BK equation aforementioned and is equivalent
to the resummation proposed in Ref.~\cite{Liou:2013qya}. 
This equation, that can be described as `the DLA equation supplemented with a saturation boundary',
is different from `the BFKL equation with a saturation boundary' alluded to before --- it is 
actually simpler and in particular it does not lead to a (non--perturbative) anomalous dimension.
That is, the scattering amplitude obtained by solving this equation is still proportional to $r^2$
up to a slowly varying function, as it would be at tree--level\footnote{But a {\em perturbative} 
anomalous dimension, of $\order{g}$, can be generated by the all--order
resummation of the double--logarithmic corrections; this is the `saturation exponent'
to be discussed in Sect.~\ref{sec:gluon}.} (see  Sect.~\ref{sec:SSA} for details). 
This in turn implies that the dominant 
corrections to $\langle p_\perp^2\rangle$ can be absorbed into a
renormalization $\delta\hat q(L)$ of the jet quenching parameter, which thus becomes mildly non--local.

Given the central role played by the DLA, it is interesting to understand the emergence of this
approximation on physical grounds.  As  we explain
in Sect.~\ref{sec:gluon}, this is related to the specificity of the high--energy evolution of the gluon 
distribution in the medium, that we here address for the first time. Namely,  
we show that the non--linear effects in the generalized BK equation \eqref{DeltaHeq} 
can be also understood as gluon saturation in the medium, but with a saturation scale 
$Q_s^2(x)$ which increases very fast when decreasing $x\equiv \lambda/\tau$ (the longitudinal 
momentum fraction carried by the gluons) --- much faster than the corresponding 
scale in a shockwave. Specifically, one has $Q_s^2(x)\propto 1/x$ already at {\em tree--level}
and this growth becomes even faster after including the effects of the small--$x$ evolution.
The physical explanation is quite simple:  the quantity $Q_s^2(x)$ is proportional
with the longitudinal size of the region where the gluons can overlap with each other. For
gluons inside the medium, this region is their wavelength $\tau\simeq \lambda/x$~; hence,
$Q_s^2\propto \tau \sim 1/x$, as anticipated. In turn, this
rapid growth of $Q_s^2(x)$ with $1/x$ widely opens the {\em transverse} phase--space and
thus favors a {\em double--logarithmic evolution}~:
when increasing $L/\lambda$, one opens not only the longitudinal phase--space at  $\lambda\ll \tau\ll L$, 
but also the transverse one at $Q_s^2(x)\ll p_\perp^2\ll Q_s^2$. The upper limit $Q_s^2=\hat q L$
(the conventional `saturation momentum' in the literature on jet quenching) is simply the largest value
of $Q_s^2(x)$, corresponding to $x_{\rm min}= \lambda/L$. This situation should be contrasted to the 
more familiar case of a shockwave, where the variation of $Q_s^2(x)$ with $1/x$ is a parametrically 
small effect, of order $\alpha_s$ (a pure effect of the evolution), 
so the transverse phase--space increases much slower 
than the longitudinal one in the approach towards saturation. Incidentally, this explains
why, in that context, 
the DLA is generally not a good approximation  \cite{Iancu:2002tr,Mueller:2002zm,Munier:2003vc}.

Such considerations will allow us to recover the double--logarithmic corrections 
of Ref.~\cite{Liou:2013qya} from a more fundamental perspective 
and with a transparent physical interpretation.
An additional  clarification refers to the phase--space for the high--energy evolution: as we shall
discuss in Sect.~\ref{sec:DLAPS}, the original argument in that sense in Ref.~\cite{Liou:2013qya} must be supplemented with the kinematical constraint $p^0>p_\perp$. This has consequences when
the medium is a weakly--coupled QGP (more generally, whenever $\hat q \lambda^3\ll 1$):  
in that case, the validity of the high--energy approximations requires the stronger constraint 
${L}/{\lambda}\gg {1}/{\alpha_s^{2}\ln(1/\alpha_s)}$ (and not just ${L}/{\lambda}\gg {1}$).
In more suggestive terms, the necessary condition can be written as $Q_s^2 \gg T^2$.

As a further application, we consider in Sect.~\ref{sec:eloss} the evolution of
the radiative energy loss, within the framework of the BDMPSZ mechanism for medium--induced
gluon radiation \cite{Baier:1996kr,Baier:1996sk,Zakharov:1996fv,Zakharov:1997uu,Baier:1998yf,Baier:1998kq,Wiedemann:2000za,Wiedemann:2000tf,Arnold:2001ba,Arnold:2001ms,Arnold:2002ja}. 
Within the approximations of interest, this problem is closely related
to that of the transverse momentum broadening and in Sect.~\ref{sec:eloss} 
we shall merely emphasize the differences. Once again, the cross--section (and its evolution) 
can be related to the dipole $S$--matrix, which obeys the equations established in Sect.~\ref{sec:JET}.
The new feature is that, now, the eikonal approximation fails not only for the soft gluon fluctuation
responsible for the evolution, but also for its relatively hard parent gluon, which is 
responsible for the energy loss. Yet, this failure poses no difficulty for the calculation of the
high--energy evolution, because of the strong separation in lifetime between
the fluctuations and the radiation. In particular, to double--logarithmic accuracy, the evolution of the
radiative energy loss is obtained by simply using the renormalized value of $\hat q$ 
(the solution to \eqn{DLAint}) within the respective formula at tree--level. Similar conclusions
have been independently reached in
Ref.~\cite{Blaizot:2014}, where \eqn{DLAint} has been obtained via a different method
(namely, via the direct calculation of the relevant Feynman graphs to DLA accuracy).

Finally, Sect.~\ref{sec:conc} summarizes our results and conclusions, together
with some open problems. 

\section{The evolution Hamiltonian in the high--energy approximation}
\label{sec:H}

Throughout this paper, we shall consider the high--energy evolution of the scattering 
amplitude for the collision between a dilute projectile and a dense target. The projectile
is a set of partons in an overall color singlet state (the prototype being a color dipole),
while the target can be either a large nucleus, or the dense partonic medium created
in the intermediate stage of an ultrarelativistic heavy ion collision. In both cases, the
target is characterized by a dense gluon distribution, which for the present
purposes will be described in the spirit of the CGC formalism, that is, as a 
random distribution of strong, classical, color fields. The interactions between the
projectile and the target will be treated in a generalized eikonal approximation,
which allows one to resum the multiple scattering between the partons in the
projectile and the strong color fields in the target to all orders, via Wilson lines,
while also keeping trace of the transverse motion of the partons.

One step in the high--energy evolution consists
in the emission of a relatively soft gluon by one of the partons in the projectile and in the
background of the target field. Such an emission modifies the partonic content
of the projectile and hence the $S$--matrix for the elastic scattering between the projectile 
and the target. In this section we shall present and motivate a rather compact expression 
for the Hamiltonian which `generates this evolution', that is, which describes the change 
in the $S$--matrix induced by one soft gluon emission. A more formal derivation of this 
Hamiltonian from the QCD path integral is given in Appendix~\ref{sec:deriv}.

\subsection{The evolution Hamiltonian}
\label{sec:DeltaH}

To be specific, let us assume that the projectile propagates in the positive $x^3$ direction 
and introduce light--cone (LC) vector notations: $x^\mu=(x^+,x^-,\bx)$, with
$x^+=(x^0+x^3)/\sqrt{2}$, $x^-=(x^0-x^3)/\sqrt{2}$, and $\bx=(x^1,x^2)$. Each parton
in the projectile has a color current oriented in the LC `plus' direction, which couples to
the $A^-_a$  component of the target color field. If the parton energy is sufficiently
high (see below for the precise condition), its transverse coordinate $\bx$ is
not affected by the interaction. Then the only effect of the latter is a rotation of 
the parton color state, as encoded in the Wilson line :
 \begin{align}\label{Udef}
 U^{\dagger}(\bx) = \rmP \exp\left\{\rmi g \int \dif x^+ \,A^-_a(x^+, {\bx}) T^a\right\}.
 \end{align}
The $T^a$'s are the color group generators in the appropriate representation
and P stands for path ordering w.r.t. $x^+$ (the LC `time' for the projectile) : 
with increasing $x^+$, matrices are ordered from right to left. 
The integral over $x^+$ formally extends along the whole real axis, but in practice it is limited 
to the support of the target field. The $x^-$ coordinate has been omitted in \eqn{Udef} since it is 
understood that $x^-\simeq 0$ for the ultrarelativistic projectile, by Lorentz contraction. 

The elastic $S$--matrix for a color--singlet projectile involves the trace of a product of such 
Wilson lines, one for each parton (quark, antiquark, or gluon) in the projectile.
For more clarity, in what follows we shall keep the notations $T^a$ and $U^{\dagger}$ 
for the color group generators and the Wilson lines in the adjoint representation, 
and use $t^a$ and respectively $V^{\dagger}$ for quarks in the fundamental representation.
As anticipated, most of the examples below will refer to a color dipole,
for which the $S$--matrix reads (in the fundamental representation, for definiteness)
 \begin{align}\label{Sdip} \hat{S}_{\bx\by} \equiv  \frac{1}{N_c} \,
\rmtr\big[ V^{\dagger}_{\bx}{V}_{\by} \big]\,,
\end{align}
where $\bx$ and $\by$ are the transverse coordinates of the quark and the antiquark, respectively,
and $V^{\dagger}_{\bx}\equiv V^{\dagger}({\bx})$, etc.
This dipole enters the calculation of a variety of physical processes, like the total cross-section 
for deep inelastic scattering, the cross--section for single inclusive hadron production in
proton--nucleus ($pA$) collisions, 
or the transverse momentum broadening of a `hard probe' (here an energetic quark) propagating
through the dense partonic medium (`quark--gluon plasma')
created at the intermediate stages of a nucleus--nucleus ($AA$) collision. 

Below we shall refer to \eqn{Udef} as the {\em strict} eikonal approximation. 
For a quantum particle, this
is correct only so long as the target is `sufficiently thin' --- namely, so long as the  
duration of the interaction (which is the same as the extent $L$ of the target in the $x^+$ direction)  
is small enough for the effects of the quantum diffusion to remain negligible. Indeed, a high
energy particle with longitudinal momentum $p^+$ is similar to a non--relativistic quantum
particle with mass equal to $p^+$ and living in two spatial dimensions, in that it undergoes a Brownian 
motion in the transverse plane: the dispersion $\Delta x_\perp^2$  in its transverse position 
grows with time according to $\Delta x_\perp^2\simeq \Delta x^+/2p^+$. 
(This transverse dynamics is explicit in \eqn{Geq} below.) The dispersion thus accumulated during the
interaction time $\Delta x^+=L$ can be neglected so long as it remains smaller than the 
respective quantum uncertainty $1/p_\perp^2$ (with $p_\perp=|\bp|$ the particle 
transverse momentum). This requires\footnote{In evaluating
the coherence time $\tau_{coh}$ one should use the maximal value of $p_\perp$ accumulated 
by the particle via rescattering in the target, that is, the saturation momentum $Q_s$ to
be later introduced.} $L < \tau_{coh}\equiv 2p^+/p_\perp^2$, a condition which is indeed 
satisfied when the target is a shockwave, but not also for the case of an extended medium. 
Hence, in the case of the medium, we shall need the generalization of \eqn{Udef} to an arbitrary
trajectory $\bx(t)$ in the transverse plane, where $t\equiv x^+$ is the LC time. This reads
 \begin{align}\label{Ugen}
 U^{\dagger}_{t_2t_1}[\bx(t)] = \rmP \exp\left\{\rmi g \int_{t_1}^{t_2} \dif t \,A^-_a\big(t, {\bx}(t)\big) T^a\right\},
 \end{align}
and is a functional of the trajectory. As compared to \eqn{Udef}\, 
we have also generalized the definition
in \eqn{Ugen} to trajectories which start at some generic (light--cone)
time $t_1$ and end up at a later time $t_2$.

We are now in a position to present the operator which generates the emission of a soft 
gluon by the dilute projectile in the presence of the strong color field of the target.
This operator acts on gauge--invariant operators built with products of Wilson lines, like that 
in \eqn{Sdip}, and reads
 \begin{align}\label{DeltaH}
\hspace*{-.5cm}
 \Delta H \,=\,\frac{1}{2}\int\limits_{\rm strip}\frac{\rmd p^+}{2\pi}\int \rmd t_1
 \int \rmd t_2 \int \rmd^2\br_2  \int \rmd^2\br_1 \ J^a(t_2,\br_2)\,G^{--}_{ab}(t_2,\br_2; t_1,\br_1; p^+)
 \,J^b(t_1,\br_1)\,,
 \end{align}
in notations to be explained below.

The variable $p^+$ is the LC longitudinal momentum of the
emitted gluon; by assumption this is much smaller than the respective momentum of the parent
parton (to be below denoted as $\Lambda$), but much larger than any `plus' component 
that can be transferred by the
target in the collision process. Accordingly, the component $p^+$ is conserved by the interactions, 
which makes it useful to use the mixed Fourier representation $(t,\bx, p^+)$, as we did above.
The `strip integral' in \eqn{DeltaH} runs over an interval in $p^+$ which
is symmetric around $p^+=0$~:
 \begin{align}\label{strip}
 \int\limits_{\rm strip}\frac{\rmd p^+}{2\pi}\,f(p^+)\,\equiv\,\left(\ \int\limits_{x\Lambda}^\Lambda+
 \int\limits_{-\Lambda}^{-x\Lambda}\ \right)\frac{\rmd p^+}{2\pi}\,f(p^+)\,=\,
 \int\limits_{x\Lambda}^\Lambda\frac{\rmd p^+}{2\pi}\,\big[f(p^+)+f(-p^+)\big]\,, \end{align}
Here $\Lambda$ is the typical `plus' momentum of the emitters, which is the relevant `hard' scale,
whereas $x$, with $x\ll 1$, is the smallest longitudinal fraction of the emitted, `soft', gluon. 
In what follows, we shall be mostly concerned with situations where the above integral is logarithmic,
$\int (\rmd p^+/p^+)$; in such a case, the evolution operator takes of the form $\Delta H = H_{\rm evol}
\ln(1/x)$, with $H_{\rm evol}$ playing the role of a Hamiltonian for the evolution with `time' $Y\equiv
\ln(1/x)$ (the rapidity difference between the valence partons in the projectile and the softest
evolution gluons).
 
Furthermore, $J^a(t,\br)$ denotes the functional derivative w.r.t. the component $A^-_a(t, {\br})$
of the gauge field and plays the role of the color charge density operator.
When acting on a Wilson line like that in \eqn{Ugen}, this operator generates the emission
of a soft gluon from the parton represented by that Wilson line:
  \begin{align}\label{Jdef}
  J^a(t,\br) U^{\dagger}_{t_2t_1}[\bx] &\,\equiv\,\frac{\delta }{\delta A^-_a(t, {\br})}
  U^{\dagger}_{t_2t_1}[\bx]\nn
 &   \,=\,\rmi g\theta(t_2-t)\theta(t-t_1)\delta^{(2)}\big(\br-\bx(t)\big)\,
   U^{\dagger}_{t_2 t}[\bx] \,T^a\,U^{\dagger}_{t t_1}[\bx]\,.
 \end{align}
As  visible on this equation, each functional derivative brings a factor of $g$, so
$\Delta H$ starts at order $g^2=4\pi\alpha_s$ (but in general includes effects of higher
order in $g$, via the background field; see below). The operator $J^a(t,\bx)$ is also the generator
of the infinitesimal color rotations. Using \eqref{Jdef}, one can check the following equal--time 
commutation relation (with $f^{abc}$ the structure constants for SU$(N_c)$ and $\delta_{\bx\by}
\equiv \delta^{(2)}(\bx-\by)$)
 \begin{align}\label{Liecom}
[ J^a(t,\bx),\, J^b(t,\by)]=- g \delta_{\bx\by} f^{abc}J^c(t,\bx)\,,
\end{align}
which confirms that these operators obey the color group algebra, as they should.

The last ingredient in \eqn{DeltaH} is the background field propagator $G^{--}$ 
of the emitted gluon. This is a functional of the target field $A^-$, via Wilson lines.
Its construction is well documented in the literature and will
be briefly discussed in Appendix \ref{sec:G}, where we show that
 \begin{align}\label{G--}
G^{--}_{ab}(x^+,\bx; y^+,\by; p^+)=\,\frac{1}{(p^+)^2}\,\del^i_{\bx}\del^i_{\by}\,
G_{ab}(x^+,\bx; y^+,\by; p^+)+\,\frac{\rmi}{(p^+)^2}\,\delta_{ab}
\delta(x^+-y^+)\delta_{\bx\by} \,.
\end{align}
Here, $G_{ab}$ is the `scalar' propagator, defined as the
solution to the following equation 
 \begin{align}\label{Geq}
\big[2\rmi p^+\big(\del^-_{x}-\rmi gA^-(x)\big)+\nabla^2_{\perp\,x}\big]_{ac} 
G_{cb}(x^+,\bx; y^+,\by; p^+)\,=\,
\rmi \delta_{ab}\delta(x^+-y^+)\delta_{\bx\by}\,,
\end{align}
with Feynman prescription for the pole at the mass--shell.
This prescription ensures that modes with positive (negative) values of $p^+$ 
propagate forward (backward) in time (see e.g. \eqn{G0m}).
For definiteness, we shall refer to the two pieces in the r.h.s. of \eqn{G--} 
as the `radiative piece' and respectively the `Coulomb piece' of the gluon propagator.

\comment{The latter becomes
important for temporal gradients $\del^-\sim \lap/2p^+$, i.e. when the characteristic time scale for 
variations in the system is comparable with the coherence time $\tau_{coh}= 2p^+/p_\perp^2$.
This is in agreement with the discussion above \eqn{Ugen}.}

\eqn{Geq} exhibits the eikonal coupling between the large component $p^+$ of the
4--momentum of the gluon and the conjugate component $A^-$ of the color field of the target,
and also the transverse dynamics responsible for quantum diffusion.
Given the formal analogy between this equation and the Schr\"odinger equation 
for a non--relativistic particle in two spatial dimensions, it is clear that
its solution can be written as a path integral. 
Namely, for $p^+>0$ and hence $x^+>y^+$, one has\footnote{The `reduced propagator' $\mcal{G}$ is formally the same as the non--relativistic evolution operator.}
  \begin{align}
  \label{Gmedium}
  G(x^+,\bx; y^+,\by; p^+)&\,=\,\frac{1}{2p^+}\,\mcal{G}(x^+,\bx; y^+,\by; p^+)\,,\nn
  \mcal{G}(x^+,\bx; y^+,\by; p^+)&\,=
  \int\big[\mcal{D}\br(t)\big]
  \ \exp\bigg\{\rmi \,\frac{p^+}{2}
  \int_{y^+}^{x^+}\rmd t \,\dot\br^2(t)\bigg\}\,U^{\dagger}_{x^+y^+}[\br(t)] \,,\end{align}
where one integrates over paths $\br(t)$ with boundary conditions $\br(y^+)=\by$ and $\br(x^+)=\bx$. 
For $p^+<0$ (and hence $x^+<y^+$), the propagator can be computed
by using the following symmetry property, which follows from \eqn{Geq} together with the Feynman prescription:
  \begin{align}\label{Gsym}
G_{ab}(x^+,\bx; y^+,\by; p^+)\,=\,G_{ba}(y^+,\by; x^+,\bx; -p^+)
 \,,\end{align}  
By exploiting the above properties, one can limit the time
integrals in \eqn{DeltaH} to $-\infty < t_1 < t_2 <\infty$, while simultaneously restricting
the $p^+$ integral to the positive side of the strip, $x\Lambda < p^+ < \Lambda$, and
multiplying the result by two. More precisely, we have here in mind the integral
over the  `radiation' piece of the propagator \eqref{G--}, which is non--local in time.
The local, Coulomb, piece must be treated separately.

Note finally that  there is no ambiguity concerning the ordering of the various factors
within the integrand of \eqn{DeltaH}  :  \texttt{(i)}
the two charge operators act at different times, $t_1$ and $t_2$, so they commute with each other; 
\texttt{(ii)} the radiation piece of the propagator involves the background 
field $A^-(t)$ only at intermediate times $t$, between $t_1$ and $t_2$, so it commutes
with any of the two functional derivatives; \texttt{(iii)} the Coulomb piece is local not only in
time, but also in color.

The structure of the evolution Hamiltonian \eqref{DeltaH} looks both simple and intuitive: 
this operator does precisely what it is expected to do, namely, it generates the evolution of 
an $S$--matrix like \eqref{Sdip} via the emission and the reabsorption of a soft gluon by any 
of the color sources within the projectile. But this apparent simplicity hides several
subtleties which show up when trying to use this Hamiltonian in practice.
These subtleties will be discussed in the next subsection, where we shall 
derive an alternative form for the evolution Hamiltonian --- more precisely, for its
action on a generic operator $\hat{\mcal{O}}[A^-]$ --- which is more
convenient in practice, especially for an extended target.

\subsection{Virtual corrections and probability conservation}
\label{sec:prob}

The purpose of this subsection is to render the Hamiltonian \eqref{DeltaH} `less formal'.
First, we shall argue that, in order to be well defined, this operator must be supplemented 
with an adiabatic prescription for switching off the interactions at large times. Second,
we shall discuss a sum--rule for the free LC gauge propagator, which ensures probability
conservation and also the cancellation of ultraviolet and infrared divergences between 
the `radiative' piece and the `Coulomb' piece of the Hamiltonian. 
Finally, we shall derive an alternative expression
for the action of $\Delta H$ where this cancellation occurs locally in time and
probability conservation becomes manifest.

Throughout this paper, we shall assume that the target is localized in $x^+$, within the
longitudinal\footnote{An interval $\Delta x^+$
is `longitudinal' from the viewpoint of the target (a left
mover), but `temporal' from that of the projectile (a right mover). In what follows, we shall
often mix the two viewpoints and the respective terminologies.
The precise meaning should be clear from the context.}
strip at $0<x^+<L$, so the collision has a finite duration $\Delta x^+\sim L$.
The scattering amplitude can only be affected by gluon emissions
which occur sufficiently close to this interaction region, within a time interval $\Delta x^+
\sim \tau_{coh}$. (We recall that the `coherence time' 
$\tau_{coh}\equiv 2p^+/p_\perp^2$ is the typical lifetime of the fluctuation.)  Vice--versa,
virtual fluctuations in the wave function of the projectile which occur very far away from
the interaction region, either in the remote past or the remote future, should have no influence
on the evolution of the $S$--matrix. As we shall see, this property is correctly encoded
in the present formalism, but it involves delicate cancellations between various terms,
which might be invalidated by careless manipulations at intermediate stages.
It turns out that a proper way to deal with this problem is to adiabatically switch off
the interactions at very large times $|x^+| \gg \tau_{coh}$
\cite{Chen:1995pa,Mueller:2012bn}. (Other, less smooth, prescriptions, like a sharp cutoff 
on  $|x^+|$, could induce spurious radiation and thus alter the Fock space
of the projectile.) 
To that aim, we shall supplement each functional derivative within $\Delta H$ 
with an exponential attenuation factor,
\begin{align}
J^a(t,\br)\,\to\,J^a(t,\br)\,\rme^{-\epsilon |t|}\,,\end{align}
where $\epsilon$ should be much smaller than $1/\tau_{coh}$.
The physical predictions will not be sensitive to the precise value of $\epsilon$ because
the limit $\epsilon\to 0$ of the final results, as obtained after performing the integrals 
over the emission times $t_1$ and $t_2$,  is indeed well defined. 

With this adiabatic switch--off, the free LC gauge propagator $G_0^{--}$, \eqn{G--0}, obeys the
following sum--rule, with paramount consequences for what follows:
\begin{align}\label{ident}
 \int \rmd t_1
 \int \rmd t_2\, G_{0}^{--}(t_2-t_1,\br; p^+)\,\rme^{-\epsilon(|t_1|+|t_2|)}
 \,=\,0\,.
 \end{align}
This will be demonstrated in Appendix~\ref{sec:ident}, where we show that the l.h.s. 
of \eqn{ident} is a quantity of $\order{\epsilon}$ and hence vanishes when $\epsilon\to 0$.
A simple way to understand this cancellation is to notice that
the integral over $\Delta t\equiv t_2-t_1$ isolates the Fourier component with $p^-=0$,
which vanishes because $G_{0}^{--}(p)\propto p^-$, cf. \eqn{G--0}. But this
property holds only for the {\em complete} propagator, $G_{0}^{--}=G_{0,rad}^{--}+G_{0, Coul}^{--}$,
as obtained after adding its radiative and Coulomb pieces. In the presence of a background field,
we have to distinguish between these two pieces, since they are differently dressed by
the background, cf. \eqn{G--}. Taken separately, the  radiative piece $G_{0,rad}^{--}$ and the  
Coulomb piece $G_{0,Coul}^{--}$ generate contributions $\propto 1/\epsilon$ to the l.h.s. of
\eqn{ident}, which however cancel, together with the finite terms of $\order{1}$, in their sum
(see Appendix~\ref{sec:ident}).

In view of the above, the sum--rule \eqref{ident} is expected to be important for the limit
$A^-\to 0$ of our formalism. In that limit, it ensures an important property, that we now explain.
As previously mentioned, quantum
fluctuations which are not measured by the collision should not matter for the evolution
of the $S$--matrix. Consider in particular the situation where,
after acting with $\Delta H$ on some generic $S$--matrix
$ \hat{\mcal{O}}$ (to produce the fluctuation), one sets $A^-=0$, so that there is no scattering. 
Without scattering, the evolution cannot be  measured (the $S$--matrix must be equal to one
both before and after the evolution), hence the action of $\Delta H$ must vanish~:
 \begin{align}\label{nullDH}
 \Delta H\, \hat{\mcal{O}} \,\big|_{A^-=0}\,=\,0\,.\end{align}
This is precisely ensured by the identity \eqref{ident}, as it can be easily seen: 
the action of the functional derivatives
on $\hat{\mcal{O}}$ becomes independent of time after we set $A^-= 0$ (since all the Wilson
lines are replaced by the unity matrix).
Accordingly, the result of first acting with  $\Delta H$ on any $ \hat{\mcal{O}}$ and then letting
$A^-\to 0$ is indeed proportional to the integral in the l.h.s. of \eqn{ident}.

These properties, Eqs.~\eqref{ident} and \eqref{nullDH}, allows one to compute the action
of $\Delta H$ on $ \hat{\mcal{O}}$ in an alternative way, where the Coulomb piece of the
propagator is not explicitly present anymore and the cancellation of would--be
divergent contributions occurs quasi--locally in time. Namely, \eqn{nullDH} implies,
with obvious notations,
  \begin{align}\label{DH2}
 \Delta H_{Coul}\, \hat{\mcal{O}} \,\big|_{A^-=0}\,=\,-
 \Delta H_{rad}\, \hat{\mcal{O}} \,\big|_{A^-=0}\,.\end{align}
Also, as we shall shortly demonstrate, the action of the 
Coulomb piece of the Hamiltonian on any observable $ \hat{\mcal{O}}$ amounts to
 \begin{align}\label{DHCoul}
 \Delta H_{Coul}\, \hat{\mcal{O}} \,=\,\Big(\Delta H_{Coul}\, \hat{\mcal{O}} \,\big|_{A^-=0}
 \Big)\, \hat{\mcal{O}}\,=\,-\Big(\Delta H_{rad}\, \hat{\mcal{O}} \,\big|_{A^-=0}
 \Big)\, \hat{\mcal{O}}\,,\end{align}
where the second equality follows after using \eqn{DH2}. By using the above, one can write
 \begin{align}\label{DH)}
 \Delta H\, \hat{\mcal{O}} \,=\,&\big[\Delta H_{rad}+\Delta H_{Coul}\big]\, \hat{\mcal{O}}
 \,=\,\Big[\Delta H_{rad} - \Big(\Delta H_{rad}\, \hat{\mcal{O}} \,\big|_{A^-=0}
 \Big)\Big]\, \hat{\mcal{O}}\,, \end{align}
or, less formally,
 \begin{align}\label{DH}
 \Delta H\, \hat{\mcal{O}} [A^-]\,=\,&
 \int\limits_{x\Lambda}^\Lambda\frac{\rmd p^+}{2\pi} 
   \int_{-\infty}^\infty\rmd t_2 \int_{-\infty}^{t_2}\rmd t_1 \ \rme^{-\epsilon(|t_1|+|t_2|)}
   \int \rmd^2\br_2  \int \rmd^2\br_1 \,
   \Big[\mathcal{H} \,-\, \Big(\mathcal{H}\, \hat{\mcal{O}} \,\big|_{A^-=0}
 \Big)\Big]\hat{\mcal{O}}
 \,, \end{align}
where $\mathcal{H}$ is a
Hamiltonian density built with the `radiation' piece of the propagator alone:
\begin{align}\label{calH}
\mathcal{H}(t_2,\br_2; t_1,\br_1; p^+)[A^-]\,\equiv\,
\frac{1}{(p^+)^2}\,J^a(t_2,\br_2)\,\Big[{\del^i_{\br_2}\del^i_{\br_1}}\, G_{ab}(t_2,\br_2; t_1,\br_1; p^+)
\Big]\,J^b(t_1,\br_1)\,.\end{align}
In \eqn{calH}, the transverse derivatives act only on the `scalar' propagator.
In particular,
\begin{align}\label{subtract}
\mathcal{H}\, \hat{\mcal{O}} \,\big|_{A^-=0}\,=\,
\frac{1}{(p^+)^2}\,\Big[{\del^i_{\br_2}\del^i_{\br_1}}\, G_{0}(t_2-t_1,\br_2-\br_1; p^+)
\Big]\,\Big(J^a(t_2,\br_2)J^a(t_1,\br_1)\hat{\mcal{O}} \,\big|_{A^-=0}\Big)\,,
\end{align}
with $G_0$ the free propagator  \eqref{G0m}.
Notice that the r.h.s. of \eqn{DH} cannot be written as the action of a linear operator
on $\hat{\mcal{O}}$. Hence, this equation does not provide an alternative expression for
the Hamiltonian $ \Delta H$, but rather a new method for computing
its action on a generic observable.  

Using $\hat{\mcal{O}}|_{A^-=0} =1$, one sees that
the property \eqref{nullDH} is now satisfied {\em locally} in time, that is, it is already verified by 
the integrand in \eqn{DH}. This allows for a natural probabilistic interpretation: the term
$\mathcal{H}\hat{\mcal{O}}$ describes the change in the $S$--matrix associated with a {\em real}
emission which occurrs during the time interval from $t_1$ to $t_2$; the {\em virtual} term 
$- \big(\mathcal{H}\, \hat{\mcal{O}} \,|_{A^-=0}\big)\hat{\mcal{O}}$ represents the 
{\em reduction} in the probability that the projectile remain in its original state during that
time interval. The local (in time) version of  \eqref{nullDH} is then  the expression
of probability conservation.

To better appreciate the advantages of \eqn{DH}
over the direct use of \eqn{DeltaH}, let us consider the action of $\Delta H_{Coul}$
in more detail. (This will also allow us to verify the first equality in \eqn{DHCoul}.)
What we would like to show is that any operator $\hat{\mcal{O}}$ is an eigenstate
of $\Delta H_{Coul}$, but with an ill--define eigenvalue, which suffers from both
infrared (large time and small $p^+$) and ultraviolet (small $|\br_2-\br_1|$, or high $p_\perp$)
divergences. Chosing $\hat{\mcal{O}}=\hat{S}_{\bx\by}$ for definiteness
(this brings no loss in generality), we can write (cf. \eqn{G--})
\begin{align}\label{CoulO}
 \Delta H_{Coul}  \hat{S}_{\bx\by} \,=\,& \int\limits_{x\Lambda}^\Lambda\frac{\rmd p^+}{2\pi} 
 \int\limits_{t_1, t_2}
 \int\limits_{\br_1,\br_2} 
  \rme^{-\epsilon(|t_1|+|t_2|)}
  \frac{\rmi}{(p^+)^2}
  \,\delta_{t_2t_1}\delta_{\br_1\br_2}\,J^a(t_2,\br_2)\,J^a(t_1,\br_1)
  \,\hat{S}_{\bx\by}\,\nn
  \,=\,&
  -\frac{\rmi g^2 C_F}{2\pi} \int\limits_{x\Lambda}^\Lambda\frac{\rmd p^+}{(p^+)^2}
  \int\rmd t\,\rme^{-2\epsilon |t|}  \int \rmd^2\br \big(\delta_{\br\bx}+\delta_{\br\by}\big)
  \,\delta_{\br\br}
    \,\hat{S}_{\bx\by}\,\nn
    \,=\,&
  -\frac{\rmi g^2 C_F}{\pi}\bigg[\delta_{\br\br}\,\frac{1}{\epsilon} \,
   \int\limits_{x\Lambda}^\Lambda\frac{\rmd p^+}{(p^+)^2}\bigg] \,\hat{S}_{\bx\by}\,.
     \end{align}
Because of the ultra--local nature of the Coulomb propagator $\propto  \delta_{t_2t_1}\delta_{\br_1\br_2}$, the two functional derivatives must act
on a same Wilson line within $\hat{S}_{\bx\by}$, either the quark one at $\bx$ or the antiquark 
one at $\by$. This feature, together with identities like
 \beq
 J^a(t,\br_2)\,J^a(t,\br_1)\,V^{\dagger}_{\bx}\,=\,-g^2C_F\delta_{\br_1\bx}\delta_{\br_2\bx}
 \,V^{\dagger}_{\bx}\,,\eeq
explains why the result is again proportional to $\hat{S}_{\bx\by}$.
But for the very same reason,  the proportionality coefficient exhibits several types of
divergences, as anticipated: a large--time divergence as $\epsilon\to $, a small--$p^+$ 
divergence when $x\to 0$, and a transverse `tadpole' $\delta_{\br\br}=
 \int [\rmd^2\bp/(2\pi^2)]$.
Being independent of $A^-$, this coefficient is necessarily the same as the limit $A^-\to 0$
of  $\Delta H_{Coul}  \hat{S}_{\bx\by}$, in agreement with  \eqn{DHCoul}. 
Clearly, a similar argument holds for any observable $\hat{\mcal{O}}$.
     
The above discussion shows that the action of the Coulomb piece of $\Delta H$
generates severe divergences. By virtue of \eqn{ident}, there divergences are 
guaranteed to cancel against similar ones generated by the 
radiation piece, but only after performing the two time integrations. 
This cancellation can be explicitly verified whenever one is able to perform the time integrations,
as in the case of a shockwave target to be discussed in Sect.~ \ref{sec:JIM}. But even in such a
case, the calculation of the finite terms is quite subtle and relies in an essential way
on the use of the adiabatic prescription (see e.g. Sect.~ \ref{sec:JIMtime}).
By contrast, the calculations based on \eqn{DH} are more robust, because the potential
divergences cancel between the `real' and `virtual' terms quasi--locally in time, so one
is not sensitive to the regularization prescription used for the time integrations.
This second method becomes particularly useful in those cases 
where one is not able to explicitly perform the time integrals,
like that of an extended target to be discussed in Sect.~\ref{sec:JET}.

\section{A shockwave target: recovering the JIMWLK Hamiltonian}
\label{sec:JIM}

In this section, we shall specialize the general formalism developed so far to the case
where the target is a `shockwave'. By this, we more precisely mean a target which looks 
localized in $x^+$ on the resolution scale set by the lifetime of the quantum 
fluctuations. For this case, we will be able to explicitly 
perform the time integrations which appear in \eqn{DeltaH} and thus recover the
JIMWLK Hamiltonian \cite{JalilianMarian:1997jx,JalilianMarian:1997gr,JalilianMarian:1997dw,
Kovner:2000pt,Weigert:2000gi,Iancu:2000hn,Iancu:2001ad,Iancu:2001md,Ferreiro:2001qy}, as expected. Besides giving us more confidence with the use of \eqn{DeltaH} in practice,
the subsequent manipulations  will also illustrate some of the subtleties discussed
in Sect.~\ref{sec:prob}, notably the role of the adiabatic prescription and the cancellation
of the ill--defined contributions between the `radiation' piece
and the `Coulomb' piece of $\Delta H$.


More precisely, the physical problem that we here have in mind is  `dense--dilute'
(e.g. proton--nucleus) scattering in the high--energy regime where the longitudinal extent 
$\Delta x^+\equiv L$  of the dense target is much smaller than the coherence time
$\tau_{coh}={2p^+}/{p_\perp^2}$ of the typical gluons fluctuations associated with the
evolution of the projectile : $\tau_{coh}\gg L$.  This condition involves
both the `energy' (actually, LC longitudinal momentum) $p^+$ and the transverse
momentum $p_\perp$ of the gluon fluctuations. In practice,
$p_\perp$ is at least as large as the target saturation momentum $Q_s$, since this is
the typical transverse momentum acquired by either the soft gluon, or its parent
parton, via interactions with the target (see e.g.  \cite{Mueller:2001fv,Iancu:2002xk,Gelis:2010nm}). Hence, the `shockwave condition' can be
written as a lower limit on the gluon energy : $p^+\gg\omega_c$, with
\begin{align}\label{omegac0}
 \omega_c \,\equiv\,Q^2_s L\,.\end{align}
This limiting energy $ \omega_c$ is an intrinsic scale of the target and grows with the target 
size like $\omega_c\sim L^2$ (since $Q^2_s\propto L$). 
To have a significant phase--space for the high--energy evolution,
the energy $p_0^+\equiv E$ of the incoming projectile must be considerably larger 
than  $\omega_c$, namely such that $\abar\ln(E/\omega_c)\gtrsim 1$ with
$\abar\equiv \alpha_s N_c/\pi$ assumed to be small ($\abar\ll 1$).  


\subsection{Performing the time integrations}
\label{sec:JIMtime}

What is special about the shockwave (SW) target, is that the probability for a 
gluon to be emitted or absorbed inside the target is negligible\footnote{Strictly
speaking, this statement is gauge--dependent, but it is indeed correct in the
gauge $a^+=0$ that we currently use; see e.g. the discussion in \cite{Blaizot:2004wu}.}, 
since suppressed by a factor $L/\tau_{coh}\ll 1$.
This physical statement is boost invariant, but the mathematics becomes simpler by
working in the `target infinite momentum frame', i.e. a frame in which the nucleus is
ultrarelativistic and it looks like a `pancake' (our 
intuitive representation of a SW). In such a frame, the target can be effectively
treated as a $\delta$--function at $x^+=0$. This drastically simplifies 
the structure of the background field propagator and the action of the
functional derivatives on the Wilson lines.

Namely, assuming the SW to be localized near $x^+=0$, one can easily show that the
path integral in \eqn{Gmedium} reduces to (for $p^+>0$ and hence $x^+>y^+$~;
see Appendix~\ref{sec:G} for details)
 \begin{align}\label{GSW}
&G(x^+,\bx; y^+,\by; p^+>0)\,=\, G_{0}(x^+-y^+,\bx-\by; p^+)\big[\theta(x^+)\theta(y^+)
+\theta(-x^+)\theta(-y^+)\big]\nn
 &\qquad\qquad\qquad\qquad +2 p^+\theta(x^+)\theta(-y^+)
  \int_{\bz} G_{0}(x^+,\bx-\bz; p^+) \,U^\dagger_{\bz}\,G_{0}(-y^+,\bz-\by; p^+) \,,
 \end{align}
where $G_0$ is the free propagator \eqref{G0m}, $U^\dagger_{\bz}$ is the adjoint
Wilson line introduced in \eqn{Udef}, and $ \int_{\bz} 
\equiv  \int\rmd^2{\bz}$.
The physical interpretation of \eqn{GSW} is quite transparent: 
when $x^+$ and $y^+$ are both positive, or both negative, the gluon does not cross the SW,
so it propagates freely; when $x^+$ and $y^+$ are on opposite sides of the SW, 
the gluon propagates freely from the initial point up to the SW, then it crosses 
the latter at some transverse position $\bz$, thus accumulating a color precession 
represented by the Wilson line $U^\dagger_{\bz}$, then it
moves freely again, up to the final point. 

Furthermore, since gluons cannot be emitted or absorbed inside the SW,
the action of the functional derivative $J^a_{\bx} (t)$ on the Wilson lines is piecewise 
independent of time.
Indeed for any negative value of the time argument, one has (compare to \eqn{Jdef})
 \begin{align}\label{JR}
  J^a_{\bx} (t < 0)U^{\dagger}_{\bz}  &   \,=\,\rmi g\delta_{\bz\bx}\,
   U^{\dagger}_{\bz}(\infty, t) \,T^a\,U^{\dagger}_{\bz}(t,-\infty)
    \,=\,\rmi g\delta_{\bz\bx}\,
   U^{\dagger}_{\bz} \,T^a\,\equiv\,R^a_{\bx} \,U^{\dagger}_{\bz}
   \,,
 \end{align}
where we have used $U^{\dagger}_{\bz}(t,-\infty)=1$ and $U^{\dagger}_{\bz}(\infty, t)
=U^{\dagger}_{\bz}(\infty,-\infty)\equiv U^{\dagger}_{\bz}$ for $t<0$ and a target field 
localized at $x^+=0$.  Similarly, for a positive value $t>0$, one can write
  \begin{align}\label{JL}
  J^a_{\bx} (t>0)U^{\dagger}_{\bz}  &   \,=\,\rmi g\delta_{\bz\bx}\,
   U^{\dagger}_{\bz}(\infty, t) \,T^a\,U^{\dagger}_{\bz}(t,-\infty)
    \,=\,\rmi g\delta_{\bz\bx}\,
   T^a\,U^{\dagger}_{\bz} \,\equiv\,L^a_{\bx} \,U^{\dagger}_{\bz}
   \,,
 \end{align}
The  above equations have introduced the `right' and `left' functional derivatives, 
$R^a_{\bx} $ and respectively  $L^a_{\bx}$, which act  on the Wilson lines as
infinitesimal color rotations of the right, respectively on the left, and 
measure the color charge density in the projectile prior, respectively 
after, the collision. They are related by the condition
$L^a_{\bx}= U^{\dagger ab}_{\bx}R^b_{\bx}$, which expresses the color
rotation acquired by a color current which crosses the shockwave.

The fact that the r.h.s.'s of Eqs.~\eqref{JR}--\eqref{JL} are independent of time allows
us to perform the time integrations  
{\em directly at the level of the evolution Hamiltonian}
\eqref{DeltaH}, that is, before acting with $\Delta H$ on the observable. 
To that aim, we need to distinguish three regions for the time integrations:

\subsubsection*{(i) {\em $-\infty < t_1 < 0$ and $0 < t_2 < \infty$ : the evolution gluon crosses the 
SW}}

After using Eqs.~\eqref{JR}--\eqref{JL} for the action of the functional derivatives,
one sees that the respective contribution to $\Delta H$, denoted as  $\Delta H_{RL}$, simplifies to
 \begin{align}\label{DeltaHRL}
 \Delta H_{RL} \,=\, \int\limits_{\bx,\by}\,L^a_{\bx} \,R^b_{\by} 
 \int\limits_{x\Lambda}^\Lambda\frac{\rmd p^+}{2\pi}
  \frac{1}{(p^+)^2}\int_{-\infty}^0\rmd t_1 \int_0^{\infty}\rmd t_2\ 
  \del^i_{\bx}\del^i_{\by}\,
G_{ab}(t_2,\bx; t_1,\by; p^+)\,,
 \end{align}
where the adiabatic prescriptions are implicit (they will be exhibited when needed) and,
cf. \eqn{GSW}, 
  \begin{align}\label{GRL}
\del^i_{\bx}\del^i_{\by}\,
G_{ab}(t_2,\bx; t_1,\by; p^+)\,=\,
2 p^+ \int_{\bz} \del^i_{\bx}G_{0}(t_2,\bx-\bz; p^+) \,\big(U^{\dagger}_{\bz}\big)_{ab}
\,\del^i_{\by} G_{0}(-t_1,\bz-\by; p^+)\,.
  \end{align}
Due to the factorized structure of the background field propagator \eqref{GRL}, the two time integrations are independent of each other. To be specific, consider the integral over $t_2$.
This involves
\begin{align}\label{int2}
 \int_0^{\infty}\rmd t_2\, \del^i_{\bx}G_{0}(t_2,\bx-\bz; p^+)&\,=\,
 \frac{-\rmi}{2p^+}\int\frac{\rmd^2\bp}{(2\pi)^2}\,p^i\,\rme^{\rmi \bp\cdot(\bx-\bz)}
 \int_0^{\infty}\rmd t_2\ \rme^{-\rmi \frac{p_\perp^2}{2p^+}t_2}
 \ \rme^{-\epsilon t_2} \nn
 &\,=\,-\int\frac{\rmd^2\bp}{(2\pi)^2}\,\frac{p^i}
 {p_\perp^2}\ \rme^{\rmi \bp\cdot(\bx-\bz)}\,=\,\frac{\rmi}{2\pi}\,
 \frac{x^i-z^i}{(\bx-\bz)^2}\,.
  \end{align}
The final result is recognized as the Weizs\"acker--Williams field created at $\bz$
by a point-like source at $\bx$.  Note that the complex exponential in the integral over $t_2$
has restricted the respective phase--space to an interval of order
$\tau_{coh}={2p^+}/{p_\perp^2}$ after the SW.
A similar conclusion holds for the emission time $t_1$, which is restricted to an interval
$\sim \tau_{coh}$ before the SW. The respective integral yields
 \begin{align}\label{int1}
 \int_{-\infty}^0\rmd t_1\, \del^i_{\by}G_{0}(-t_1,\bz-\by; p^+)&\,=\,
 \frac{\rmi}{2\pi}\,
 \frac{y^i-z^i}{(\by-\bz)^2}\,.
  \end{align}
Importantly, the final results in Eqs.~\eqref{int2}--\eqref{int1} are independent
of $p^+$. In both cases, this is due to a cancellation between the factor $1/p^+$ implicit in the 
free propagator $G_0$ and the phase--space factor ${2p^+}/{p_\perp^2}$ produced
by the time integral. As a consequence, the ensuing integral over $p^+$ 
in \eqn{DeltaHRL} is {\em logarithmic}~: $\int ({\rmd p^+}/p^+)=\ln(1/x)$. Putting all together,
one finds
 \begin{align}\label{HRL}
 \Delta H_{RL} \,=\,-\ln\frac{1}{x} \ \frac{1}{(2\pi)^3}
 \int_{\bx\by\bz}
 \mcal{K}_{\bx\by\bz}\,\big(2 L^a_{\bx} \,U^{\dagger\,ab}_{\bz}\,R^b_{\by}\big)
\,,
  \end{align}
with the following notations:
   \begin{align}
 \mcal{K}_{\bx\by\bz}\,\equiv\, \mathcal{K}^i_{\bx\bz}\, \mathcal{K}^i_{\by\bz}\,,\qquad
  \mathcal{K}^i_{\bx\bz} \,\equiv\, \frac{(\bx - \bz)^i}{(\bx - \bz)^2}\,.
 \end{align} 

 \subsubsection*{(ii)
  $-\infty < t_1 \le t_2 < 0$ : {\em the evolution gluon is emitted and reabsorbed prior 
to the SW}}

In this case, both functional derivatives within $\Delta H$ act as `right' derivatives, 
cf. \eqn{JR}. Also, the gluon propagator reduces to the free propagator $G^{--}_{0}$, 
as shown in \eqn{G--0}. Consider first the `radiation' piece of this propagator, which gives
   \begin{align}
  \label{DeltaHRR}
 \Delta H_{RR}^{rad} \,=\,& \int\limits_{\bx,\by}\,R^a_{\bx} \,R^a_{\by} 
 \int\limits_{x\Lambda}^\Lambda\frac{\rmd p^+}{2\pi}
  \frac{1}{(p^+)^2}\int_{-\infty}^0\rmd t_2 \int_{-\infty}^{t_2}\rmd t_1\ 
  \del^i_{\bx}\del^i_{\by}\,
G_{0}(t_2-t_1,\bx-\by; p^+)\,.
  \end{align}
The time integrations involve (with
the shorthand notation $p^-\equiv {p_\perp^2}/{2p^+}$)
  \begin{align}\label{timeRR}
 \int_{-\infty}^0\rmd t_2 \int_{-\infty}^{t_2}\rmd t_1\ 
 \,\rme^{-\rmi p^-(t_2-t_1)}\,\rme^{\epsilon(t_1+t_2)}&\,=\,
  \int_{-\infty}^0\rmd t_2  \,\rme^{-\rmi p^- t_2+\epsilon t_2}\,
  \frac{\rme^{\rmi p^- t_2+\epsilon t_2}}{\rmi p^-+\epsilon}\nn
  &\,=\,\frac{1}{2\epsilon}\,
  \frac{1}{\rmi p^-+\epsilon}\,=\,\frac{1}{2\epsilon}\,
  \frac{1}{\rmi p^-}\,+\,\frac{1}{2(p^-)^2}\,+\,\order{\epsilon}\,.
  \end{align}
The use of the adiabatic prescription has been essential in obtaining the above result,
as we now explain. The  time separation 
$t_2-t_1$ is restricted by the oscillatory phase $\rme^{-\rmi p^-(t_2-t_1)}$ to values of order $\tau_{coh}={2p^+}/{p_\perp^2}$, but the central value $(t_2+t_1)/2$
is only restricted by the adiabatic switch--off, so the corresponding integral yields an `infrared' divergence proportional to $1/\epsilon$. By itself, this divergence is pretty harmless,
since ultimately cancelled by a similar contribution due to the Coulomb piece,
as we shall see. What is more subtle though, is the obtention of
the finite term accompanying the divergence (namely, the term $\propto 1/(p^-)^2$
in \eqn{timeRR}) : this term is
correctly computed when using the adiabatic prescription, as above, but it would be mistreated
by other regularizations, like a sharp cutoff on $|t_2+t_1|$ 
\cite{Chen:1995pa,Mueller:2012bn}. Importantly, this finite contribution, 
which is the actual physical result, has been generated by values $t_1$ and $t_2$ 
which both lie in the vicinity of the interaction time $x^+=0$, 
within a distance of order $\tau_{coh}$.

By using \eqn{timeRR} together with simple manipulations, one finds
  \begin{align}
  \label{DeltaHRR1}
 \Delta H_{RR}^{rad} \,=\,& \int\limits_{\bx,\by}\,R^a_{\bx} \,R^a_{\by} 
 \int\limits_{x\Lambda}^\Lambda\frac{\rmd p^+}{2\pi}
  \frac{1}{(p^+)^2}\left\{-\frac{\rmi}{2\epsilon}\,\delta_{\bx\by}\,+p^+
   \int\frac{\rmd^2\bp}{(2\pi)^2}\ \frac{\rme^{\rmi \bp\cdot(\bx-\by)}}{p_\perp^2}
   \right\}
 \,.
  \end{align}
The first term within the braces exhibits all types
of divergences previously identified in relation with the Coulomb piece, 
cf. \eqn{CoulO}. As demonstrated by the above calculation (and anticipated in 
Sect.~\ref{sec:prob}), such divergences are also generated by the `radiation' piece
after performing the time integrations.
We shall shortly check that this singular term is cancelled by the respective
`Coulomb' contribution, in agreement with the discussion in Sect.~\ref{sec:prob}.
Keeping only the second term in \eqn{DeltaHRR1},  one finds that the respective 
integral over $p^+$ is again logarithmic and yields
 \begin{align}\label{HRR}
 \Delta H_{RR} \,=\,\ln\frac{1}{x} \ \frac{1}{(2\pi)^3}
 \int_{\bx\by\bz}
 \mcal{K}_{\bx\by\bz}\,R^a_{\bx} \,R^a_{\by}
\,,
  \end{align}
 where we have also used the identity
 \begin{align}\label{intK}
 \int\frac{\rmd^2\bp}{(2\pi)^2}\ \frac{\rme^{\rmi \bp\cdot(\bx-\by)}}{p_\perp^2}\,=\,
  \frac{1}{(2\pi)^2}\int_{\bz}\,\mcal{K}_{\bx\by\bz}\,.\end{align}
The above integral over $p_\perp$  develops a logarithmic infrared ($p_\perp\to 0$)
divergence, which is however harmless, as it disappears in the evolution of gauge--invariant
quantities (see e.g. Sect.~\ref{sec:BK} below).
  
For completeness, let us also consider the respective Coulomb contribution:
  \begin{align}
  \label{HRRC}
 \Delta H_{RR}^{Coul} &\,=\,\rmi \int\limits_{\bx,\by}\,R^a_{\bx} \,R^a_{\by} 
 \int\limits_{x\Lambda}^\Lambda\frac{\rmd p^+}{2\pi}
  \frac{1}{(p^+)^2}\int_{-\infty}^0\rmd t_2 \int_{-\infty}^{0}\rmd t_1\ 
  \,\delta(t_2-t_1)\,\rme^{\epsilon(t_1+t_2)}\,\delta_{\bx\by}\nn
  &\,=\,\frac{\rmi}{2\epsilon} \int\limits_{\bx}\,R^a_{\bx} \,R^a_{\bx} 
 \int\limits_{x\Lambda}^\Lambda\frac{\rmd p^+}{2\pi}
  \frac{1}{(p^+)^2}\,.
  \end{align}
This precisely cancels the divergent piece in \eqn{DeltaHRR1}, as anticipated.
This cancellation illustrates a general argument developed in Sect.~\ref{sec:prob},
namely the fact that emissions which occur at large distances $\gg \tau_{coh}$ from
the interaction region cannot affect the scattering amplitude.

 \subsubsection*{(iii)
 $0 < t_1 \le t_2 < \infty$ : {\em the evolution gluon is emitted and reabsorbed after 
 the SW} }

\comment{
\bigskip
\texttt{(iii)} $0 < t_1 \le t_2 < \infty$ : {\em the evolution gluon is emitted and reabsorbed after 
 the SW} 
\bigskip}

The calculation is entirely similar to that in the previous case, so we can
write the final result without further discussion:
 \begin{align}\label{HLL}
 \Delta H_{LL} \,=\,\ln\frac{1}{x} \ \frac{1}{(2\pi)^3}
 \int_{\bx\by\bz}
 \mcal{K}_{\bx\by\bz}\,L^a_{\bx} \,L^a_{\by}
\,,
  \end{align}

By combining the previous results \eqref{HRL}, \eqref{HRR}, and \eqref{HLL},
one finds $\Delta H=\ln(1/x) H_{_{\rm JIMWLK}} $, with the JIMWLK Hamiltonian 
\cite{JalilianMarian:1997jx,JalilianMarian:1997gr,JalilianMarian:1997dw,
Kovner:2000pt,Weigert:2000gi,Iancu:2000hn,Iancu:2001ad,Iancu:2001md,Ferreiro:2001qy}
(see also \cite{Mueller:2001uk,Jeon:2013zga,Caron-Huot:2013fea,Binosi:2014xua} for more recent derivations).
 \begin{align}\label{JIMWLK}
 H_{_{\rm JIMWLK}} \,=\,\frac{1}{(2\pi)^3}
 \int_{\bx\by\bz}
 \mcal{K}_{\bx\by\bz}\,\big[R^a_{\bx} \,R^a_{\by}\,+\,L^a_{\bx} \,L^a_{\by}\,-\,
 2 L^a_{\bx} \,U^{\dagger\,ab}_{\bz}\,R^b_{\by}\big]
\,. \end{align}
By using the unitarity of the Wilson
lines together with the condition $L^a_{\bx}= U^{\dagger ab}_{\bx}R^b_{\bx}$, one can
rewrite the color structure in  \eqn{JIMWLK} in the following form
 \begin{align}\label{HW}
 R^a_{\bx} \,R^a_{\by}\,+\,L^a_{\bx} \,L^a_{\by}\,-\,
 2 L^a_{\bx} \,U^{\dagger\,ab}_{\bz}\,R^b_{\by} & \,=\,
  \big[L^a_{\bx} - U^{\dagger ab}_{\bz} R^b_{\bx}\big]
 \big[L^a_{\by} - U^{\dagger ac}_{\bz} R^c_{\by}\big]\nn
 & \,=\,
  \big[U^{\dagger ab}_{\bx} - U^{\dagger ab}_{\bz}\big] R^b_{\bx}
 \big[U^{\dagger ac}_{\by} - U^{\dagger ac}_{\bz}\big]R^c_{\by}\,.
  \end{align}
This makes it obvious that  $H_{_{\rm JIMWLK}}$ vanishes when $A^-=0$,
in agreement with \eqn{nullDH}.


\subsection{The Balitsky--Kovchegov equation}
\label{sec:BK}

The simplest among the evolution equations generated by the JIMWLK Hamiltonian is
the Balitsky--Kovchegov (BK) equation \cite{Balitsky:1995ub,Kovchegov:1999yj}, that is,
the equation obeyed by the average $S$--matrix for a $q\bar q$ dipole.
In what follow we shall present two different derivations for this equation:
the standard one in the literature, where one directly acts with  $H_{_{\rm JIMWLK}}$ on the dipole
operator $\hat{S}_{\bx\by}$, and the alternative one based on \eqn{DH}, which distinguishes
between `real' and `virtual' corrections. Clearly, the final result will be the same, but the 
comparison between these two methods will shed more light on the reorganization of the
perturbation theory performed by the sum--rule \eqref{ident} and also on the
origin of the virtual terms in the B--JIMWLK equations.

\subsubsection*{(i) {\em The standard approach}}

Consider the dipole--nucleus scattering in a Lorentz frame where the nuclear target carries
most of the rapidity separation $Y$, so that the projectile is a {\em bare} dipole --- a quark--antiquark
pair without additional gluons. In this frame, the average $S$--matrix is computed as
\cite{Iancu:2002xk,Gelis:2010nm}
\begin{align}
 \label{aveS}
  \lan \hat{S}_{\bx\by} \ran_Y = \int [DA^-]\, W_{Y}[A^-]\,  \frac{1}{N_c} \,
\rmtr\big( V^{\dagger}_{\bx}{V}_{\by} \big)\,,
 \end{align}
where the `CGC weight function' $W_{Y}[A^-]$ is a functional probability density
describing the distribution of the color fields in the target (including its evolution
up to rapidity $Y$). Let us now increase the rapidity separation, $Y\to Y+\Delta Y$,
by giving an additional boost $\Delta Y$ to the projectile. Then the dipole evolves by emitting a 
soft gluon (from either the quark, or the antiquark), with longitudinal momentum fraction $x_1$ within
the range $x< x_1<1$, where $\Delta Y=\ln 1/x$. The ensuing evolution of the $S$--matrix is
obtained by acting with the JIMWLK Hamiltonian on the bare scattering operator:
\beq
 \Delta\big\langle\hat{S}_{\bx\by}\big\rangle_Y
 \,=\,\Delta Y\big\langle H_{_{\rm JIMWLK}}\hat{S}_{\bx\by}\big\rangle\,\equiv\,
 \int [DA^-]\, W_{Y}[A^-]\, H_{_{\rm JIMWLK}}\hat{S}_{\bx\by}\,.
 \eeq
Using \eqn{JIMWLK} together with the differentiation rules in Eqs.~\eqref{JR}--\eqref{JL},
one can easily deduce 
 \begin{align}\label{virt}
(H_{RR}+H_{LL})\, \hat{S}_{\bx\by} = -\frac{\abar}{2\pi}\, \left(1 -
\frac{1}{N_c^2} \right) \int_{\bz} \mcal{M}_{\bx\by\bz}
\hat{S}_{\bx\by}.
\end{align}
for the contribution of the `non--crossing' terms and, respectively,
\begin{align}\label{real}
 H_{RL}\, \hat{S}_{\bx\by}
 &= \,\frac{\alpha_s}{\pi^2}\,
 \int_{\bm{z}}
 \mcal{M}_{\bx\by\bz}\
 U_{\bz}^{\dagger ab}
 \,\frac{1}{N_c}\,
 \rmtr \Big(V_{\bx}^{\dagger} t^b\, V_{\by} t^a  \Big)
 \nn
 &=
 \atpi\,
 \int_{\bm{z}}
 \mcal{M}_{\bx\by\bz}
 \Big(\hat{S}_{\bx\bz} \hat{S}_{\bz\by}
 -\frac{1}{N_c^2}\, \hat{S}_{\bx\by}\Big),
 \end{align}
for that of the `crossing' one. In these equations, $ \mcal{M}_{\bx\by\bz}$ is the `dipole kernel',
 \begin{align}\label{Mdef}
 \mcal{M}_{\bx\by\bz}\,\equiv\,  \mcal{K}_{\bx\bx\bz}+ \mcal{K}_{\by\by\bz}
 -2 \mcal{K}_{\bx\by\bz}\,=\,
 \frac{(\bm{x}-\bm{y})^2}
 {(\bm{x}-\bm{z})^2(\bm{z}-\bm{y})^2}\,.\end{align}
In the linear combination above, the positive terms $\mcal{K}_{\bx\bx\bz}$ and $\mcal{K}_{\by\by\bz}$
correspond to self--energy corrections, i.e. graphs where both emissions are attached to a same fermion
(the quark at $\bx$ or the antiquark at $\by$), whereas the negative term $-2 \mcal{K}_{\bx\by\bz}$
summarizes the two exchange graphs, where the gluon is emitted by the quark and absorbed by
the antiquark, or vice--versa (see also Fig.~\ref{fig:dipole} for similar graphs). Note that
the leading behavior at large $z_\perp$, which is proportional to $1/z_\perp^2$ for each of these
individual graphs, has cancelled in their linear combination, with the result that
$\mcal{M}_{\bx\by\bz}\sim 1/z_\perp^4$ when $z_\perp\to\infty$. This decay is sufficiently fast 
 to guarantee that the integral over $\bz$ is convergent in this limit. 
Similar cancellations occur for
any projectile which is a color singlet and ensure that the respective evolution equation is free
of infrared problems \cite{Hatta:2005as}.

 \comment{These results are easy to understand.
The `non--crossing' contributions in \eqn{virt} are proportional to the $S$--matrix
$\hat{S}_{\bx\by}$ of the original dipole because the gluon fluctuation
does not exist at the time of scattering $x^+=0$. The `crossing' piece as written 
in the first line of \eqn{real} describes the scattering of the three parton system
made with the original quark--antiquark pair and the evolution gluon. }

The second line in \eqn{real}
follows after reexpressing the adjoint Wilson line in terms of fundamental ones, 
according to $U_{\bz}^{\dagger ab}t^b=V_{\bz} t^a V_{\bz}^{\dagger}$,
and then using the Fierz identity
 \begin{align}\label{Fierz1}
 \rmtr \big(t^a A \,t^a B \big)
 =\frac{1}{2}\,\rmtr A \, \rmtr B
 -\frac{1}{2 N_c}\, \rmtr(AB).
 \end{align}
By adding together the above results, one sees that the terms 
proportional to $1/N_c^2$ 
exactly cancel between `crossing' and `non--crossing' contributions\footnote{This
cancellation too can be recognized as a consequence of
the identity \eqref{ident}.}, so the net result reads
 \begin{align}\label{BK}
 \frac{\del \lan \hat{S}_{\bx\by} \ran_Y}{\del Y}=
 \frac{\abar}{2\pi}\, \int_{\bz}
 \mcal{M}_{\bx\by\bz}
 \big\lan \hat{S}_{\bx\bz} \hat{S}_{\bz\by}
 -\hat{S}_{\bx\by} \big\ran_Y\,,
 \end{align}
where we have also taken the average over the target. Formally,
this equation depicts the evolution as the splitting of the original
dipole $(\bx,\by)$ into a system of two dipoles, $(\bx,\bz)$ and
$(\bz,\by)$, which have a common leg at $\bz$. This would be the
actual physical picture at large $N_c$, but it formally holds also for
finite $N_c$, due to the `accidental' cancellation of the terms 
suppressed by $1/N_c^2$.

In deriving \eqn{BK} as above, it has been convenient to work in a frame where the projectile was a
bare dipole prior to the evolution step under consideration. But the ensuing equation is valid in any
frame (so long as the projectile remains dilute, of course). In a generic frame, where the projectile
also includes an arbitrary number of soft gluons, the quantity  $\lan \hat{S}_{\bx\by} \ran_Y$ describes
the scattering of that complicated partonic system,
for a given rapidity separation $Y$ between the projectile and the target.
Similarly, the quantity $\big\lan \hat{S}_{\bx\bz} \hat{S}_{\bz\by}\big\ran_Y$ describes the scattering
of a projectile which at low energy consists in only two dipoles, but in general also involves additional
soft gluons, as produced by the evolution of the `valence' dipoles.

\subsubsection*{(ii) {\em The manifestly probabilistic approach}}

In applying \eqn{DH} to a SW target, one must perform
manipulations similar to those in Sect.~\ref{sec:JIMtime} --- that is, 
distinguish between `crossing' and `non--crossing' contributions and then compute the
respective time integrals. In doing that, it is essential to keep together
the `real' and `virtual' pieces in \eqn{DH}, for each of the three integration ranges.
Then the calculations simplify since \texttt{(a)} there are no divergences
in the intermediate stages, and \texttt{(b)} the full result comes from
the `crossing' pieces  (`real' plus `virtual') alone.
Moreover, the associated manipulations have a clear probabilistic interpretation, 
in agreement with the discussion in Sect.~\ref{sec:prob}.

To demonstrate this, notice the following identities for the action of the functional derivative
on the dipole $S$--matrix:
 \begin{align}\label{JJS}
  R^a_{\br_1}R^a_{\br_2}\hat{S}_{\bx\by}\,=\,
 L^a_{\br_1}L^a_{\br_2}\hat{S}_{\bx\by}\,=\,&\Big[
 -g^2C_F\big(\delta_{\br_1\bx}-\delta_{\br_1\by}\big)
 \big(\delta_{\br_2\bx}-\delta_{\br_2\by}\big)\Big]
 \hat{S}_{\bx\by}\nn
 \,=\,& \Big(J^a(t_2,\br_2)J^a(t_1,\br_1)\hat{S}_{\bx\by}\,\big|_{A^-=0}\Big)\hat{S}_{\bx\by}
 \,,\end{align}
 where the equality in the second line holds for any $t_1$ and $t_2$.
 These identities imply that the `virtual' and `real' contributions mutually
cancel within the `non--crossing' terms, as anticipated. 

Consider now the  respective 
`crossing' contributions. For the `real' term, this has been already computed in \eqn{real}.
For the `virtual' term, we also need the following integral
\begin{align}\label{virtint}
&\int_{-\infty}^0\rmd t_1 \int_0^{\infty}\rmd t_2\ 
  \del^i_{\br_1}\del^i_{\br_2}\,
G_{0}(t_2-t_1,\br_2-\br_1; p^+)\,=\,-\frac{2}{p^+}
\int\frac{\rmd^2\bp}{(2\pi)^2}\ \frac{\rme^{\rmi \bp\cdot(\br_2-\br_1)}}{p_\perp^2}\,.
 \end{align}
By using this, together with Eqs.~\eqref{JJS}, \eqref{intK}, and \eqref{Mdef}, one finds
 the following `virtual--crossing'  contribution:
  \begin{align}\label{virtcross}
-\Big(\Delta H_{rad}\, \hat{S}_{\bx\by}\,\big|_{A^-=0}
 \Big)\,\hat{S}_{\bx\by} = -\frac{\alpha_s C_F}{\pi^2}
  \int_{\bz} \mcal{M}_{\bx\by\bz}
\hat{S}_{\bx\by}\,.
 \end{align}
This is the same as the contribution \eqref{virt} of the `non--crossing' terms in
the JIMWLK Hamiltonian. By adding this to the `real--crossing'  piece
in \eqn{real}, one finally recovers the BK equation \eqref{BK}.

The probabilistic interpretation is now manifest. The quantity
$ (\abar/2\pi)\mcal{M}_{\bx\by\bz}\rmd Y\rmd^2\bz$ is the differential probability for emitting
a gluon at transverse coordinate $\bz$ out of the quark--antiquark dipole $(\bx,\by)$.
The `real--crossing' piece
represents the process where the evolved partonic system (quark, antiquark, and gluon)
exists at the time of scattering $x^+=0$. The `virtual--crossing' piece measures the decrease
in the probability to find the original $q\bar q$ dipole at $x^+=0$.
This decrease is associated with evolution processes which occur either before ($x^+<0$), 
or after ($x^+>0$), the scattering. So, the `virtual--crossing' contribution must be equal
to that of such genuinely `non--crossing' processes. This is indeed what we have found
in \eqn{virtcross}. 

The validity of this interpretation is also comforted by the fact the `real' and `virtual' terms 
in the BK equation separately develop logarithmic `ultraviolet' divergences, which precisely
cancel in their sum. These divergences, coming from the poles of the dipole kernel
at $\bz=\bx$ and $\bz=\by$, correspond to self--energy corrections where
the gluon lies arbitrarily close to its parent quark in the transverse
plane. But such short--distance emissions should not affect the $S$--matrix since the
scattering cannot distinguish between a bare quark and a bare quark accompanied
by its radiation gluon, so long as the two partons are very close to each other.
And indeed, by inspection of  \eqref{BK} one sees that the pole of $\mcal{M}_{\bx\by\bz}$
at, say, $\bz=\bx$ is compensated by the linear combination of Wilson line correlators,
 due to `color transparency' ($\hat{S}_{\bx\bz}\to 1$ as $\bz\to\bx$).
 
 Notice that, in order for such cancellations to work, it has been essential to have the
 right relative coefficient between the `virtual' term and the `real' one
 (or, equivalently, between `crossing' and  `non--crossing' contributions). In turn, this
 emphasizes the importance of using the adiabatic prescription when computing 
 the time integrals in  Sect.~\ref{sec:JIMtime} (cf. the discussion after \eqn{timeRR}).
 
 Note finally that \eqn{BK}  is not a closed equation ---
its r.h.s. also involves the  $S$--matrix 
$\big\lan \hat{S}_{\bx\bz} \hat{S}_{\bz\by}\big\ran_Y$ for a system of two dipoles ---,
so it cannot be solved as it stands. This is truly the first equation from an infinite hierarchy,
the B--JIMWLK hierarchy, which describes the coupled evolution of scattering 
amplitudes for dilute systems with increasing complexity in terms of partonic structure.
This hierarchy  simplifies in the large $N_c$ limit, where expectation values of
gauge invariant operators can be factorized from each other. In particular, for $N_c\to
\infty$, \eqn{BK}
reduces to a closed equation  for the dipole $S$--matrix, as originally derived
by Kovchegov \cite{Kovchegov:1999yj}~:
 \begin{align}\label{BKmf}
 \frac{\del \lan \hat{S}_{\bx\by} \ran_Y}{\del Y}=
 \frac{\abar}{2\pi}\, \int_{\bz}
 \mcal{M}_{\bx\by\bz}\Big\{
 \big\lan \hat{S}_{\bx\bz}  \big\ran_Y\,\big\lan\hat{S}_{\bz\by} \big\ran_Y\,
 -\, \big\lan\hat{S}_{\bx\by} \big\ran_Y\Big\}\,,
 \end{align}
In the next section, we shall generalize this equation to the case
of an extended target. 
 
 
\comment{
Note that the above construction has been explicitly performed in a frame in
which the dipole was elementary prior to the last step in the evolution (indeed,
we acted
with the evolution Hamiltonian $H_{_{\rm JIMWLK}}$ on the {\em bare} dipole operator
$\hat{S}_{\bx\by}$), meaning that all the evolution in the previous steps has been
implicitly encoded in the target CGC weight function. Yet, the final equation \eqref{BK}
is boost invariant, within the limits of the `dense--dilute' set--up (i.e. so long as
one can ignore the effects of Pomeron loops). In an arbitrary frame, the 
high--energy evolution is divided between the target and the projectile, so the latter 
can be considerably more complicated than just a bare dipole. Yet, in any such a frame,
the solution $\lan \hat{S}_{\bx\by} \ran_Y$ to \eqn{BK} represents the 
forward $S$--matrix for a rapidity separation equal to $Y$.
}


\section{The high--energy evolution of transverse momentum broadening}
\label{sec:JET}

Starting with this section, we address the main physical problem of interest for us here,
namely the high energy evolution of a `hard probe' (energetic parton) which crosses
a dense QCD medium, like a weakly--coupled quark--gluon plasma (QGP).
The distinguished feature of the `medium' for the present purposes is the
fact that its longitudinal extent in $x^+$, to be denoted as $L$, is much larger than the
coherence time $\tau_{coh}={2p^+}/{p_\perp^2}$ of the soft gluons generated by the evolution.
This is the source of several complications that we here anticipate.

First, the eikonal approximation for the emitted gluon is not appropriate anymore.
Rather one has to use the general, but formal, expression of the gluon propagator
as a path integral, cf. \eqn{Gmedium}. Accordingly, the Wilson lines attached to the
gluon fluctuations become {\em functionals} of the gluon trajectories, which are
themselves random.

Second, the time integrations in \eqn{DeltaH} cannot be
performed {\em directly at the level of the Hamiltonian}, i.e. before acting with $\Delta H$
on the relevant scattering operator. This can be understood by inspection
of \eqn{Jdef} : if the time argument $t$ of the functional derivative $J^a(t,\br)$
lies inside the medium, then the Wilson lines in the r.h.s. of \eqn{Jdef} are
explicitly time--dependent, 
in contrast to what happened for a shockwave (compare to Eqs.~\eqref{JR}--\eqref{JL}).  
So one cannot disentangle the integrations over the emission times $t_1$ and $t_2$
from the Wilson lines. 
This also implies that, in order to make progress, one needs to
specify the projectile.

Third, even after choosing a projectile and using $\Delta H$ to construct the associated 
evolution equation, this equation is still too complicated to be dealt with in full generality.
Not only this is not a closed equation (a feature that we are already familiar with from the
example of the B--JIMWLK hierarchy), but this equation is also {\em functional} 
(the unknown correlators enter under the path integral representing the trajectory of
the fluctuation) and
{\em non--local in time} (the correlators depend upon the integration variables $t_1$ 
and $t_2$). Our strategy to render such equations
tractable in practice will be to perform additional simplifications, notably
concerning the structure of the medium correlations, and also to study limiting cases
which are simpler but still interesting.

For simplicity, we shall focus on the evolution of a color dipole. This is pertinent indeed,
since the corresponding scattering amplitude enters the calculation of two important 
observables --- the transverse momentum broadening and the energy loss by an energetic 
parton --- that we shall discuss in this and the next coming sections.
Also, we shall use a similar set--up as in the previous section~: the quark is 
approaching the medium from very far away and its interactions are adiabatically switched off at 
large times, $|x^+|\to\infty$. This is not necessarily the actual situation in a nucleus--nucleus
collision, where the quark can also be created inside the medium, via some hard
process. This would lead to additional radiation which could mix with the quantum
fluctuations triggered by the interactions in the medium. The simplest way to avoid such a
mixing is to assume that the quark was on--shell prior to the collision.

But even though the general set--up looks similar to the shockwave set--up discussed in 
Sect.~\ref{sec:JIM}, the physical problem that we shall consider from now on
is fundamentally different, 
because of the different kinematics. In Sect.~\ref{sec:JIM}, we have assumed
that the energy $E\equiv p_0^+$ of the incoming projectile is much higher than the 
characteristic target scale $\omega_c$, cf. \eqn{omegac0}, in such a way to ensure a 
large phase--space, at $\omega_c\ll p^+ \ll E$, for long--lived fluctuations to which the target 
looks like a shockwave. Here, we are instead interested in the complimentary 
situation, where $E$ and $\omega_c$ are comparable with each other. This is indeed the
interesting situation for jet production in $AA$ collisions at the LHC, where the measured
jets have energies of the order of 100~GeV, whereas the medium scale  $\omega_c$
is in the ballpark of 50~GeV. Then the soft quantum fluctuations of the projectile,
with energies $p^+\ll\omega_c$, have
relatively short lifetimes, which are typically much smaller than the medium size $L$. 
In what follows, we shall describe the evolution encompassing these fluctuations.

\subsection{The tree-level approximation}
\label{sec:tree}

In preparation for the quantum evolution to be discussed in the next sections,
we shall first briefly review the tree--level calculation of the transverse momentum 
broadening. This will give us the opportunity to introduce
the relevant scales and notations and, moreover, it will inspire some of the
approximations to be performed later on.

At tree--level, the problem of transverse momentum broadening for an energetic quark
which enters the medium is formally similar to that of quark production in $pA$ collisions
at forward rapidities. In both cases, the cross--section $\rmd N/\rmd^2\bp$ for finding the 
quark in a state with transverse momentum $\bp$ after the collision can be computed,
within the limits of the eikonal approximation,
as the Fourier transform of a dipole forward amplitude (below, $\br\equiv \bx - \by$)~:
\begin{align}
 \label{ptbroad}
 \frac{\rmd N}{\rmd^2\bp} \,=\,
 \frac{1}{(2\pi)^2} \int_{\br} \rme^{-\rmi \bp \cdot \br} 
\langle \hat{S}_{\bx\by}  \rangle.
\end{align}
The `dipole' here is merely a mathematical construction: the `quark leg at $\bx$' 
represents the physical quark in the direct amplitude, whereas 
the `antiquark leg at $\by$' is the physical quark in the complex conjugate
amplitude. As usual, the brackets within $\langle \hat{S}_{\bx\by}  \rangle$
denotes  the target average over the configurations of the color field
$A^-_a(x)$.  The target is a weakly--coupled QCD medium with longitudinal support at $0<x^+<L$.
For simplicity, we assume this medium to be homogeneous (on the average)
in the transverse plane. Accordingly, the average $S$--matrix depends only upon
the dipole size $\br= \bx - \by$ and we can write $\langle \hat{S}_{\bx\by}  \rangle \equiv \mathcal{S}(\br)$.
Using $\mathcal{S}(0)=1$ (`color transparency'), one sees that
the cross--section \eqref{ptbroad} is properly normalized: $\int \rmd^2\bp\, (\rmd N/\rmd^2\bp)=1$.


A weakly--coupled medium, such as a QGP with sufficiently high temperature $T$,
can be described as an incoherent collection of independent
color charges, `quarks' and `gluons'. These charges will be assumed to be
point--like and to have no other mutual interactions,
except for those responsible for the screening of the color interactions over a (transverse)
distance $r\sim 1/m_D$, with $m_D$ the `Debye mass'. 
Under these assumptions, the only non--trivial correlator of the target 
field $A^-$ is the respective 2--point function, which has the following structure
 \begin{align}\label{correlmed}
\left\langle A^-_a(x^+,x^-,\bx)\,A^{-}_b(y^+,y^-,\by)\right\rangle_0\,=\,\delta_{ab}\delta(x^+-y^+)\,n(x^+)
\gamma(\bx-\by)\,,
\end{align}
where $n$ is the number  density of the medium constituents (more
precisely, a linear combination of the respective densities for quarks and gluons,
weighted with appropriate color factors). 
As indicated in \eqn{correlmed}, this density can generally depend upon $x^+$ 
(e.g. for an expanding medium), but here we shall mostly work with a medium 
which is uniform in $x^+$ (within its longitudinal support at  $0 < x^+ < L$, of course).  Also
  \begin{align}\label{Coul}
 \gamma(\bk)\equiv \int \rmd^2\br\ \rme^{\rmi \bk\cdot\br}\,\gamma(\br)\,\simeq\,
 \frac{g^2}{\bk^4}\,,\end{align}
with the approximate equality holding for $k_\perp\gg m_D$,
is the square of the 2--dimensional Coulomb propagator. It is understood
that \eqn{Coul} must be used with an infrared cutoff $k_\perp\simeq m_D$. 

The correlator \eqref{correlmed} is local
in the color indices, by gauge symmetry. It is furthermore independent of the light--cone variables 
$x^-$ and $y^-$, and it is local in $x^+$, because of the high energy kinematics.
These properties can be best understood in a frame where the medium
is a ultrarelativistic left mover: then, the dynamics in $x^-$ (the light--cone
`time' for a left mover) is frozen by Lorentz time dilation, whereas the correlation length in
$x^+$ gets squeezed by Lorentz longitudinal contraction. The locality in $x^+$ is clearly an
idealization, whose limitations will be discussed in Sect.~\ref{sec:DLAPS}.
The shockwave counterpart of \eqn{correlmed} is the description of a large nucleus in 
the McLerran--Venugopalan (MV) model, which employs a Gaussian CGC weight function
 \cite{McLerran:1993ni,McLerran:1994vd}.

For the Gaussian field distribution in \eqn{correlmed}, it is a straightforward exercise to
compute the average $S$--matrix for a quark--antiquark dipole.  One finds
  \begin{align}\label{Sdip0}
 \mcal{S}_0(\br)\,=\,\exp\Biggl\{-g^2 C_F\int\limits_{0}^{L}\rmd x^+
 n(x^+)\int \frac{\rmd^2\bm{k}}{(2\pi)^2}\,\gamma(\bk)\left(1-
 \rme^{\rmi\bm{k}\,\cdot\,\bm{r}}\right)\Biggr\}\,.\end{align}
The quantity within the braces is (minus) the amplitude for a single scattering between
the dipole and the medium. The fact that the multiple scattering series exponentiates reflects
the lack of non--trivial medium correlations: successive collisions proceed
independently from each other.
 
Using \eqn{Coul}, one sees that the integral over $\bk$ in \eqn{Sdip0} is logarithmically
sensitive to the IR cutoff $m_D$. We shall be mostly interested in
small dipole sizes $r\equiv |\br|\ll 1/m_D$. 
Then, there is a large  logarithmic phase--space, at $m_D\ll k_\perp\ll 1/r$. 
To leading logarithmic accuracy, the integral can be evaluated by expanding the complex
exponential $\rme^{\rmi\bm{k}\,\cdot\,\bm{r}}$ to second order (the linear term vanishes
after angular integration). One thus finds (with $n(x^+)=n_0$ from now on)
  \begin{align}\label{Sdip1}
 \mcal {S}_{0}(\br)\,\simeq\,\exp\left\{-\frac{1}{4}\,L \hat q(1/r^2)\,\br^2\right\}\,,
  \end{align}
where $\hat q$ is the {\em jet quenching parameter} for an incoming quark  :
   \begin{align}\label{qhat}
   \hat q(Q^2)\,\equiv\,g^2C_F n_0\int^{Q^2}
    \frac{\rmd^2\bm{k}}{(2\pi)^2}\,\bk^2\,\gamma(\bk)\,\simeq\,
   4\pi\alpha_s^2 C_F n_0 \ln\frac{Q^2}{m_D^2}\,.\end{align}
In the above integral, the differential cross--section $g^2C_F\gamma(\bk)$ 
for the scattering between the quark and the medium is weighted by the transverse 
momentum squared $k_\perp^2$ transferred in the collision. 
Accordingly, $\hat q(Q^2)$ is proportional to a {\em transport} cross--section
--- the total cross--section for collisions which are accompanied by a relatively hard transfer 
of transverse momentum, within the range $m_D^2\ll k_\perp^2 \ll Q^2$.
For a weakly coupled QGP, one has, parametrically, $n_0\sim T^3$, 
$m_D^2\sim \alpha_s T^2$, and hence $\hat q\sim \alpha_s^2 N_c T^3\ln(1/\alpha_s)$.
(See Refs.~\cite{Arnold:2008iy,Arnold:2008vd,CaronHuot:2008ni} for detailed calculations.)
 
The dipole scattering becomes strong 
when the exponent in \eqn{Sdip1} is of order one, or larger. 
This happens when $r\gtrsim 1/Q_s$, with $Q_s$ a characteristic
transverse momentum scale,  defined as
  \begin{align}\label{Qs}
 Q_s^2 \,=\,  L \hat q(Q_s^2)\,=\,
 4\pi\alpha_s^2 C_F n_0 L \ln\frac{Q_s^2}{m_D^2}\,.\end{align}
(We implicitly assume that $Q_s$ is much larger than $m_D$, which in turn requires 
the medium size $L$ to be large enough; see Sect.~\ref{sec:DLAPS} for details.)
This scale $Q_s$ is generally referred to as the `target saturation momentum', because
the physics responsible for the unitarization of the dipole amplitude  --- the
multiple scattering between the dipole and the color charges in the target ---
can also be viewed, in a suitable frame where the target is highly boosted, 
as the result of non--linear phenomena 
in the gluon distribution in the target, leading to gluon
saturation \cite{Mueller:2001fv,Iancu:2002xk,Gelis:2010nm}. In Sect.~\ref{sec:gluon},
we shall argue that this profound relation between jet quenching and gluon saturation,
which here has been observed at tree--level, is also preserved by the high--energy evolution.


Using \eqn{Sdip1}, one can now estimate the Fourier transform in \eqn{ptbroad} for $p_\perp
\gg m_D$. Consider first the case $p_\perp\lesssim Q_s$~; then, the integral in \eqn{ptbroad} 
is cut off by the dipole $S$--matrix at a value $r\simeq 1/Q_s$. To the accuracy of interest,
one can ignore the slow dependence of the jet quenching parameter upon $r$, and thus deduce
 \begin{align}
 \label{ptGauss}
 \frac{\rmd N}{\rmd^2\bp} \,\simeq\,\frac{1}{\pi Q_s^2}\,\rme^{-{\bp^2}/{Q_s^2}}\,.\end{align}
This Gaussian distribution is the hallmark of a diffusive process --- a random walk in the
transverse momentum space, leading to a momentum broadening
 $\langle p_\perp^2\rangle\simeq Q_s^2$ ---, which is induced by a succession
of independent collisions in the medium.

Consider also the high--momentum limit $p_\perp\gg Q_s$~; then, the integral in 
\eqn{ptbroad} is cut off by the complex exponential at a value $r\sim 1/p_\perp\ll 1/Q_s$,
so it is appropriate to expand the dipole $S$--matrix to linear order in its exponent. This 
gives
\begin{align}
 \label{highpt}
 \frac{\rmd N}{\rmd^2\bp} \,\simeq\,&
 \frac{\alpha_s^2 C_F}{4\pi}  n_0 L \int_{\br} \rme^{-\rmi \bp \cdot \br}\, (-r^2)
  \ln\frac{1}{r^2m_D^2}
  \,=\,\frac{4\alpha_s^2 C_F n_0 L}{p_\perp^4}
  \nn \,=\,&
  \frac{1}{\pi Q_s^2\ln(Q_s^2/m_D^2)}\,\frac{Q_s^4}{p_\perp^4}
  \,.
\end{align}
The logarithmic scale dependence of $\hat q(1/r^2)$ has been essential
in deriving this result. As clear from its above derivation, the $1/p_\perp^4$
tail in the spectrum at  high $p_\perp$ is produced via a single, hard, scattering. This
represents a rather rare event, as visible from the fact that the integral of \eqref{highpt} over
$p_\perp> Q_s$ is suppressed by a large logarithm:
 \begin{align}
 \label{highptint} \int_{Q_s} \rmd^2\bp\, 
 \frac{\rmd N}{\rmd^2\bp} \,\simeq\,\frac{1}{\ln(Q_s^2/m_D^2)}\,\ll\,1\,.
 \end{align}
This is furthermore in agreement with the fact that the probability sum rule 
$\int \rmd^2\bp\, (\rmd N/\rmd^2\bp)=1$ is already exhausted (to the leading--logarithmic
accuracy of interest) by the contribution \eqref{ptGauss} of relatively soft
($k_\perp\lesssim Q_s$) multiple scattering.

In what follows, we shall be mostly interested in typical events, in which the final spectrum is
the result of multiple soft scattering and has the Gaussian form in
 \eqn{ptGauss}. Accordingly, we shall focus on a quark--antiquark
dipole with transverse size $r\sim 1/Q_s$. This in turn implies that the exponent
in \eqn{Sdip1} for the dipole $S$--matrix is of $\order{1}$~: on the average, the dipole undergoes
{\em one inelastic scattering} while crossing the medium. This more precisely means that the
dipole may undergo a single hard collision, with a transferred momentum $k_\perp\sim Q_s$,
or a large number of softer collisions, but in such a way that the total transferred momentum
squared is again of order $Q_s^2$. The present calculation cannot distinguish between such
scenarios, so in that sense the jet quenching parameter $\hat q(Q_s^2)$ is not really a {\em local}
transport coefficient, but rather a measure of the average
properties of the medium coarse--grained over a longitudinal distance of order $L$. 
(This is also visible in the fact that the quantity $\hat q(Q_s^2)$ `knows' about 
the overall size $L$ of the medium, via its logarithmic dependence upon  $Q_s^2 \propto L$.)
This should
be kept in mind when interpreting the radiative corrections to be computed in what follows.

\subsection{The dipole evolution equation}
\label{sec:dipole}

The relation \eqref{ptbroad} between the cross--section for $p_\perp$--broadening and the
forward dipole amplitude is known to be preserved by the high--energy evolution
up the next--to--leading logarithmic accuracy (in the sense of large energy logarithms, 
like $\ln(\omega_c/p^+)$ in the present context)
\cite{Kovchegov:2001sc,Mueller:2012bn}. Here, we shall merely
work at leading logarithmic accuracy, so we can safely use this dipole picture for the purpose
of understanding the high energy evolution of the momentum broadening.
As before, we shall construct an evolution equation for
the average dipole $S$--matrix by first acting with the evolution Hamiltonian
$\Delta H$ on the dipole operator
$\hat{S}_{\bx\by}$ and then taking the average over the target. As explained
in Sect.~\ref{sec:BK}, this procedure assumes that the projectile is a `bare' dipole prior
to the evolution step under consideration, which implicitly means that the effects of the earlier
steps have been incorporated in the distribution of the color fields in the target
(the CGC weight function $W_{Y}[A^-]$). Hence, in general, this distribution can be more
complicated than the Gaussian introduced in the previous subsection and which
applies at tree--level.

In the present context, it becomes
advantageous to use the alternative form \eqref{DH} for the action of $\Delta H$,
since this permits to avoid spurious divergences already before integrating over the
emission times $t_1$ and $t_2$ (an operation that we shall not be able to 
perform in general). In particular, the `virtual' term required by probability
conservation is automatically built in.
Let us first compute the coefficient of this  `virtual' term, which is independent of
the background field and thus immediately follows from Eqs.~\eqref{subtract}
and \eqref{JJS}~:
  \begin{align}\label{virtdip}
  \hspace*{-.6cm}
-\Delta H_{rad}\, \hat{S}_{\bx\by}\,\big|_{A^-=0}
  = \frac{g^2C_F}
 {2\pi}\!\int\limits_{\omega}^{\omega_c}\frac{\rmd p^+}{(p^+)^2}
   \int_{-\infty}^\infty\rmd t_2 \int_{-\infty}^{t_2}\rmd t_1 
\Big[{\del^i_{\br_2}\del^i_{\br_1}}G_{0}(t_2-t_1,\br_2-\br_1; p^+)\Big]
\Big|^{\br_2=\bx}_{\br_2=\by}\,
    \Big|^{\br_1=\bx}_{\br_1=\by}\,.
 \end{align}
As compared to  \eqn{DeltaH}, we have changed the integration limits for $p^+$,
to emphasize that the maximum energy of the gluon fluctuations is now of the order
of the medium scale $\omega_c$,  which reads 
 \begin{align}\label{omegac}
 \omega_c \,=\,Q^2_s L\,=\,\hat q(Q^2_s) L^2\,.\end{align}
 We assume here that the energy $E$ of the incoming quark is larger than $\omega_c$,
 albeit not {\em much} larger. 
The minimal energy $\omega$ is at this stage generic, but
it is understood that it is small enough, $\omega\ll\omega_c$,  to leave a sufficiently large
phase--space for the high energy evolution (see Sect.~\ref{sec:DLAPS}). 

\begin{figure}
\begin{center}
\begin{minipage}[b]{0.95\textwidth}
\begin{center}
\includegraphics[width=1.\textwidth,angle=0]{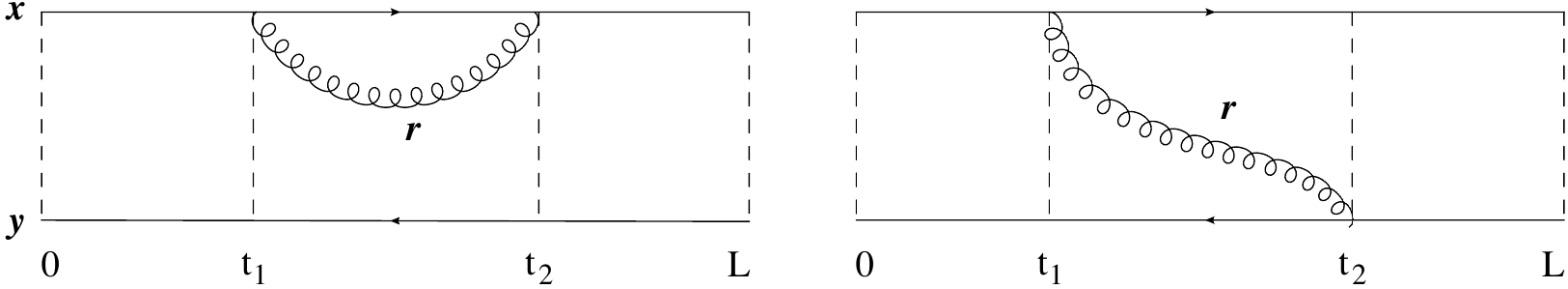}
\end{center}
\end{minipage}
\end{center}
\caption{\small\sl \label{fig:dipole}Two diagrams illustrating the dipole evolution described by \eqn{DeltaHrad}
(they correspond to the first and respectively the fourth term in the r.h.s. of \eqn{DeltaHrad}).
It is understood that all the partonic lines (quark, antiquark, and gluon) are accompanied by
Wilson lines describing scattering off the medium.}
\end{figure}

Consider now the `real' term in \eqn{DH}, which involves the
radiation piece of the background field propagator, cf. \eqn{calH}. 
By repeatedly using \eqn{Jdef}, one finds 
    \begin{align}
  \label{DeltaHrad}
 \Delta H_{rad}  \hat{S}_{\bx\by}  \,=\,& -g^2
  \int\limits_{\omega}^{\omega_c}\frac{\rmd p^+}{2\pi} \frac{1}{(p^+)^2}
  \int_{-\infty}^\infty\rmd t_2 \int_{-\infty}^{t_2}\rmd t_1\ 
  \del^i_{\br_1}\del^i_{\br_2}\,G_{ab}(t_2,\br_2; t_1,\br_1; p^+)\,\nn
  &\qquad\quad\bigg\{\delta_{\br_1\bx}\delta_{\br_2\bx}\,\frac{\rmtr}{N_c}
  \left[\Big(V_{\infty, t_2}^{\dagger}t^a\, V_{t_2t_1}^{\dagger} t^b
  V_{t_1, -\infty}^{\dagger}  \Big)_{\bx}{V}_{\by}\right]\nn
  &\qquad \qquad+\,\delta_{\br_1\by}\delta_{\br_2\by}\,\frac{\rmtr}{N_c}
 \left[V^{\dagger}_{\bx}\Big(V_{t_1, -\infty}t^b\, V_{t_2t_1} t^a
  V_{\infty , t_2} \Big)_{\by}\right]\nn
  &\qquad \qquad-\,\delta_{\br_1\by}\delta_{\br_2\bx}\,\frac{\rmtr}{N_c}
 \left[\Big(V_{\infty, t_2}^{\dagger}t^a\, V_{t_2,-\infty}^{\dagger}  \Big)_{\bx}
 \Big(V_{t_1, -\infty}t^b\, V_{\infty,t_1} \Big)_{\by}\right]\nn
  &\qquad \qquad-\,\delta_{\br_1\bx}\delta_{\br_2\by}\,\frac{\rmtr}{N_c}
 \left[\Big(V_{\infty, t_1}^{\dagger}t^b
  V_{t_1, -\infty}^{\dagger}  \Big)_{\bx}\Big(V_{t_2,-\infty} t^a
  V_{\infty , t_2} \Big)_{\by}\right]\bigg\}\,.
  \end{align}
Note that $V_{t_2t_1}\equiv\big(V_{t_2t_1}^{\dagger}\big)^\dagger$ truly describes
backward propagation in time, from $t_2$ to $t_1$ (recall that $t_2>t_1$). The
physical interpretation of the four terms within the braces in the r.h.s. of \eqn{DeltaHrad}
is quite transparent (see also Fig.~\ref{fig:dipole}): 
the first two terms describes processes in which the soft gluon is
emitted and then reabsorbed by a same leg of the dipole (either the quark, or the antiquark);
the last two terms, which have the opposite sign, correspond to exchange processes
where the gluon is emitted by the quark and then absorbed by the antiquark, or vice-versa.

To obtain an evolution equation, one needs to average \eqn{DeltaHrad} over the 
target field distribution and also perform the integrations over the emission
times $t_1$ and $t_2$. (Note that the target average should also include the adjoint
Wilson line implicit in the structure of the gluon propagator, cf. \eqn{Gmedium}.) 
We here meet with one of the difficulties anticipated at the beginning of this section:
unlike in the corresponding discussion for a shockwave, 
the support of the Wilson lines in \eqn{DeltaHrad} is now truly
dependent upon the emission times $t_1$ and $t_2$. 
Accordingly, the time integrations cannot be disentangled from the target 
correlations anymore: the Wilson line correlators which enter
$\langle \Delta H  \hat{S}_{\bx\by}\rangle$ are explicitly time dependent.
So, it seems impossible to make further progress {\em in full generality} --- 
i.e., without additional assumptions about the medium correlations.

Inspired by the situation at tree--level and also by the mean field approximation to the 
B--JIMWLK equations \cite{Kovner:2001vi,Iancu:2002xk,Iancu:2002aq,Blaizot:2004wv,Kovchegov:2008mk,Dominguez:2011wm,Iancu:2011ns,Iancu:2011nj}, 
which appears to be remarkably successful \cite{Dumitru:2011vk,Iancu:2011nj}, we shall 
from now on assume that the background field distribution remains approximatively 
Gaussian after including the effects of the high energy evolution. That is, the only non--trivial 
background field correlator is the respective 2--point function, which has the general structure
(compare to \eqn{correlmed})
\begin{align}\label{Gaussian}
\left\langle A^-_a(x^+,x^-,\bx)\,A^{-}_b(y^+,y^-,\by)\right\rangle_\omega\,=\,\delta_{ab}\delta(x^+-y^+)\,
\bar\Gamma_\omega(x^+,\bx-\by)\,.
\end{align}
This depends upon the energy scale $\omega$ down to which one has
integrated out the soft gluons, since this fixes the longitudinal resolution
on which one probes the medium correlations. The $x^+$--dependence of the correlator
$\bar\Gamma_\omega(x^+,\bx-\by)$ reflects the evolution of the 
time inhomogeneity introduced at tree--level by the charged particles density $n(x^+)$,
cf. \eqn{correlmed}.
Vice--versa, if the latter is independent of time, $n(x^+)=n_0$, then so is also the function
$\bar\Gamma_\omega$ --- except, of course, for the fact that its longitudinal support is 
restricted\footnote{This restriction is not affected by the evolution since
one can neglect the fluctuations occurring near the edges of the medium; see
the discussion below.} to $0<x^+<L$.

The main characteristic of the 2--point function \eqref{Gaussian} 
is to be {\em local in time}. This was an approximation already at tree--level and it is 
even more so after including the effects of the radiative corrections associated with 
soft gluon emissions, which are delocalized over a time interval $t_2-t_1\sim \tau_{coh}$.
If this approximation makes nevertheless sense, it is because, as we shall see, 
the most interesting emissions have very short lifetimes, which are 
much shorter than the medium length $L$.
The locality of  \eqn{Gaussian} in $x^+$ allows one to factorize the Wilson correlations within
$\langle \Delta H  \hat{S}_{\bx\by}\rangle$ according to their time arguments. To be specific, consider
the first among the four terms within the braces in \eqn{DeltaHrad}. After also including the adjoint
Wilson line from the gluon propagator, cf. \eqn{Gmedium},
we are led to the following medium correlation function:
 \begin{align}
  \label{DeltaH1}
&  \hspace*{-1.cm}
 \left\langle U^{\dagger\,ab}_{t_2t_1}[\br]\,\frac{\rmtr}{N_c}
 \left[\Big(V_{\infty, t_2}^{\dagger}t^a\, V_{t_2t_1}^{\dagger} t^b
  V_{t_1, -\infty}^{\dagger}  \Big)_{\bx}{V}_{\by}\right]\right\rangle=\nn
&  \hspace*{-1.cm}
\quad  =  \left\langle \frac{\rmtr}{N_c} \Big(V_{\infty, t_2}^{\dagger}(\bx)
  V_{\infty, t_2}(\by)\Big)\right\rangle\,
  \left\langle U^{\dagger\,ab}_{t_2t_1}[\br]\, \frac{\rmtr}{N_c}  
   \Big(t^a\, V_{t_2t_1}^{\dagger}(\bx) t^b\,V_{t_2t_1}(\by) 
 \Big)\right\rangle\,\left\langle \frac{\rmtr}{N_c}
 \Big(V_{t_1, -\infty}^{\dagger}(\bx) V_{t_1, -\infty}(\by) \Big)\right\rangle
 \nn
&  \hspace*{-1.cm}
\quad  = \frac{N_c}{2}\left\{ \big\langle  \hat{S}_{\infty, t_2}(\bx,\by)\big\rangle
\big\langle  \hat{S}_{t_2,t_1}(\bx,\br) \hat{S}_{t_2,t_1}(\br,\by)\big\rangle
 \big\langle  \hat{S}_{t_1, -\infty}(\bx,\by)\big\rangle\,-\,\frac{1}{N_c^2} 
 \big\langle  \hat{S}(\bx,\by)\big\rangle
 \right\}\,,
\end{align}
where the first equality is obtained by using the Gaussian Ansatz \eqref{Gaussian} 
for the medium averages and the second one follows via the Fierz identity \eqref{Fierz1}.
We have defined the time--dependent dipole operator $\hat{S}_{t_2,t_1}(\bx,\by)$ via
the obvious generalization of \eqn{Sdip}.  To simplify writing, we keep implicit the dependence
of the various correlations upon the renormalization scale $\omega$.
Also, in the very last term we have reconstructed the average of the global $S$--matrix according to
\begin{align}\label{global}
\big\langle  \hat{S}(\bx,\by)\big\rangle=
\big\langle  \hat{S}_{\infty, t_2}(\bx,\by)\big\rangle
\big\langle  \hat{S}_{t_2,t_1}(\bx,\by)\big\rangle
\big\langle  \hat{S}_{t_1, -\infty}(\bx,\by)\big\rangle\,.\end{align}
This last term, which is explicitly suppressed at large $N_c$, vanishes against
the respective contribution of the `virtual' term.
A similar  cancellation has been noticed in Sect.~\ref{sec:BK}
at the level of the BK equation. As in that case though, 
it is nevertheless convenient to consider the large--$N_c$ limit, which
allows us to factorize the two--dipoles $S$--matrix during the lifetime of the fluctuation:
\begin{align}\label{factNc}
 \big\langle  \hat{S}_{t_2,t_1}(\bx,\br) \hat{S}_{t_2,t_1}(\br,\by)\big\rangle\,\simeq\,
  \big\langle  \hat{S}_{t_2,t_1}(\bx,\br)\big\rangle  \big\langle  \hat{S}_{t_2,t_1}(\br,\by)\big\rangle
  \quad\mbox{at large $N_c$}\,.\end{align}
 One should keep in mind that $\hat{S}_{t_2,t_1}(\bx,\br)\equiv
\hat{S}_{t_2,t_1}\big(\bx, [\br(t)]; \omega\big)$ is truly a {\em functional} of the path
$\br(t)$, with $t_1 < t < t_2$, and also a function of $\omega$, although such features are
kept implicit, to simplify the notations.
In what follows, we shall often use the more compact notations
\begin{align}
\mcal{S}_{t_2,t_1}(\bx,\by)\equiv \big\langle  \hat{S}_{t_2,t_1}(\bx,\by)\big\rangle\,,\qquad
\mcal{S}(\bx,\by)\equiv \mcal{S}_{\infty, -\infty}(\bx,\by)\,.\end{align}

Before we proceed, let us open here a parenthesis on the generalization of the
present results to finite $N_c$~: within the Gaussian approximation at hand, it is in fact
possible to also evaluate the finite--$N_c$ corrections.  (See e.g. 
Refs.~\cite{Kovner:2001vi,Blaizot:2004wv,
Kovchegov:2008mk,Dominguez:2011wm,Dumitru:2011vk,Iancu:2011nj} for corresponding
discussions in the framework of the CGC.) 
To keep the presentation as simple as possible, we shall stick to the large--$N_c$ limit 
in all the intermediate steps, but indicate the generalization of the final results to finite $N_c$.
Some formul{\ae} which are useful to that purpose are summarized in
Appendix \ref{sec:Nc}.

We now close the parenthesis  and return to the evaluation of  \eqn{DeltaHrad} in the
Gaussian approximation and at large $N_c$. The first term in the r.h.s. has been already
discussed in Eqs.~\eqref{DeltaH1}--\eqref{factNc}.
The remaining three terms can be similarly manipulated.
Under the present assumptions, they all involve the same product of Wilson line correlators, as
written down in the last line of \eqn{DeltaH1}. Thus, they differ from each other (and from the 
first term) only via the actual values taken by the endpoints $\br_1$ and $\br_2$ of the emitted gluon.
By putting together the previous results and adding the contribution of the
`virtual' piece (i.e. \eqn{virtdip} times $\mcal{S}(\bx,\by)$), one obtains
    \begin{align}
  \label{DeltaHav}
\left\langle \Delta H \hat{S}_{\bx\by}\right\rangle & \,=\, -\frac{\alpha_sN_c}{2}
  \int\limits_{\omega}^{\omega_c}\frac{\rmd p^+}{(p^+)^3}
  \int_{-\infty}^\infty\rmd t_2 \int_{-\infty}^{t_2}\rmd t_1\ 
  \del^i_{\br_1}\del^i_{\br_2}\Bigg\{
  \int\big[\mcal{D}\br(t)\big]
  \ \rme^{\rmi \,\frac{p^+}{2}
  \int\limits_{t_1}^{t_2}\rmd t \,\dot\br^2(t)}\nn
 &  \times \mcal{S}_{\infty, t_2}(\bx,\by)\Big[\mcal {S}_{t_2,t_1}(\bx,\br) \mcal{S}_{t_2,t_1}(\br,\by)
 - \mcal{S}_{t_2,t_1}(\bx,\by)\Big]
    \mcal{S}_{t_1, -\infty}(\bx,\by)\Bigg\}\Bigg|^{\br_2=\bx}_{\br_2=\by}\,
    \Bigg|^{\br_1=\bx}_{\br_1=\by}\,.\nn
    \end{align}
This equation can be recognized as
a generalization of the BK equation \eqref{BKmf}, to which it 
reduces in the limit where the target is a shockwave. 
Namely, for a target localized near $x^+=0$, the
BK equation is obtained from \eqn{DeltaHav}
by integrating over positive values for $t_2$ and negative values
for $t_1$. (When both $t_1$ and $t_2$ have the same sign, one has
$\mcal{S}_{t_2,t_1}=1$ for a SW target and then the r.h.s. of \eqn{DeltaHav} 
simply vanishes.) In that case, $\mcal{S}_{\infty, t_2}=\mcal{S}_{t_1, -\infty}=1$,
whereas $\mcal{S}_{t_2,t_1}(\bx,\by)=\mcal{S}(\bx,\by)$ is independent of time. Moreover,
the $S$--matrices for the two daughter dipoles, $\mcal {S}(\bx,\br)$ and  $\mcal{S}(\br,\by)$,
do not depend upon the detailed trajectory $\br(t)$ of the soft gluon, but only upon its
position $\br(0)$ at the interaction time $t=0$. Hence one can write (compare to \eqn{USW})
 \begin{align}\label{SSW}
 \mcal {S}\big(\bx, [\br(t)]\big)\,\simeq\,\mcal {S}(\bx,\br(0))\,=\,\int\rmd^2{\bz}\
 \delta^{(2)}\big(\bz-
 {\br}(0)\big)\,\mcal {S}(\bx,\bz)\,.
 \end{align}
After also using the factorization property \eqref{fact} for the free path integral, one can perform
the time integrations as in Eqs.~\eqref{int2}, \eqref{int1} and \eqref{virtint}, 
then recognize the logarithmic
enhancement of the ensuing integral over $p^+$, and finally reconstruct the BK equation 
\eqref{BKmf}, as anticipated.

Returning to the case of an extended target, we notice that the time integrations
can still not be done in full generality at the level of  \eqn{DeltaHav}, although the latter looks
both simpler and physically more transparent than the original expression
in \eqn{DeltaHrad}. Yet, as we shall shortly see, \eqn{DeltaHav} is a convenient starting point for
further studies: it allows for explicit calculations in limiting situations of interest and also for
general physics conclusions. 
Before we proceed with more specific studies, it is convenient
to recast this expression in a more suggestive form.

First, one can interpret \eqn{DeltaHav} as an equation for the evolution 
w.r.t. the longitudinal momentum (`energy') $p^+\equiv \omega$ of the emitted gluon. 
To that aim, we shall write
    \begin{align}
 \label{DeltaS}
\left\langle \Delta H \hat{S}_{\bx\by}\right\rangle
\,= \, \Delta \mcal{S}(\bx,\by)\,\equiv\, \int\limits_{\omega}^{\omega_c} {\rmd \omega_1}\,\frac{\del
 \mcal{S}(\bx,\by)}{\del \omega_1}\,,
  \end{align}
which by comparison with  \eqn{DeltaHav} allows us to deduce the following evolution equation
 \begin{align}
\label{DeltaHeq}
 - \frac{\del
 \ln\mcal{S} (\bx,\by)}{\del \omega} \,=\,\frac{\alpha_sN_c}{2} &
  \frac{1}{\omega^3} 
  \int_{-\infty}^\infty\rmd t_2 \int_{-\infty}^{t_2}\rmd t_1\ 
  \del^i_{\br_1}\del^i_{\br_2}\Bigg\{
  \int\big[\mcal{D}\br(t)\big]
  \ \rme^{\rmi \,\frac{\omega}{2}
  \int\limits_{t_1}^{t_2}\rmd t \,\dot\br^2(t)}\nn
 &  \times \Big[\mcal {S}_{t_2,t_1}(\bx,\br) \mcal{S}_{t_2,t_1}(\br,\by)\mcal{S}_{t_2,t_1}^{-1}(\bx,\by)
 \,-\, 1\Big]
    \Bigg\}\Bigg|^{\br_2=\bx}_{\br_2=\by}\,
    \Bigg|^{\br_1=\bx}_{\br_1=\by}\,.
  \end{align}
(Recall that $0 < \mcal {S} \le 1$, so $\ln\mcal{S} < 0$.)
\eqn{DeltaHeq} is somewhat formal because, so far, we have not
demonstrated the logarithmic enhancement $\int (\rmd \omega/\omega)$ of the soft gluon emission
for the case of an extended target. Yet, as we shall see starting
with Sect.~\ref{sec:SSA}, such enhancement shows up indeed in all the cases where we
will be able to perform the time integrations.

Furthermore, we anticipate that the dominant corrections in the regime of interest
are associated with very soft gluons, with energies $p^+\ll\omega_c$. Such
gluons have small lifetimes $\tau_{coh}\ll L$, hence they are
typically emitted and absorbed {\em deeply inside the medium}: boundary effects, i.e. emissions
which occur within a distance $\sim \tau_{coh}$ from the medium edges at $x^+=0$ 
or $x^+=L$, are comparatively suppressed by a factor $\tau_{coh}/L\ll 1$.
Accordingly,
it is justified to restrict the time integrations in \eqn{DeltaHeq} to $0<t_1 <t_2< L$. In this
range, one can exploit the Gaussian nature of the medium correlations, cf. \eqn{Gaussian},  
to express the dipole $S$--matrix as  (compare to \eqn{Sdip1})
 \begin{align}\label{SdipEv}
 \mcal {S}_{t_2,t_1}(\bx,\by; \omega)\,=\,\exp\biggl\{-g^2 C_F\int_{t_1}^{t_2}\rmd t\,
 \Gamma_\omega(t,\bx-\by)\biggr\}\,=\,\exp\biggl\{-g^2 C_F (t_2-t_1)
 \Gamma_\omega(\bx-\by)\biggr\}
 \end{align}
 where $\Gamma_\omega(t,\bx-\by)\equiv \bar\Gamma_\omega(t, \bm{0}) -
 \bar\Gamma_\omega(t,\bx-\by)$.
The second equality in \eqn{SdipEv} holds for a medium which is homogeneous in time,
a case to which we shall restrict ourselves in what follows. In particular,
$\mcal{S} (\bx,\by)$ is obtained by replacing $t_2-t_1\to L$ in \eqn{SdipEv}. 

After using  \eqn{SdipEv} and restricting the time integrations to the support of the target,
\eqn{DeltaHeq} can be rewritten as an equation for $ \Gamma_\omega(\bx-\by)$ :
 \begin{align}
\label{DeltaHGamma}
  \hspace*{-.8cm}L\, \frac{\del
 \Gamma_{\omega} (\bx,\by)}{\del \omega} \,=\,&
  \frac{1}{4\pi \omega^3} 
  \int_{0}^L\rmd t_2 \int_{0}^{t_2}\rmd t_1\ 
  \del^i_{\br_1}\del^i_{\br_2}\Bigg\{  \int\big[\mcal{D}\br\big]
  \ \rme^{\rmi \,\frac{\omega}{2}
  \int\limits_{t_1}^{t_2}\rmd t' \,\dot\br^2(t')}\nn
 &  \hspace*{-.8cm} \times 
 \Bigg[\exp\bigg(\! -\frac{g^2N_c}{2}\!  \int_{t_1}^{t_2}\rmd t\,
\big[\Gamma_{\omega} (\bx,\br(t))+
 \Gamma_{\omega} (\br(t),\by)-\Gamma_{\omega} (\bx,\by)\big]\bigg)
 \,-\, 1\Bigg]
    \Bigg\}\Bigg|^{\br_2=\bx}_{\br_2=\by}\,
    \Bigg|^{\br_1=\bx}_{\br_1=\by}\ .\nn
  \end{align}
We have also used  $C_F\simeq N_c/2$, as appropriate at large $N_c$.

It is in fact easy to generalize the above results to {\em finite} $N_c$ and also
to a {\em generic representation} $R$ for the original color dipole 
(see Appendix \ref{sec:Nc} for details): within the limits of the
Gaussian approximation \eqref{Gaussian}, the average $S$--matrix for an 
$R\bar R$--dipole and for finite $N_c$ 
is given by \eqn{SdipEv} with $C_F\to C_R$ (the second Casimir for the respective
representation) and with the function $\Gamma_\omega(\bx-\by)$ 
obeying {\em exactly the same equation} as above, i.e.  \eqn{DeltaHGamma}. 
Such a simplification has been previously noticed in Ref.~\cite{Kovchegov:2008mk},
where the analog of \eqn{DeltaHGamma} has been proposed in the context of
a shockwave target (that is, as a mean field approximation to JIMWLK evolution
at finite $N_c$).

Note finally that \eqn{DeltaHGamma} can be viewed as a generalization
(and also a corrected version) of a corresponding result in Ref.~\cite{Liou:2013qya}, as
shown in Eqs.~(11--12) there\footnote{For the sake of this comparaison, 
note that the quantities denoted as $S(x_\perp)$
and $N(x_\perp, \omega)$ in  \cite{Liou:2013qya} correspond to our present quantities
$\Delta \mcal{S}(\bx,\by)$ and respectively $\omega(\del \mcal{S}/{\del \omega})$, cf.
 \eqn{DeltaS}. Hence,  Eq.~(12) in \cite{Liou:2013qya} must be compared to
the equation obtained by multiplying both sides of our \eqn{DeltaHGamma} by a factor
$[-\omega(g^2 N_c/2)\mcal{S}(\bx,\by)]$.}. \eqn{DeltaHGamma} is more general because
it is an {\em evolution equation}, whose solution (say, as obtained via successive iterations) 
would resum an infinite series of radiative corrections of arbitrary loop order
(within the high--energy approximations at hand).  
By comparison, Eq.~(12) in \cite{Liou:2013qya} is a one--loop result,
which can be viewed as the first iteration of our \eqn{DeltaHGamma} --- the limit
in which the r.h.s. of the latter is evaluated in the tree--level 
approximation (i.e., by using the expression \eqref{Sdip1} for the average $S$--matrix).
Moreover, even at one--loop order, Eq.~(12) in \cite{Liou:2013qya} mistreats
the `virtual' corrections:  instead of subtracting a term proportional to $\mcal{S}(\bx,\by)$, 
as required by the correct prescription in \eqn{DH}, the authors of Ref.~\cite{Liou:2013qya} 
have merely subtracted the vacuum limit ($A^-\to 0$) of the corresponding `real' term\footnote{In
our present set--up, the procedure of Ref.~\cite{Liou:2013qya} would amount to
subtracting just the {\em coefficient}  \eqref{virtdip} of the virtual term, and not the
product between that coefficient and the average $S$--matrix $\mcal{S} (\bx,\by)$
of the unevolved dipole.}. This imprecision has consequences to
leading logarithmic accuracy, as we shall see in the next subsection. (In Ref.~\cite{Liou:2013qya}, the
proper virtual term has been heuristically added in the calculation of the single--logarithmic corrections,
where it was indeed needed.)

\subsection{The single scattering approximation}
\label{sec:SSA}

In this section, we shall discuss two approximate versions of  
\eqn{DeltaHGamma}, the BFKL equation and the DLA equation,
 which, besides being simple enough to
allow for explicit solutions, have also the virtue to 
capture the most interesting physical regime for the high energy evolution of jet quenching.

\subsubsection{The BFKL equation}
\label{sec:BFKL}

\begin{figure}
\begin{center}
\begin{minipage}[b]{0.85\textwidth}
\begin{center}
\includegraphics[width=0.8\textwidth,angle=0]{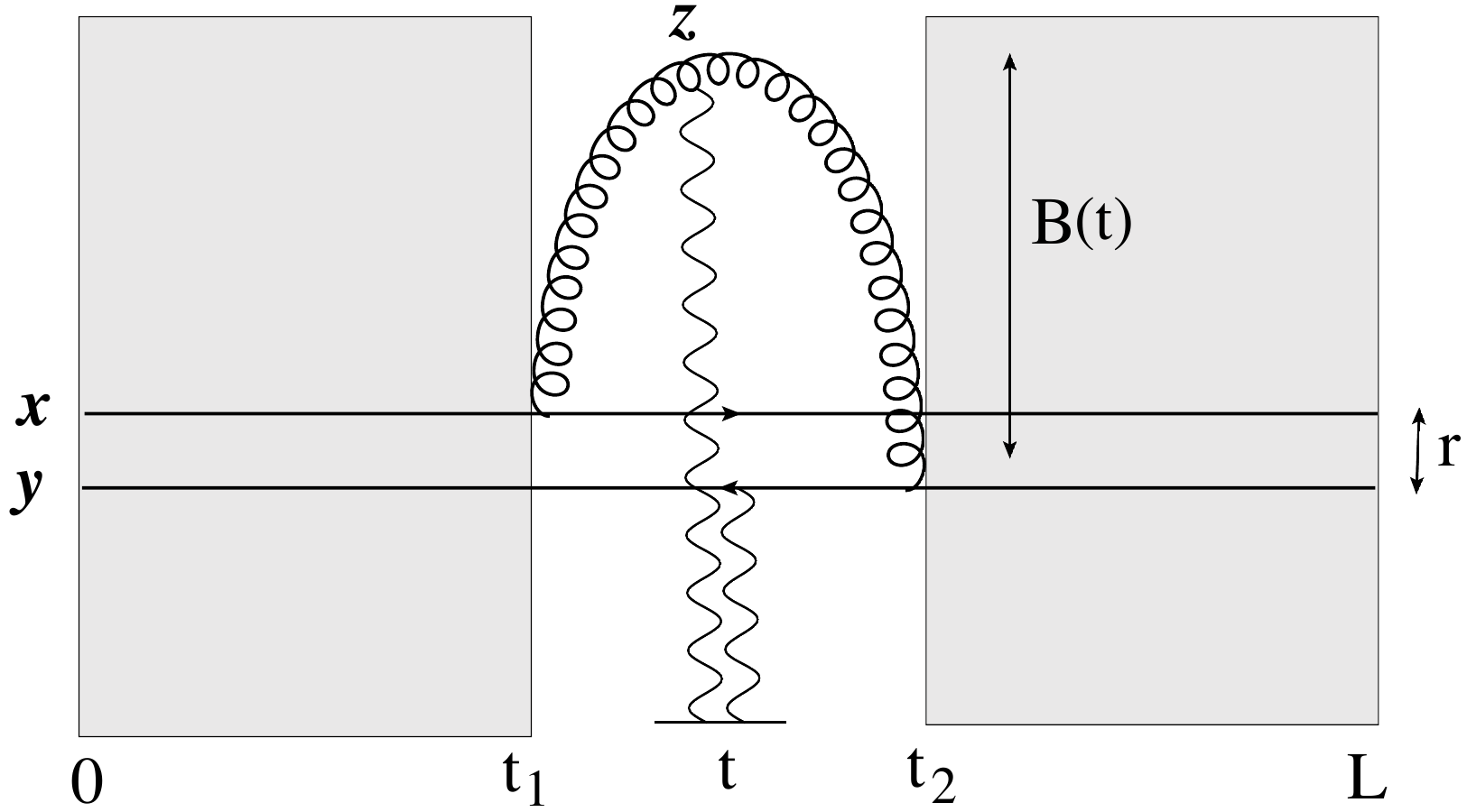}
\end{center}
\end{minipage}
\end{center}
\caption{\label{fig:dipoleDLA} \small\sl A diagram for dipole evolution in the single scattering
approximation, cf. Eqs.~\eqref{EqGamma1} and \eqref{BFKL}.
The grey areas prior and after the fluctuation are regions of multiple scattering.
During the fluctuation, the 3--parton ($q\bar q g$) system scatters only once, 
at some intermediate time $t$. The lifetime
$\Delta t=t_2-t_1$  of the fluctuation is considerably smaller than $L$; its 
transverse size $B_\perp$ is typically much larger than the size $r$ of the original dipole,
but much smaller than the `saturation' size ${2}/k_{_{\rm br}}(\omega)$ introduced
by multiple scattering.}
\end{figure}

We first recall our basic working assumptions: the 
transverse size $r\equiv |\bx-\by|$ of the original dipole is parametrically of order $1/Q_s$,
with $Q_s^2= \hat q L$,
and the typical energies of the emitted gluons obey $\omega \ll \omega_c$,
with $\omega_c = \hat q L^2$. 
Under these assumptions, we shall focus on the regime where the quark--gluon 
(`two--dipole') fluctuation living during 
the time interval $\Delta t=t_2-t_1$ undergoes a single collision with the medium,
as illustrated in Fig.~\ref{fig:dipoleDLA}.
We shall refer to this regime as the `single scattering approximation', but one should keep in 
mind that multiple scattering is still allowed prior to, and after, the fluctuation, that is, during the
comparatively large time intervals between $0$ and $t_1$ and, respectively, between
$t_2$ and $L$. Roughly speaking, this approximation is justified provided the transverse
separation between the soft gluon and the parent dipole (or, equivalently, the transverse
sizes of the two daughter dipoles) is small enough (see \eqn{range} below).

Within this  `single scattering' regime, one can expand the exponential within the 
square brackets in \eqn{DeltaHGamma} to linear order in its exponent. This gives
 \begin{align}
\label{EqGamma}
 L\, \frac{\del
 \Gamma_{\omega} (\bx,\by)}{\del \omega} \,=\,&-\frac{\alpha_sN_c}{2}
  \frac{1}{\omega^3} 
  \int_{0}^L\rmd t_2 \int_{0}^{t_2}\rmd t_1\ 
  \del^i_{\br_1}\del^i_{\br_2}\Bigg\{  \int\big[\mcal{D}\br\big]
  \ \rme^{\rmi \,\frac{\omega}{2}
  \int\limits_{t_1}^{t_2}\rmd t' \,\dot\br^2(t')}\nn
 &  \times  \int_{t_1}^{t_2}\rmd t\,
\Big[\Gamma_{\omega} (\bx,\br(t))+
 \Gamma_{\omega} (\br(t),\by)-\Gamma_{\omega} (\bx,\by)\Big]
    \Bigg\}\Bigg|^{\br_2=\bx}_{\br_2=\by}\,
    \Bigg|^{\br_1=\bx}_{\br_1=\by}\ .
  \end{align}
Once the solution to this linear equation is known, it can be also used to compute the effects of
{\em multiple} scattering, via  \eqn{SdipEv} which holds for generic values
of its exponent. A main virtue of \eqn{EqGamma} is that the time integrals
over $t_1$, $t_2$, and $t$, can be explicitly performed, as we now explain. To that aim, it is
convenient to reverse the order of these integrations, as follows:
\begin{align}
 \int_{0}^L\rmd t_2 \int_{0}^{t_2}\rmd t_1\int_{t_1}^{t_2}\rmd t\,=\,
 \int_{0}^L\rmd t \int_{0}^{t}\rmd t_1  \int_{t}^L\rmd t_2\,.\end{align}
After introducing the identity in the form 
$1=\int\rmd^2{\bz}\, \delta^{(2)}\big(\bz- {\br}(t)\big)$
and using \eqn{fact} with $\br(0)\to\br(t)$, \eqn{EqGamma} becomes
 \begin{align}
\label{EqGamma1}
 L\, \frac{\del
 \Gamma_{\omega} (\bx,\by)}{\del \omega} &\,=\,-\frac{\alpha_sN_c}{2}
  \frac{1}{\omega^3} \int\rmd^2{\bz}\, \Big[\Gamma_{\omega} (\bx,\bz)+
 \Gamma_{\omega} (\bz,\by)-\Gamma_{\omega} (\bx,\by)\Big]\nn
 &\times  \int_{0}^L\rmd t \int_{0}^{t}\rmd t_1  \int_{t}^L\rmd t_2\, 
  \del^i_{\br_1}\del^i_{\br_2}\Bigg\{ 
\mcal{G}_0(t_2-t,\br_2-\bz; \omega)\,\mcal{G}_0(t-t_1,\bz-\br_1; \omega)  
    \Bigg\}\Bigg|^{\br_2=\bx}_{\br_2=\by}\,
    \Bigg|^{\br_1=\bx}_{\br_1=\by}\ .\nn
  \end{align}
The time integrations can now be performed as in Eqs.~\eqref{int2}--\eqref{int1}.
For instance,
 \begin{align}\label{int2L}
\frac{1}{2\omega} \int_t^{L}\rmd t_2\, \del^i_{\br_2}\mcal{G}_0(t_2-t,\br_2-\bz; \omega)&\,=\,
\frac{-\rmi}{2\omega}
\int\frac{\rmd^2\bp}{(2\pi)^2}\,p^i\,\rme^{\rmi \bp\cdot(\br_2-\bz)}
 \int_t^{L}\rmd t_2\ \rme^{-\rmi \frac{p_\perp^2}{2\omega}(t_2-t)}\nn
 &\,=\,- \int\frac{\rmd^2\bp}{(2\pi)^2}\,\frac{p^i}
 {p_\perp^2}\ \rme^{\rmi \bp\cdot(\br_2-\bz)}\bigg[1-
 \rme^{-\rmi \frac{p_\perp^2}{2\omega}(L-t)}\bigg]\nn
 &\,\simeq\,- \int\frac{\rmd^2\bp}{(2\pi)^2}\,\frac{p^i}
 {p_\perp^2}\ \rme^{\rmi \bp\cdot(\br_2-\bz)}\,=\,\frac{\rmi}{2\pi}\,
 \frac{r_2^i-z^i}{(\br_2-\bz)^2}\,.
  \end{align}
The crucial approximation that has been performed here, was to neglect
the rapidly--oscillating complex exponential in the second line. This is indeed justified 
for the problem at hand (even beyond the single scattering approximation), 
because the lifetime $\tau_{coh}=2\omega/p_\perp^2$ of the interesting gluon
fluctuations is much smaller than $L$, as we shall see.
Accordingly, the final result in \eqn{int2L} is independent of $t$.
A similar reasoning applies to the integral over $t_1$, whose result can be read off
\eqn{int1}. The final integral over $t$ then simply yields a factor of $L$, which cancels
the similar factor in the l.h.s. of \eqn{EqGamma}.
Note that the dominant contributions come from times $t_1$ and $t_2$
which are distributed within a distance $\sim \tau_{coh}$ around the interaction time $t$~;
accordingly, one has $\Delta t\equiv t_2-t_1 \sim \tau_{coh}$, as expected from
the uncertainty principle.

We are thus lead to the following, relatively simple, equation for $\Gamma_{\omega}$,
 \begin{align}
\label{BFKL}
 \omega\, \frac{\del
 \Gamma_{\omega} (\bx,\by)}{\del \omega} &\,=\,\frac{\abar}{2\pi}
  \int_{\bm{z}}
 \mcal{M}_{\bx\by\bz}\, \Big[\Gamma_{\omega} (\bx,\bz)+
 \Gamma_{\omega} (\bz,\by)-\Gamma_{\omega} (\bx,\by)\Big]\,,
  \end{align}
where $\abar= \alpha_sN_c/\pi$ and the `dipole' kernel  $\mcal{M}_{\bx\by\bz}$
has been defined in \eqn{Mdef}. \eqn{BFKL} is recognized as the BFKL equation 
\cite{Lipatov:1976zz, Kuraev:1977fs,Balitsky:1978ic} (up to 
issues related to the integration limits, to be shortly discussed). In particular, the differential
operator in its l.h.s. is $\omega(\del/\del\omega) = \del/\del Y$, which demonstrates the logarithmic
enhancement of the respective radiative corrections.

Let us now clarify the validity limits for  \eqn{BFKL}. From its above
derivation, it is clear that this equation holds so long as the argument of the exponential within 
the square brackets in \eqn{DeltaHGamma} --- the amplitude for having a single 
scattering during $\Delta t$ for any of the two daughter dipoles --- 
remains much smaller than one. This condition can be written as
 \begin{align}\label{sat1}
  2g^2 C_F \Delta t\, \Gamma_{\omega}(B_\perp)\,\ll\,1\,,\end{align}
where $B_\perp$ is the maximal size of any of the daughter dipoles
during $\Delta t$ --- that is,  the largest among the distances $|\bx-\br(t)|$
and $|\br(t)-\by|$ for $t_1 < t < t_2$ ---  and the overall factor of 2 stands
for the two daughter dipoles (see also Fig.~\ref{fig:dipoleDLA}). As we shall shortly check, this 
$B_\perp$ is typically much larger than the size $r\sim 1/Q_s$ of the original dipole, 
hence it is indeed justified to indistinguishably treat the two daughter dipoles. 

To be more specific,
let us estimate the fluctuation time via the uncertainty principle, 
$\Delta t\simeq 2\omega/p_\perp^2\simeq \omega B_\perp^2/2$, 
and use the tree--level estimate for $\Gamma$ in \eqn{Sdip1}:  
$ g^2 C_F \Gamma(B_\perp)\simeq \hat q(1/B_\perp^2)\,B_\perp^2/4$.
Then the condition \eqref{sat1} can be rewritten as an energy--dependent upper limit 
on $B_\perp^2$~:
\begin{align}\label{sat2}
 B_\perp^2\,\ll\, \frac{4}{2\sqrt{\omega\hat q}}\,\equiv\,\frac{4}{k_{_{\rm br}}^2(\omega)}
 \,,\end{align}
with $\hat q$ itself evaluated on the scale set by this limit :
$\hat q= \hat q(k_{_{\rm br}}^2)$. 
This constraint will be confirmed by the analysis in Sect.~\ref{sec:multiple}, which 
also shows that the very large fluctuations, with sizes $B_\perp\gtrsim 2/k_{_{\rm br}}$, 
are efficiently suppressed by multiple scattering.
On the other hand, the very small fluctuations, with transverse sizes much smaller than 
$r\sim 2/Q_s$, do not contribute to the evolution, since their effects cancel 
between  the `real' and `virtual' terms in \eqn{BFKL}. 
To conclude, the  phase--space for the single scattering approximation reads 
 \begin{align}\label{range}
r\simeq \frac{2}{Q_s}\ \lesssim \ |\bx-\bz|\,,\, |\bz-\by|\ \ll\ 
\frac{2}{k_{_{\rm br}}(\omega)}
 \,.\end{align}
Via the uncertainty principle, \eqn{range}
implies that the transverse momentum $p_\perp$ of the emitted gluons 
lies within the range $k_{_{\rm br}}(\omega)\ll p_\perp\lesssim Q_s$.
This phase--space depends upon the energy $\omega$ of the emitted gluons, 
so it is important to recall that the interesting fluctuations have
$\omega \ll \omega_c$.  When $\omega$ approaches the upper limit $\omega_c$, 
one has $k_{_{\rm br}}(\omega_c)\sim Q_s$ and then the phase--space
in \eqn{range} shrinks to zero.
Vice--versa, for the soft gluons with $\omega \ll \omega_c$, one has 
$k_{_{\rm br}}^2(\omega)\ll Q_s^2$ and the transverse phase--space for linear evolution
is significantly large.

The previous discussion also explains why the present BFKL approximation is fundamentally 
different from that emerging in the context of the BK--JIMWLK evolution. In that case, the single
scattering approximation is obtained by linearizing the BK equation \eqref{BK} w.r.t. the dipole 
amplitude $\mathcal{T}_Y(\bx,\by)\equiv 1-\lan \hat{S}_{\bx\by} \ran_Y$. The ensuing equation
for $\mathcal{T}_Y(\bx,\by)$, which is formally similar to our \eqn{BFKL} above, is valid so long
as the scattering amplitude remains small, $\mathcal{T}_Y\ll 1$. In our present notations, this
condition is tantamount to $g^2 C_F L\, \Gamma_{\omega}(B_\perp)\ll 1$, or, equivalently,
$B_\perp^2\ll 1/Q_s^2$, with $Q_s^2=\hat q L$. This condition is independent of the energy
$\omega$ of the fluctuation (up to high--order effects introduced by the evolution) and implies that
the transverse phase--space for the evolution is roughly independent of the respective longitudinal
phase--space. As a consequence, the high--energy evolution is almost exclusively controlled by the 
longitudinal phase--space (which increases with $Y$) and it is faithfully described, so long as 
$B_\perp^2\lesssim 1/Q_s^2$, by the BFKL equation supplemented with the saturation boundary
$\mathcal{T}_Y < 1$ \cite{Iancu:2002tr,Mueller:2002zm,Munier:2003vc}. But for the present 
problem, the longitudinal and transverse phase--spaces are equally important and the
ensuing evolution is qualitatively different, as we shall shortly see.

\subsubsection{The double logarithmic approximation}
\label{sec:DLA}

The discussion around \eqn{range}
shows that, when increasing the {\em longitudinal} phase--space for the
evolution by decreasing $\omega$ below $\omega_c$, one simultaneously increases
the corresponding {\em transverse} phase--space, by decreasing the lower limit $k_{_{\rm br}}(\omega)$ on the transverse momentum $ p_\perp$  of the
fluctuations (or, equivalently, increasing the upper limit on their transverse size, cf. \eqn{range}). 
In view of this and of the well--known
fact that the BFKL evolution admits a double--logarithmic regime \cite{Kovchegov:2012mbw},
it is clear that the radiative corrections can be 
enhanced not just by the large energy logarithm
$\ln(\omega_c/\omega)$, but also by the even larger  {\em double logarithm}
$\ln(\omega_c/\omega) \ln(Q_s^2/k_{_{\rm br}}^2(\omega)=(1/2)
\ln^2(\omega_c/\omega)$. This enhancement has been previously recognized in 
Ref.~\cite{Liou:2013qya}, where the respective correction to the transverse momentum 
broadening has been first computed (see also \cite{Wu:2011kc,Blaizot:2013vha}). 

In what follows, we will use \eqn{BFKL} to identify, compute, and resum the corrections 
enhanced by a double logarithm. To that aim, we need to focus
on the relatively large fluctuations with $|\bx-\bz|\simeq |\bz-\by|\gg r$. 
 Since the dipole scattering amplitude $\Gamma_{\omega} (\bx,\bz)$
is a rapidly growing function of the dipole size $|\bx-\bz|$ (see below), it is quite clear that,
in this regime, one can neglect the `virtual' term $\propto \Gamma_{\omega} (\bx,\by)$ in \eqn{BFKL}.
Then this equation simplifies to
  \begin{align}
\label{DLA0}
 \omega\, \frac{\del
 \Gamma_{\omega} (r)}{\del \omega} &\,\simeq\,{\abar}
  \int \rmd B_\perp^2\, \frac{r^2}{B_\perp^4} \,
  \Gamma_{\omega} (B_\perp)\,,
  \end{align}
where we have used $B_\perp$ to denote the size of any of the two daughter dipoles. 
The initial condition for this equation at $\omega\simeq \omega_c$, i.e. the tree--level result
in \eqn{Sdip1}, is roughly proportional to the dipole size squared.  Remarkably,
 \eqn{DLA0} shows that
this property is preserved by the evolution under the approximations of interest. Hence, we can
write
 \begin{align}\label{GammaDLA}
 g^2 C_F \Gamma_{\omega}(B_\perp)\,\equiv\, \frac{1}{4}\,
 \hat q_{\omega}(1/B_\perp^2)\,B_\perp^2
 \,,
 \end{align}
where the function $ \hat q_{\omega}(1/B_\perp^2)$ has only a weak
dependence upon $B_\perp^2$ (for $\omega\simeq \omega_c$, it reduces to
the zeroth order result in  \eqn{qhat}).  Then \eqn{DLA0} implies the following
equation for $ \hat q_{\omega}$~:
  \begin{align}
\label{DLA}
 \omega\, \frac{\del
\hat q_{\omega}(1/r^2)}{\del \omega} &\,\simeq\,{\abar}
  \int_{\, r^2}^{{4}/{k_{_{\rm br}}^2(\omega)} }
  \frac{\rmd B_\perp^2}{B_\perp^2} \,\hat q_{\omega}(1/B_\perp^2)
  \,,
  \end{align}
where the limits in the integral over $B_\perp^2$ follow from the previous discussion.
\eqn{DLA} looks similar to the familiar `double--logarithmic approximation' (DLA)
to the BFKL equation (see e.g. \cite{Kovchegov:2012mbw}). 
But one should keep in mind that the upper limit in the above integral,
which is energy--dependent, 
is specific to the problem at hand and reflects the non--linear physics
of multiple scattering. 
For this particular problem, and unlike in more standard applications of the BFKL equation to 
high--energy scattering \cite{Kovchegov:2012mbw,Iancu:2002tr,Mueller:2002zm,Munier:2003vc}, 
the DLA  encompasses
the {\em dominant} radiative corrections in the high--energy limit $\omega_c/\omega \gg 1$~:
its solution is equivalent to the resummation of the double--logarithmic corrections singled out
in \cite{Liou:2013qya} (see below).

Importantly, the approximation \eqref{GammaDLA} for the dipole amplitude
automatically implies that the radiative corrections that we are about to
compute can be absorbed into a redefinition of $\hat q$. Indeed, with this approximation
for $ \Gamma_{\omega}$, the evolved dipole $S$--matrix in \eqn{SdipEv} has the same formal
structure as at tree--level, that is (compare to \eqn{Sdip1})
\begin{align}\label{SDLA}
 \mcal {S}_{\omega}(\br)\,\simeq\,\exp\left\{-\frac{1}{4}L \hat q_\omega(1/r^2)\,\br^2\right\}.
 \end{align}
In turn, this implies that the quark spectrum has the Gaussian form in \eqn{ptGauss}, but with a
{\em renormalized}, energy--dependent, saturation momentum, defined by
$Q_s^2=\hat q_\omega(Q_s^2) L$.

Consider now the first iteration to \eqn{DLA} and assume for simplicity that
the respective zeroth order result (to be denoted as $\hat q^{(0)}$ in what follows)
is scale--independent. The first order correction implied by \eqn{DLA} reads\footnote{To
the accuracy of interest, one can replace $\hat q\simeq \hat q^{(0)}$ within 
the argument of the logarithm.}
  \begin{align}
\label{DLA1}
\delta \hat q_{\omega}^{(1)}(Q_s^2)\,=\,&{\abar} \hat q^{(0)}\int_{\omega}^{\omega_c}
\frac{\rmd\omega_1}{\omega_1}
  \int^{Q_s^2}_{k_{_{\rm br}}^2(\omega_1)} \frac{\rmd p_\perp^2}{p_\perp^2} \,
  =\,{\abar} \hat q^{(0)}\int_{\omega}^{\omega_c}
\frac{\rmd\omega_1}{\omega_1}\,\ln\frac{Q_s^2}{k_{_{\rm br}}^2(\omega_1)}\nn
 \,\simeq\,&\frac{\abar}{2}\, \hat q^{(0)}\int_{\omega}^{\omega_c}
\frac{\rmd\omega_1}{\omega_1}\,\ln\frac{\omega_c}{\omega_1}\,=\,\frac{\abar}{4}\, \hat q^{(0)}
 \ln^2\frac{\omega_c}{\omega}
    \,,
  \end{align}
where the approximate equality sign refers to the double--logarithmic accuracy and
we preferred to use the transverse momentum variable $p_\perp^2\equiv 4/B_\perp^2$
as an integration variable, instead of $B_\perp^2$. As expected,
this correction is of order $\abar$, but it is enhanced by the potentially large double logarithm
$\ln^2 ({\omega_c}/{\omega})$. To understand how large can this actually be, 
one needs to know what is the minimal value for $\omega$ which is allowed on physical grounds.
This issue has been previously addressed in Ref.~\cite{Liou:2013qya}, where one has argued
that this minimal value is controlled by a lower limit $\lambda$ on the lifetime 
$\tau_{coh}=2\omega/p_\perp^2$ of the fluctuations, which is
independent of $L$. In the next subsection, we shall revisit and
complete the arguments in Ref.~\cite{Liou:2013qya} and thus clarify the physical origin and the value
of $\lambda$. But for the time being, it suffices to know that such a cutoff exists and examine its
consequences.  As we now explain, this implies the existence of a large phase--space
for double--logarithmic evolution in the regime where $L\gg\lambda$.

Specifically, the condition $\tau_{coh} > \lambda$ implies
a lower limit on the gluon energy, $\omega> \lambda p_\perp^2/2$, which
also depends upon its transverse momentum $p_\perp$. 
This last feature
forces us to modify our previous analysis leading to \eqn{DLA1}.
Indeed, in the integral over $p_\perp^2$ within that equation, we have assumed
that the maximal limit is equal to $Q_s^2$, but in reality this cannot exceed a value
$p^2_{\perp {\rm max}}=2\omega_1/ \lambda$ which also depends upon 
$\omega_1$ (the other integration variable there).
That is, the proper integration range in $p_\perp^2$ for a given value of $\omega$ is
 \begin{align}\label{rangep}
 k_{_{\rm br}}^2(\omega)\,\ll\,p_\perp^2\,\ll\,{\rm min}\bigg(\frac{2\omega}{\lambda}, Q_s^2\bigg)
 \,.\end{align}
 The upper limit introduces two constraints. First, it implies the necessary condition
 $2\omega/ \lambda\gg  k_{_{\rm br}}^2(\omega)$, or equivalently
 $\omega \gg \omega_0\equiv \hat q \lambda^2$,
showing that $\omega_0$ is the absolute lower limit on the energy $\omega$ of the fluctuation.
 Second, it means that, when performing the integrals over the phase--space,
 one must distinguish between two different ranges in $\omega$,
 namely  $\omega_0 < \omega <\omega_*$ and $\omega_* < \omega <\omega_c$,
 where $\omega_*\equiv Q_s^2\lambda/2$ is the energy for which $Q_s^2$
 becomes equal to $2\omega/ \lambda$ and obeys $\omega_* \gg \omega_0$ for
$L\gg\lambda$.

 To summarize, the DLA phase--space is defined as the range \eqref{rangep} in $p_\perp^2$
 together with the energy range at $ \omega_0 \ll \omega \ll \omega_c$.
 The whole analysis becomes more transparent if, instead
 of the original variables $\omega$ and $p_\perp^2$, one uses the variables 
 $\tau\equiv 2\omega/p_\perp^2$ (the lifetime of the fluctuation) and  $p_\perp^2$.
 As an intermediate step, notice
 that \eqn{rangep} is tantamount to 
  \begin{align}\label{rangetau2}
{\rm max}\bigg(\lambda,\,  \frac{2\omega}{Q_s^2}\bigg)
\ \ll \ \tau \ \ll \ \tau_{_{\rm br}}(\omega)\,\equiv\,
  \frac{2\omega}{k_{_{\rm br}}^2(\omega)}\,=\,\sqrt{\frac{\omega}{\hat q}}\,.\end{align}
 Then it is easy to see that, in terms of the new variables
$\tau$  and  $p_\perp^2$, the DLA phase--space can be simply characterized as
(see also Fig.~\ref{fig:phasespace} for an illustration)
 \begin{align}\label{DLAps}
 \lambda\,\ll\,\tau\,\ll\,L\,,\qquad
 2\hat q\tau\,\ll\,p_\perp^2\,\ll\, Q_s^2\,.\end{align}
 Then the  first order correction is computed as (compare to \eqn{DLA1})
 \begin{align}
\label{DLAfin}
\delta \hat q^{(1)}\,=\,&{\abar} \hat q^{(0)}\int_{\lambda}^{L}
\frac{\rmd\tau}{\tau}
  \int^{Q_s^2}_{2\hat q\tau} \frac{\rmd p_\perp^2}{p_\perp^2} \,
  =\,{\abar} \hat q^{(0)} \int_{\lambda}^{L}
\frac{\rmd\tau}{\tau}\, \ln\frac{L}{2\tau} 
\,\simeq\,\hat q^{(0)}\,
\frac{\abar}{2}
 \ln^2\frac{L}{\lambda}  \,,
  \end{align}
where we have used $Q_s^2=\hat q L$. (As before, we ignore the difference between 
$\hat q$ and $\hat q^{(0)}$, or the factors of 2, in the argument of the logarithm.)
The above result, which is in agreement with Refs.~ \cite{Liou:2013qya,Blaizot:2013vha},
is not the same as the result of evaluating \eqn{DLA1} 
at the minimal value of the energy $\omega=\omega_0$. The last calculation
would be naive, in that it would incorrectly treat the contribution of the low energy 
region at $\omega_0 < \omega < \omega_*$. 

More generally, it is preferable to use the new variables $\tau$  and  $p_\perp^2$
also within the  DLA equation \eqref{DLA} and to replace the latter by
an integral equation, where all the integration limits are explicit. This equation reads
 \begin{align}\label{DLAint}
  \hat q_L(Q_s^2)
  \,=\,\hat q^{(0)}+\abar \int^{L}_\lambda
   \frac{\rmd \tau}{\tau} 
 \int_{\hat q\tau}^{Q_s^2} \frac{\rmd p_\perp^2}{p_\perp^2} \,\hat q_{\tau}(p_\perp^2)
  \,.
  \end{align}
\eqn{DLAint} shows that the physical quantity $\hat q_L(Q_s^2)$ of interest
 ---  the renormalized jet
quenching parameter as obtained after including the radiative corrections to DLA accuracy ---
is obtained as the value of a function of two variables, $\hat q_{\tau}(p_\perp^2)$,
at the physical point $\tau=L$ and  $p_\perp^2=Q_s^2(L)$ in the phase--space.
(As it will be explained in Sect.~\ref{sec:gluon}, this physical point lies on the saturation line
for the gluon distribution of the medium; see also Fig.~\ref{fig:phasespace}.) In turn, the function
 $\hat q_{\tau}(p_\perp^2)$ has support at $p_\perp^2>\hat q\tau$ and is obtained
 as the solution to the following integral equation:
 \begin{align}\label{DLAgen}
  \hat q_{\tau}(p_\perp^2)\,=\,\hat q^{(0)}+\abar \int^{\tau}_\lambda
   \frac{\rmd \tau_1}{\tau_1} 
 \int_{\hat q\tau_1}^{p_\perp^2} \frac{\rmd k_\perp^2}{k_\perp^2} \,\hat q_{\tau_1}(k_\perp^2)
  \,.
  \end{align}
 As already stressed after \eqn{DLA}, the above equation differs from the standard
DLA equation which appears e.g. in studies of the jet evolution in the vacuum
\cite{Dokshitzer:1991wu,Kovchegov:2012mbw} via the $\tau$--dependence
of the lower limit in the integral over $p_\perp^2$, 
which comes from the restriction to single scattering.

\eqn{DLAgen} can be easily solved via iterations. The first iteration (with $\hat q^{(0)}$
assumed to be scale--independent, once again) gives
 \begin{align}\label{DLA1}
  \delta \hat q^{(1)}_{\tau}(p_\perp^2)\,=\,\hat q^{(0)}\,\frac{\abar}{2}\left(
\ln^2\frac{p_\perp^2}{\hat q\lambda}-\ln^2\frac{p_\perp^2}{\hat q\tau}\right)
 \,,
  \end{align}
which on the physical point $\tau=L$ and  $p_\perp^2=\hat q L$ reduces to \eqn{DLAfin}, as
it should. The second iteration yields (we only show its result at the physical point)
  \begin{align}
\label{DLAfin2}
\delta \hat q^{(2)}\,=\,\hat q^{(0)}\,
\frac{\abar^2}{2! 3!}
 \ln^4\frac{L}{\lambda}  \,.
  \end{align}
These and the subsequent terms in this iterative procedure are recognized as the Taylor expansion
of the modified Bessel function\footnote{This should be contrasted to the standard DLA solution, 
which involves the Bessel function  ${\rm I}_0(x)$
\cite{Kovchegov:2012mbw}.} 
 ${\rm I}_1(x)$ :
 \begin{align}\label{qeff}
 \hat q_L(Q_s^2)\,=\, \hat q^{(0)}
 \,\frac{1}{\sqrt{\abar}\ln\big(L/\lambda\big)}\,{\rm I}_1\left(2\sqrt{\abar}
 \ln\frac{L}{\lambda}\right)\,.
 \end{align}
The same result has been obtained in Ref.~\cite{Liou:2013qya} via a resummation of
the relevant Feynman graphs. This resummation  becomes pertinent 
when the medium is large enough, such that $\abar  \ln^2({L}/{\lambda})\gtrsim 1$.
In such a case, the radiative corrections enhance the medium--size dependence of the
(renormalized) jet quenching parameter, which thus becomes even more non--local than it 
was at tree--level.

We conclude this subsection with a few comments on the physical meaning of the radiative
corrections displayed in Eqs.~\eqref{DLAfin} or \eqref{qeff}. As obvious from the previous
calculations, these corrections are generated by the emission of soft gluons with energies
$\omega$ deeply within the range between $\omega_0=\hat q \lambda^2$ and
$\omega_c= \hat q L^2$ and with transverse momenta $p_\perp$ deeply between 
$k_{_{\rm br}}(\omega)$ and $Q_s$.
Such gluons have lifetimes considerably smaller than the medium longitudinal size $L$ and 
transverse sizes which are considerably larger than the size $r\sim 1/Q_s$ of the original dipole. 
This hierarchy is furthermore respected by the successive emissions which are summed 
up by \eqn{qeff} and whose energies are softer and softer with increasing generation. 
Because of this hierarchy, the corrections appear to be quasi--local on the longitudinal 
scale relevant for measuring the transverse momentum broadening, which is $L$.
Similarly, they do not affect the transverse resolution on
which we scrutinize the medium properties, which is set by $Q_s$. This ultimately explains why
such corrections can be simply accounted for by a renormalization of the jet quenching parameter
$ \hat q_L(Q_s^2)$.

\subsubsection{The phase--space for the high--energy evolution}
\label{sec:DLAPS}

In the previous section, we have argued that the phase--space for the double--logarithmic
approximation is given by \eqn{DLAps}, where the lower limit $\lambda$  
has not yet been specified. In this subsection, we shall first explain 
the physical origin and the value of $\lambda$, thus following a discussion in 
Ref.~\cite{Liou:2013qya}. Then we shall critically revisit the original 
arguments in Ref.~\cite{Liou:2013qya} and demonstrate
that, in general, the structure of the DLA phase--space is more 
complicated than suggested there (and shown in \eqn{DLAps}). 
The differences are unessential in the limit where the medium size $L$ is arbitrarily large,
but they become important for realistic values of $L$, in which case they limit the
validity of the DLA. Based on such considerations, we shall derive the constraint
\eqref{wQGP} for the applicability of the DLA (and, more generally, of the present 
high--energy approximations) in the case where the target is a weakly coupled QGP.

As discussed in Ref.~\cite{Liou:2013qya}, the existence of a lower limit $\lambda$ 
on the lifetime  $\tau$ of the fluctuations 
follows from energy--momentum conservation. In the
high--energy approximation of interest, the gluon fluctuation is nearly on--shell, 
so it carries a `minus' momentum $p^- = {p_\perp^2}/{2\omega}$ (recall that $\omega \equiv p^+$).
This component cannot be inherited from the parent quark, which is a right mover, so it must
be acquired via interactions with the medium. Since moreover we assume that there is only
one scattering during the fluctuation, it is clear that either the virtual gluon, or its
parent quark, must have absorbed a quanta having this momentum $p^-$. This quanta is a
(generally off--shell) gluon exchanged between the projectile and some constituent 
of the medium --- say, a thermal quark or gluon, in the case where the medium is a
finite temperature plasma. Let us denote by $k^-$ the respective 4--momentum component 
of that particular constituent and introduce the {\em longitudinal momentum fraction} $x\equiv
p^-/k^-$, with $x\le 1$ of course. (This is `longitudinal' since the medium is a left mover.)
We have
 \begin{align}\label{smallx}
 x\,=\,\frac{p_\perp^2}{2\omega k^-}\,=\,\frac{p_\perp^2}{2 p\cdot k}\,,
 \end{align}
where the second equality, which confirms that $x$ is boost invariant, follows from the high
energy kinematics. Indeed, in whatever frame we use, at least one of the two subsequent 
statements is correct:
(I)  $p^+$ is the large component of the 4--momentum of the gluon fluctuation, and
(II) $k^-$ is the large component of the 4--momentum of the medium constituent.

In particular, in the plasma rest frame, one has $k^-\simeq T$ for a typical plasma particle 
and \eqn{smallx} becomes
 \begin{align}\label{smallxT}
 x\,\simeq\,\frac{p_\perp^2}{2\omega T}\,=\,\frac{\lambda}{\tau}\,,\end{align}
where $\tau=2\omega/p_\perp^2$ and $\lambda\equiv 1/T$ 
is the typical thermal wavelength. Since $x\le 1$ and $\tau \le L$, the above relation implies the
following ranges of values for $x$ and $\tau$ :
 \begin{align}\label{rangex}
  \frac{\lambda}{L}\,\le\,x\,\le\,1\,,\qquad L\ge\,\tau \,\ge\, \lambda\,.\end{align}
This motivated the authors of  Ref.~\cite{Liou:2013qya} to choose $\lambda=1/T$ 
as the minimal value for $\tau$ in equations like \eqref{DLAint}.  This conclusion is essentially correct, 
but its validity is restricted by an addition kinematical constraint, that has been overlooked in
the analysis in  Ref.~\cite{Liou:2013qya} and that we shall now discuss.

\begin{figure}
\begin{center}
\begin{minipage}[b]{0.95\textwidth}
\begin{center}
\includegraphics[width=0.8\textwidth,angle=0]{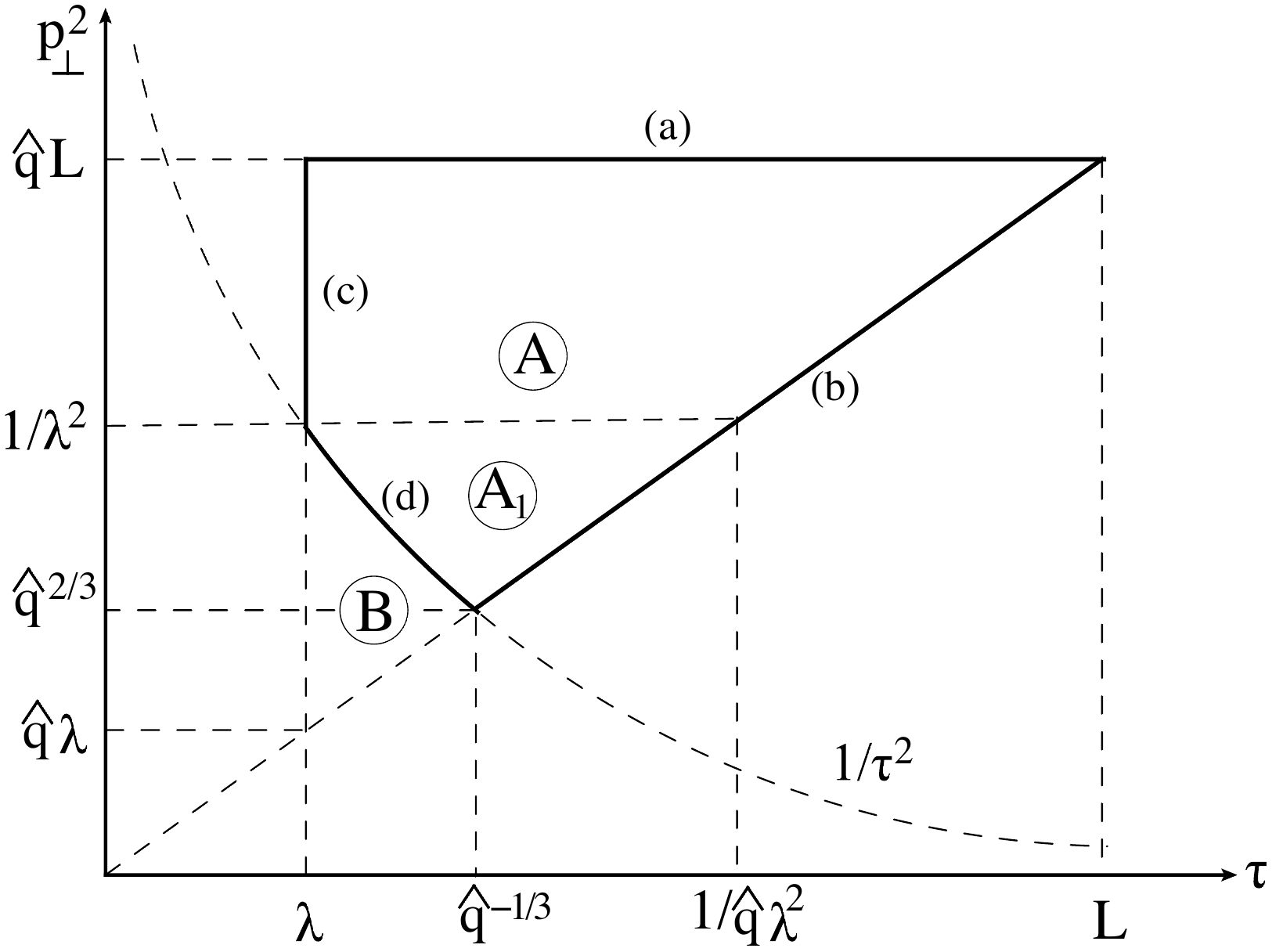}
\end{center}
\end{minipage}
\end{center}
\caption{\label{fig:phasespace} \small\sl The phase--space for the high--energy evolution of 
jet quenching, in terms of the variables $\tau$ and $p_\perp^2$ (the lifetime
and the transverse momentum squared of the gluon fluctuations). We assume 
$\hat q\lambda^3 \ll 1$ and $\hat q L \lambda^2\gg 1$.
Line (b) is the `saturation line' $p^2_\perp=\hat q\tau$ (see Sect.~\ref{sec:gluon}).
Line (d) reads $p^2_\perp=1/\tau^2$ and implements the kinematic constraint $p^+>p_\perp$.
The phase--space for the double--logarithmic evolution is region A, as delimited by the lines
(a), (b), (c), and (d). For discussions of regions B and A$_1$, see the main text.}
\end{figure}

The high--energy picture that we have developed so far is based on the assumption
that the gluon fluctuations are very energetic in the plasma rest frame, meaning that
their `energy' $\omega$ is (much) larger than their transverse momentum: $\omega\gg
p_\perp$.  In particular, this must be true for the hardest allowed fluctuations, with
energy $\sim\omega_c$ and transverse momentum $\sim Q_s$. Hence, the inequality 
$\omega_c\gg Q_s$, or equivalently $\hat q L^3\gg 1$, is a necessary condition for
the validity of our approach. This condition has been implicitly assumed 
throughout our analysis and is indeed well satisfied in practice.
Returning to generic values for $\omega$ and $p_\perp$, we observe that the
kinematical constraint $\omega\gg p_\perp$ implies the following conditions on the lifetime 
$\tau=2\omega/p_\perp^2$ of the fluctuations: $\tau\gg 1/p_\perp\gg 1/\omega$. 
For a given $\tau$,  the transverse 
momentum cannot be smaller than a value $p_\perp^{\rm min}\sim \sqrt{\hat q\tau}$ introduced 
by multiple scattering. So, clearly, the kinematical constraint $\tau> 1/p_\perp$ is
satisfied for any permitted value of $p_\perp$ provided the following condition is fulfilled:
\beq\label{cond0}
\tau\,\gtrsim\,\frac{1}{p_\perp^{\rm min}}\,\sim\,\frac{1}{\sqrt{\hat q\tau}}
\ \Longrightarrow \ \tau \,\gtrsim\,\hat q^{-1/3}\,.\eeq
If this condition was satisfied for any $\tau$ within the range
 $\lambda \lesssim \tau <  L$, then the kinematical constraint would 
play no special role for the present analysis (since automatically satisfied throughout
the phase--space). 
But for a weakly coupled QGP and with $\lambda=1/T$,
the condition \eqref{cond0} is {\em not} satisfied for sufficiently
small values $\tau\sim\lambda$~: one has indeed
$\hat q^{1/3}\lambda\sim [\alpha_s^{2}\ln(1/\alpha_s)]^{1/3}\ll 1$.

The solution to this problem is in fact quite simple\footnote{I am grateful
to Al Mueller for clarifying discussions on this point.}: it suffices to impose the additional
constraint $\tau > 1/p_\perp$ on the kinematical domain \eqref{DLAps} for the double--logarithmic
contributions. This leads to the phase--space denoted by the letter A in Fig.~\ref{fig:phasespace}.
Note that, in drawing this figure, we have chosen not only  $\hat q L^3\gg 1$ (together
with $\hat q \lambda^3\ll 1$, of course), but also the stronger condition  
$\hat q L \lambda^2\gg 1$. The reason for that should shortly become clear.
 
The domain A in Fig.~\ref{fig:phasespace}
differs from the original  phase--space in  \eqref{DLAps}  by the domain
denoted there by the letter B, whose contribution to $\delta \hat q$ can be computed as
\beq\label{hatqB}
\delta \hat q_{\rm B}\,=\,
\abar \hat q^{(0)}\,
\int^{\hat q^{-1/3}}_\lambda
   \frac{\rmd \tau}{\tau} 
 \int_{\hat q\tau}^{1/\tau^2} \frac{\rmd p_\perp^2}{p_\perp^2} \,
=\,\hat q^{(0)}\,
\frac{\abar}{6}\ln^2\frac{1}{\hat q \lambda^3}\,.\eeq
This contribution is independent of the medium size $L$, so clearly it becomes negligible
compared to the DLA result in \eqn{DLAfin} --- the contribution of
the domains A$\cup$B in  Fig.~\ref{fig:phasespace} --- for sufficiently large values of
$L$. The precise condition is
\beq\label{wQGP} 
\frac{L}{\lambda}\,\gg\,\frac{1}{\hat q \lambda^3}\,\sim\,\frac{1}{
\alpha_s^{2}\ln(1/\alpha_s)}\,,\quad\mbox{or, equivalently,}\qquad
Q_s^2\,\gg\,\frac{1}{\lambda^2}\,\sim\,T^2\,,\eeq
where the parametric estimates refer to the weakly coupled QGP. Vice--versa,
these considerations suggest the existence of $L$--independent radiative corrections 
of order $\alpha_s \ln^2(1/\alpha_s^2)$, which are quite large and are
not properly taken into account by our high--energy approximations.

Note finally that, under the same assumptions as above, cf. \eqn{wQGP}, the radiative
corrections of interest for us here can be fully attributed to the relatively hard fluctuations, with
transverse momenta within the range
$Q_s\gtrsim p_\perp\gg 1/\lambda = T$. Indeed, the respective contribution of the softer momenta
$p_\perp\le 1/\lambda$ is given by the domain A$_1$  in Fig.~\ref{fig:phasespace} (a subdomain of
region A) and can be easily estimated as
 \beq
\delta \hat q_{\rm A_1}\,=\,
\abar \hat q^{(0)}\,
 \int_{\hat q^{2/3}}^{1/\lambda^2} \frac{\rmd p_\perp^2}{p_\perp^2} 
 \int_{1/p_\perp}^{p_\perp^2/\hat q} 
   \frac{\rmd \tau}{\tau} \,
=\,\hat q^{(0)}\,
\frac{\abar}{3}\ln^2\frac{1}{\hat q \lambda^3}\,.\eeq
This contribution is comparable to that in \eqn{hatqB} and hence it is negligible under the
same conditions. Moreover, these small contributions, from domains B and A$_1$, are not even
enhanced by a {\em single} large logarithm $\ln(L/\lambda)$, so they are irrelevant for 
the high--energy evolution.

To summarize, the gluon fluctuations which control the high--energy evolution of a slice of
weakly--coupled QGP which is large enough (in the sense of \eqn{wQGP}) are characterized by large
lifetimes $\tau \gg 1/T$, large transverse momenta  $p_\perp\gg T$, and even larger energies 
$\omega\gg p_\perp$.  In particular, the phase--space for DLA can be restricted to the
`hard' region depicted as $ {\rm A}\backslash {\rm A}_1$ (i.e. the difference betweens 
the domains A and A$_1$) in Fig.~\ref{fig:phasespace}.

\subsection{Gluon evolution and saturation in the medium}
\label{sec:gluon}

In this subsection, we shall develop an alternative physical picture for the high--energy evolution 
of jet quenching in terms of the gluon distribution in the medium.
In particular, we would like to argue that the multiple scattering between the soft fluctuations
and the medium can be alternatively interpreted as saturation effects in  the gluon distribution  
at small $x$. 
To develop this new picture, we will have to use a different Lorentz frame, namely, an `infinite
momentum frame' for the medium, in which the relevant gluon fluctuations appear as
`partons' from the plasma. A similar picture has been developed for a strongly coupled
plasma described by ${\mathcal N}=4$ SYM \cite{Hatta:2007cs,Hatta:2008tx,Dominguez:2008vd},
but we are not aware of previous, related, discussions at weak coupling, except at tree--level
(cf. Sect.~\ref{sec:tree}).

Namely, consider the interaction between the projectile  (the quark, or the dipole) and
the medium (plasma) in a frame where the projectile is quite slow whereas the target
is an ultrarelativistic left mover, with a Lorentz boost factor of order\footnote{In the
plasma rest rame, a gluon fluctuation with energy $\omega$ and transverse momentum $k_\perp$
has a rapidity $\eta$ such that $\gamma\equiv \cosh \eta \simeq \omega/k_\perp$.
Since $k_\perp^2\gtrsim k_{_{\rm br}}^2=\sqrt{\hat q \omega}$, this implies $\gamma
\lesssim (\omega^3/\hat q)^{1/4}$, where the upper limit reaches a maximal value
$\gamma_{\rm max}=\omega_c/Q_s$ corresponding to $\omega=\omega_c$.}
$\gamma\simeq\omega_c/Q_s=\sqrt{\hat q L^3}$. 
In this frame, all the fluctuations that we have discussed so far
become left movers, so they are more naturally associated with the plasma. 
They cannot be a part of the thermal distribution, since that was not true in the 
plasma rest frame and the thermal distribution is boost invariant. Rather, they must be 
considered as bremsstrahlung (or Weizs\"acker--Williams) quanta emitted by the
medium constituents (thermal quarks and gluons). Thus, in this boosted frame, the relevant
fluctuations are a part of the {\em medium gluon distribution}.

The typical fluctuations carry small fractions $x\ll 1$ of the longitudinal ($k^-$) momenta
of their parent particles (which are large, $k^-\simeq \gamma T$, in the boosted frame).
Accordingly, they have large wavelengths $\Delta x^+\sim 1/(xk^-)\gg 1/k^-$, meaning that they
overlap with many medium constituents. On the other hand, they have very short lifetimes
$\Delta x^- = 1/p^+ \ll \gamma/T$, which explain why they cannot thermalize.
(Notice that $\gamma/T$ is the smallest time scale associated with the thermal distribution
in this frame.) Furthermore, the high--energy evolution that we had previously
associated with the  wavefunction  of the projectile can
be alternatively interpreted as an evolution of the medium gluon distribution with decreasing
$x$.  Interestingly, this evolution is somewhat different than it would be in a shockwave : 
the small--$x$ gluons 
cannot overlap with all the color sources within a longitudinal tube throughout the target (as they
do in a shockwave  \cite{Mueller:2001fv,Iancu:2002xk,Gelis:2010nm}), 
but only with those within a longitudinal distance 
$\Delta x^+\sim 1/(xk^-)$. This is consistent with the peculiar boundaries 
on the phase--space for linear evolution, as discussed in the previous sections.
It furthermore implies a stronger $x$--dependence of the respective saturation momentum,
as we now explain.

To make contact with the previous developments, let us
recall that the `unintegrated gluon distribution' in the target --- the number
of gluons per unit rapidity $Y\equiv \ln(1/x)$ and per unit transverse phase--space --- 
is closely related
to the cross--section  \eqref{ptbroad} for transverse momentum broadening. 
One has indeed (see e.g. the discussion in Ref.~\cite{Dominguez:2011wm})
\begin{align}
 \label{unint}
 x\,\frac{\rmd N_{\rm g}}{\rmd x\,\rmd^2\bp\, \rmd^2\bm{b}} \,=\,\frac{N_c^2-1}{4\pi^3}
 \, \frac{p_\perp^2}{g^2 C_F} \int_{\br} \rme^{-\rmi \bp \cdot \br} 
\mathcal{S}(\br)\,,\end{align}
where the subscript g stays for `gluon' and
 $\bm{b}$ denotes the position in transverse space, or `impact parameter'.
More precisely, \eqn{unint} 
is the gluon distribution `unintegrated' in the transverse phase--space,
but integrated in the longitudinal ($x^+$) direction over the whole size $L$ of the target.
Since our medium is assumed to be homogeneous in both $x^+$ and $\bx$, the 
{\em  occupation number} for gluons with 3--momentum $(p^-,\bp)$
is naturally estimated as
 \begin{align}\label{occur}
 f(p^-,\bp)\,\equiv\,
 \frac{4\pi^3}{N_c^2-1}\,\frac{1}{L}\,\frac{\rmd N_{\rm g}}{\rmd p^-\,\rmd^2\bp\, \rmd^2\bm{b}}
\,=\,\frac{p_\perp^2}{g^2 C_F\, p^- L} \int_{\br} \rme^{-\rmi \bp \cdot \br} 
\mathcal{S}(\br)\,.\end{align}

As a simple illustration, consider the high--momentum (or low occupancy) regime,
where the dipole $S$--matrix in \eqn{unint} 
can be evaluated in the single scattering approximation. At tree--level
one obtains, similarly to \eqn{highpt},
\begin{align}\label{occur0}
 f_0(p^-,p_\perp)\,\simeq\,\frac{4\pi\alpha_s n_0}{p^- p_\perp^2}\,.\end{align}
Although valid in the dilute regime, this result can be used to estimate the borderline
of the saturation region. Namely, the non--linear effects in the gluon distribution 
are expected to become important when $f\sim {1/\abar}$ 
\cite{Mueller:2001fv,Iancu:2002xk,Gelis:2010nm}. This occurs for a 
transverse momentum $p_\perp$ of the order of the {\em saturation momentum $Q_s(x)$},
which at tree--level is estimated as
 \begin{align}\label{Qs0}Q_{s0}^2(x)\,\sim\,\frac{\alpha_s^2 N_c n_0}{p^-}
 \,\sim\,\frac{\hat q^{(0)}\lambda}{x}\,,\end{align}
where we have used $p^- = xk^-$  and the second equality is written, for convenience,
in the plasma rest frame, where $k^-\simeq 1/\lambda$ and, parametrically, $\alpha_s^2 N_c n_0
\sim \hat q^{(0)}$. Using $\tau=\lambda/x$, the above equation can also
be written as $Q_{s0}^2(\tau)=\hat q^{(0)}\tau$, which is recognized as the line (b) in  
Fig.~\ref{fig:phasespace} (the borderline of the multiple scattering region).


What is remarkable about the saturation momentum in \eqn{Qs0} is that it exhibits
a strong $x$--dependence already at tree--level (unlike the corresponding scale
for a shockwave, which is independent of $x$). Clearly, this is the consequence
of the fact that the longitudinal phase--space for gluon overlapping 
is now the longitudinal wavelength $\Delta x^+ \propto 1/x$ of a gluon fluctuation, 
and not the width $L$ of the target as a whole. This quantity $Q_{s0}^2(x)$
reaches its maximal value $Q_{s0}^2= \hat q^{(0)} L$ for
$x_{\rm min} =\lambda/L$. Thus, the quantity that
we had conventionally dubbed `the saturation momentum' in our previous discussion 
(e.g. in \eqn{Qs}) is in fact
the {\em proper} saturation scale for the {\em softest}\footnote{By `softest' we here 
mean the smallest value of $x$,  as appropriate from the
viewpoint of the left--moving target; from the viewpoint of the right--moving projectile,
these are rather the {\em hardest} fluctuations, with `energy' $p^+=\omega_c$.}
fluctuations allowed by the size of the medium. This quantity is boost invariant, but its
physical interpretation  as a saturation scale holds only in a frame where the target
is highly boosted.

Going beyond tree--level, it is clear that the evolution of the dipole scattering
amplitude in the linear approximation, \eqn{BFKL}, can be interpreted as the BFKL evolution
of the gluon occupation number in the medium. Indeed, in the single scattering approximation,
Eqs.~\eqref{occur} and \eqref{SdipEv} imply
 \begin{align}\label{occurG}
 f_\omega(p^-,\bp)\,\simeq\,-\frac{p_\perp^2}{p^-} \int_{\br} \rme^{-\rmi \bp \cdot \br} 
 \Gamma_\omega(\br)\ \sim \,\frac{1}{\abar}\, \frac{\hat q_{\omega}(p_\perp^2)}{p^- p_\perp^2}
 \,,\end{align}
where the second, parametric, estimate holds to double--logarithmic accuracy,
cf. \eqn{GammaDLA}.
Moreover, the effects of multiple scattering, i.e. the non--linear terms in \eqn{DeltaHGamma},
can be interpreted as gluon saturation in the medium, as we now explain.
Indeed, such effects become important when the exponent in \eqn{DeltaHGamma} becomes
of $\order{1}$. Using the DLA estimate \eqref{occurG}  for $ f_\omega(p^-,\bp)$, this condition can
be recognized as the saturation condition  for the gluon occupation number:
 \begin{align}
1\,\sim\,\hat q_{\omega}(1/B_\perp^2)\,B_\perp^2\,\Delta t
\,\sim\,\frac{\hat q_{\omega}(p_\perp^2)}{p_\perp^2}\,\frac{\lambda}{x}
 \,\sim\,\abar\,f_\omega(x,\bp)\,.\end{align}
Solving this condition for $p_\perp^2$, one finds the saturation
momentum in the presence of radiative corrections (to double--logarithmic accuracy):
\begin{align}\label{Qsx}
Q_{s}^2(x) \,\sim\,\frac{\hat q(x)\lambda}{x}\,,\end{align}
where $\hat q(x)\equiv \hat q_{\tau}(p_\perp^2)$ with $\tau=\lambda/x$ and $p_\perp^2
= Q_{s}^2(x)$. That is, $\hat q(x)$ is the function $ \hat q_{\tau}(p_\perp^2)$ evaluated along 
the saturation line. (In particular, for $x=x_{\rm min} =\lambda/L$, this is the physical
jet quenching parameter.) In view of  \eqn{DLAgen}, this can be 
given by the following integral representation
\begin{align}\label{DLAQs}
  \hat q(x)
  \,=\,\hat q^{(0)}+\abar \int^{1}_x
   \frac{\rmd x_1}{x_1} 
 \int_{Q_{s}^2(x_1)}^{Q_s^2(x)} \frac{\rmd p_\perp^2}{p_\perp^2} \,\hat q_{x_1}(p_\perp^2)
  \,.
  \end{align}
Within the integration limits above, one can use the zeroth order estimate
$Q_{s}^2(x)\simeq \hat q^{(0)}\lambda/x$.
Note that, with decreasing $x$, both the longitudinal phase--space and the transverse 
phase--space in \eqn{DLAQs}  increase equally fast --- that is, the both increase
like $\ln(1/x)$ ---
due to the rapid increase $Q_s^2(x)\sim 1/x$ of the saturation scale.
Accordingly, the above DLA calculation correctly captures the dominant radiative corrections
to the evolution of $Q_s^2(x)$, unlike what happens in the case of a shockwave. (For the latter,
the longitudinal phase--space increases faster than the transverse one in the approach 
towards saturation, hence the
correct calculation of $Q_s^2(x)$ requires the full BFKL equation, and not just its DLA limit
\cite{Iancu:2002tr,Mueller:2002zm}.)

In particular, if one treats the zeroth order result $\hat q^{(0)}$ as a constant, one finds
(cf. \eqn{qeff})
 \begin{align}\label{qx}
 \hat q(x)\,=\, \hat q^{(0)}
 \,\frac{1}{\sqrt{\abar}\ln\big(1/x\big)}\,{\rm I}_1\left(2\sqrt{\abar}
 \ln\frac{1}{x}\right)\,.
 \end{align}
In the extreme limit where $2\sqrt{\abar}\ln(1/x)\gg 1$, one can use the asymptotic behavior
of the modified Bessel function to deduce 
\begin{align}\label{Qsasymp}
Q_s^2(x)\,\simeq\,\frac{Q_0^2}{x^{1+\gamma_s}}
\,,\end{align}
with $Q_0^2\equiv \hat q^{(0)}\lambda$ and the
 `anomalous dimension' $\gamma_s=2\sqrt{\abar}$. The overall power 
 $\lambda_s\equiv 1+\gamma_s$ in \eqn{Qsasymp} is the medium saturation 
exponent within the present approximation. This is independent of the precise nature of the 
medium, as it is fully determined by the high--energy evolution. The radiative correction
$\gamma_s$ looks like a strong effect
since $2\sqrt{\abar}\sim 1$ for $\alpha_s\approx 0.3$ and $N_c=3$. But one should keep
in mind that the present approximation is strictly valid only when $\abar\ll 1$.

This being said, it is also interesting to notice that this perturbative result 
appears as a reasonable interpolation towards the corresponding result in 
${\mathcal N}=4$ SYM at infinitely strong coupling
($g^2 N_c\to \infty$), as obtained in  \cite{Hatta:2007cs}. Namely,
Ref.~\cite{Hatta:2007cs} reported a saturation momentum $Q_s^2(x) \sim T^2/x^2$,
where the `saturation exponent' $\lambda_s=2$ can be recognized as the sum of a kinematical
contribution  $\lambda_{s0}=1$, the same as in \eqn{Qs0}, and a large
`anomalous dimension' $\gamma_s=1$, which is the intercept of the
graviton. (At strong coupling, the unitarization occurs via multiple graviton exchanges 
\cite{Hatta:2007he}.) 
Together, the present results and the previous ones in Ref.~\cite{Hatta:2007cs} suggest a
rather smooth and fast transition from a weak coupling--like  behavior to a strong coupling--like
with increasing $\abar$ (at least, in the absence of running coupling effects).


\subsection{Comments on the effects of multiple scattering}
\label{sec:multiple}

So far, we have not attempted to explicitly evaluate the non--linear terms in  \eqn{DeltaHGamma},
which encode the effects of multiple scattering. Rather, we have used them within semi--quantitative
considerations allowing us to restrict the phase--space for the linear approximation and to develop
a physical picture in terms of gluon saturation. But, clearly, it would be interesting to have a
more quantitative control on these effects, e.g. in order to understand the systematics of the
high--energy resummation. Ideally, one would like to isolate all the radiative
corrections which are enhanced by a large energy logarithm $\ln(1/x) \sim \ln({L}/{\lambda})$
and thus obtain a non--linear equation which is explicitly valid to leading logarithmic accuracy.
Unfortunately, this turns out to be very hard since the non--linear effects enter
\eqn{DeltaHGamma} via a path--integral, in which the unknown function $ \Gamma_\omega(\br)$
plays the role of the effective potential. That is, \eqn{DeltaHGamma}  is truly a {\em functional}
integro--differential equation and very little is known about how to deal with such
equations in practice.

In this section, we shall perform a limited study of the non--linear terms  in  \eqn{DeltaHGamma}, 
with two main objectives: to elucidate the systematics of the logarithmically--enhanced radiative corrections
(which turns out to be very different from the case where the target is a shockwave)
and to better justify the arguments in the previous sections concerning the kinematics of
the fluctuations and the borderlines of the single scattering regime.

Concerning the first objective above, we would like to demonstrate the following point : 
for the case of an extended target,
and unlike for a shockwave, the individual terms beyond the first one in the multiple
scattering series --- i.e. the terms describing double scattering, triple scattering etc ---
are not {\em separately} enhanced by a large energy logarithm.
This is so because the longitudinal phase--space for multiple scattering is the lifetime 
of the soft gluon fluctuations, which is itself energy--dependent.
This being said, the effects of multiple scattering are nevertheless
important for a complete calculation at leading logarithmic accuracy, in that they provide 
the physical cutoff for the respective contribution of the single scattering (which otherwise would be
infrared divergent).


To be specific, let us first recall the way how the energy logarithm
has occurred for the single--scattering contribution. This arises via the phase--space for
the three time integrations in \eqn{EqGamma1} : over the emission times $t_1$ and 
$t_2$,  with $0 < t_1 < t_2 < L$, and over the interaction time $t$. 
Namely, the integral over $t$ between $t_1$ and $t_2$ scales like $\Delta t=t_2-t_1$, 
that over $\Delta t$ scales like $\tau_{coh}=2\omega/p_\perp^2$, and the final integral over 
say $t_2$ scales like $L$. Altogether, there is a longitudinal phase--space 
$L\tau_{coh}^2\propto \omega^2$ which, when combined with the overall factor $1/\omega^3$
in \eqn{EqGamma1}, produces the logarithmic phase--space $\int (\rmd\omega/\omega)$
for the ensuing energy integration. Now, let us similarly consider the contribution of a double
scattering. As compared to the previous case, there are now two interaction times to be
integrated over between $t_1$ and $t_2$. This introduces an additional factor $\Delta t$,
so the global result scales like $L\tau_{coh}^3\propto \omega^3$, which spoils the logarithmic 
integration over $\omega$. A similar conclusion holds for the contribution of 
$n$ successive collisions, which scales like $L\tau_{coh}^{n+1}\propto 
\omega^{n+1}$.

It is further instructive to compare with the respective situation for a shockwave target,
as discussed in Sect.~\ref{sec:JIMtime}. In that case, the two integrals over 
$t_1$ and $t_2$ separately restrict each of the emission times to values of order 
$\tau_{coh}$ around  $t=0$ (the position of the shockwave). Also each scattering with the
target occurs within the longitudinal extent $L$ of the latter (with $L\ll \tau_{coh}$ in this
context), so the corresponding time integral brings in a factor of $L$.
Hence, an individual
 $n$--scattering contribution with $n\ge 1$ scales like $\tau_{coh}^2 L^n \propto \omega^2$,
and thus it is by itself accompanied by a large energy logarithm.

Returning to the case of an extended target, one should observe that the previous 
power--counting argument was a bit formal, in that it is plagued with infrared divergences:
each additional scattering brings in a factor $\tau_{coh}=2\omega/p_\perp^2$, which
becomes singular when $p_\perp\to 0$.  This leads to a logarithmic divergence in the
single--scattering contribution, as manifest on \eqn{DLA}, and to even stronger, power--like,
divergences in the terms with two or more scatterings.
We expect such divergences to be cured by the resummation of the multiple
scattering series to all orders,  but in order to verify this, 
one needs a non--perturbative calculation of this series.

To illustrate these considerations, let us consider the first iteration of \eqn{DeltaHGamma}.
That is, we shall evaluate the r.h.s. of this equation using the tree--level approximation 
for the dipole $S$--matrix, \eqn{Sdip1}, together with the `harmonic approximation' for
the jet quenching parameter --- meaning that we ignore the scale dependence of the latter : $\hat q
\simeq $~const (throughout this subsection, one has $\hat q \equiv \hat q^{(0)}$). 
The harmonic approximation is
indeed important for the present purposes, since it allows us
to explicitly perform the path integral in  \eqn{DeltaHGamma}, which  becomes
    \begin{align}\label{path} \hspace*{-1.cm}
   \mcal{I}(\br_2,\br_1,\Delta t)\equiv
\int\big[\mcal{D}\br\big]
  \ \rme^{\rmi \,\frac{\omega}{2}
  \int\limits_{t_1}^{t_2}\rmd t' \,\dot\br^2(t')}
 \exp\left\{-\frac{\hat q}{4} \int_{t_1}^{t_2}\rmd t\,
\Big[\big(\bx-\br(t)\big)^2+ \big(\br(t)-\by\big)^2 - \big(\bx-\by\big)^2\Big]\right\}
 \end{align}
with boundary conditions $\br(t_1)=\br_1$ and $\br(t_2)=\br_2$. ($\mcal{I}$ is also a function of $\bx$
and $\by$, but the respective arguments are kept implicit.) A standard calculation yields
 \begin{align}\label{Iharm} \hspace*{-1.cm}
   \mcal{I}(\br_2,\br_1,\Delta t)\,=\,&
   \frac{-\rmi}{2\pi}\,\frac{\omega\Omega}{\sinh\Omega\Delta t}\,\rme^{\frac{\hat q}{8}\Delta
   t \,(\bx-\by)^2}\nn
   &\times \exp\left\{\frac{\rmi}{2}\frac{\omega\Omega}{\sinh\Omega\Delta t}\,\left[
   \cosh\Omega\Delta t\Big((\br_2-\bR)^2 + (\br_1-\bR)^2\Big) - 2(\br_2-\bR)
   \cdot (\br_1-\bR)\right]\right\}\,,
    \end{align}
where $\bR\equiv(\bx+\by)/2$ and 
  \begin{align}\label{Omega}
 \Omega\,=\,\frac{1+\rmi}{\sqrt{2}}\,\sqrt{\frac{\hat q}{\omega}}\,=\,
 \frac{1+\rmi}{\sqrt{2}}\,\frac{1}{\tau_{_{\rm br}}(\omega)}
 \,.
 \end{align}
In the limit $\hat q\to 0$ (no scattering), this reduces to the  `non--relativistic' propagator in
the vacuum, as it should (cf. \eqn{Gmedium}): $\mcal{I}\to \mcal{G}_0$ with
\begin{align}\label{G0time}
   \mcal{G}_0(\br_2-\br_1,\Delta t)\,=\,&
   \frac{-\rmi}{2\pi}\,\frac{\omega}{\Delta t}\,
   \exp\left\{\frac{\rmi \omega\big(\br_2-\br_1\big)^2}{2\Delta t}\right\}\,.
  \end{align}
  
The perturbative (`small $\hat q$') expansion of \eqn{Iharm}, which would reconstruct the multiple scattering series, turns out to be quite tedious. However, by inspection of this equation,
it is clear that such an expansion makes sense only for sufficiently small times
$\Delta t \ll \tau_{_{\rm br}}(\omega)$. In this perturbative regime at early times,
the convergence in $\Delta t$ is
controlled by the free propagator \eqref{G0time}, which implies that the transverse
size of the gluon fluctuation grows via quantum diffusion:
$|\br_2-\br_1|\simeq \sqrt{2\Delta t/\omega}$.
However, when $\Delta t$ approaches $\tau_{_{\rm br}}(\omega)$,
the effects of the interactions become non--perturbative. Via the Gaussian in \eqn{Iharm},
they restrict  the further growth of the transverse separation 
$B_\perp\equiv {\rm max}\big(|\br_2-\bR|\,, |\br_1-\bR|\big)$
to values $B_\perp\lesssim 2/k_{_{\rm br}}(\omega)$, in agreement with \eqn{range}.
For even larger time separations $\Delta t\gtrsim \tau_{_{\rm br}}(\omega)$, we can use
 \begin{align} \frac{1}{2\sinh\Omega \Delta t}\,\simeq\,\rme^{-\Omega \Delta t}
 \,\propto\,\rme^{-{\Delta t}/{\sqrt{2}\tau_{_{\rm br}}}}\,,\end{align}
showing that the long--lived gluon fluctuations are exponentially suppressed. 
Note that this behaviour at large
$\Delta t$ is non--analytic in $\hat q$, so it indeed arises from resumming the
perturbative series to all orders. 

To summarize, the effect of multiple scattering is to limit the lifetime and the
transverse size of a gluon fluctuation with energy $\omega$ to values
$\Delta t\lesssim\tau_{_{\rm br}}(\omega)$ and respectively $B_\perp\lesssim 2/k_{_{\rm br}}(\omega)$.
Moreover, since the perturbative  expansion of \eqn{Iharm} is truly an expansion in powers of
$\Delta t/\tau_{_{\rm br}}(\omega)$ and $B_\perp k_{_{\rm br}}(\omega)$, it is clear that the
single scattering approximation --- which corresponds to the first non--trivial term in this expansion --- 
is valid only for fluctuations which are hard enough (in the sense of having a sufficiently
large transverse momentum) for $\Delta t\ll\tau_{_{\rm br}}(\omega)$ and 
$B_\perp\ll 1/k_{_{\rm br}}(\omega)$.
These conditions have been often used in the previous discussion in this section.

Using the explicit expression for the path integral in \eqn{Iharm}, it is in principle
possible the fully evaluate the r.h.s. of \eqn{DeltaHGamma} and thus compute the 
leading--order radiative correction to the dipole amplitude beyond the double--logarithmic
approximation. In general this calculation is hindered by the
complexity of the time integrations. 
In Ref.~\cite{Liou:2013qya}, this calculation has been pushed to {\em single}
logarithmic accuracy  --- that is, one has explicitly evaluated the subleading correction to 
$\hat q$ of order $\abar  \ln({L}/{\lambda})$.  Note that \eqn{Iharm} explicitly includes the
`virtual' term that has been added by hand in the respective calculation in Ref.~\cite{Liou:2013qya}.

{
\section{The high--energy evolution of the radiative energy loss}
\label{sec:eloss}

In the previous section, we have studied the high--energy evolution of the transverse
momentum broadening for an energetic quark propagating through a dense QCD medium.
As well known, this physical problem  is closely related to another one:
the energy loss by an energetic parton via {\em
medium--induced radiation}, that is, gluon emissions which are triggered by the interactions 
between the parent parton or the radiated gluon and the constituents of the medium. 
Within the high--energy kinematics
of interest, the differential cross--section for such an emission involves Wilson line
correlators which measure the color coherence between the emitter and its radiation.
This coherence is progressively washed out via rescattering in the medium and thus is sensible
to the physics of collisions, as encoded in $\hat q$. This relation is manifest at tree level,
where any Wilson line correlator --- the medium average of a gauge--invariant product of 
Wilson lines --- can be expressed in terms of the `dipole cross--section' (the exponent in \eqn{Sdip0})
and hence in terms of $\hat q^{(0)}$ (as defined in \eqn{qhat}). In what follows, we would like to
demonstrate that this relation remains valid after including the effects of the high--energy evolution
within the double--logarithmic approximation. That is,  to compute the radiative energy loss 
in the presence of radiative corrections and to DLA accuracy, one can use the
same formul{\ae} as at tree level, but with $\hat q^{(0)}$ replaced by the solution $\hat q_{L}(Q^2_s)$ 
to \eqn{DLAint}. A similar conclusion has been independently obtained in Ref.~\cite{Blaizot:2014}.

\subsection{The tree--level approximation: the BDMPSZ formalism}
\label{sec:BDMPS}

To start with, let us briefly review the relevant formalism at tree--level, namely 
the BDMPSZ calculation of medium--induced gluon radiation
\cite{Baier:1996kr,Baier:1996sk,Zakharov:1996fv,Zakharov:1997uu,Baier:1998yf,Baier:1998kq,Wiedemann:2000za,Wiedemann:2000tf,Arnold:2001ba,Arnold:2001ms,Arnold:2002ja}. 

We consider the emission of a single gluon by an asymptotic
quark and assume, for simplicity, that the incoming quark is energetic enough to be treated in the
eikonal approximation. On the other hand, the eikonal approximation cannot be used for the
emitted gluon, because the transverse diffusion plays an essential role for the gluon formation.
The energy lost by the quark is the energy taken away by the emitted gluon and can
be computed from the respective spectrum as
 \begin{align}\label{Eloss}
 \Delta E\,=\int_0^{\omega_c}\rmd k^+\, k^+\frac{\rmd N_g}{\rmd k^+}
 \,=\int_0^{\omega_c}\rmd k^+\int \rmd^2\bk\ k^+\frac{\rmd N_g}{\rmd k^+\rmd^2\bk}\,.
 \end{align}
The upper cutoff $\omega_c$ stems from the fact that only gluons with energies
$k^+ < \omega_c$ can be emitted via the mechanism at hand (see below).
Moreover, the spectrum $k^+({\rmd N_g}/{\rmd k^+})$ of the radiated gluons is such that
the integral over $k^+$ in \eqn{Eloss} is dominated by this upper cutoff. 
Accordingly, in what follows we shall focus on the emission of relatively hard gluons, 
with $k^+ \sim\omega_c$. 

As before, we assume that the medium correlations at tree--level are Gaussian and local
in $x^+$. Under these assumptions,
one can deduce the following formula for the spectrum of the medium--induced gluon 
radiation \cite{Wiedemann:2000za}
(see also Refs.~\cite{Kovner:2003zj,MehtarTani:2006xq,CasalderreySolana:2007zz,Mehtar-Tani:2013pia,
CasalderreySolana:2011rz} 
for pedagogical discussions)
 \begin{align}\label{spectrum}
 k^+\frac{\rmd N_g}{\rmd k^+\rmd^2\bk}\,=\,
 &\frac{\alpha_s C_F}{2\pi^2}\,\frac{1}{(k^+)^2} \,{\rm Re}
 \int_{-\infty}^{\infty}
  \rmd x^+\! \int_{-\infty}^{x^+} \!  \rmd y^+\,    \nn
   &\quad\times
   \int \rmd^2\bx \
   \rme^{-\rmi\bk\,\cdot\,{\bm x}}\   {\mcal S}^{^{\rm adj}}_{L, x^+}( \bx)
 \, \del^i_{\bx}\del^i_{\by}{\mcal K}(x^+,{\bm x}; y^+,{\bm y}; k^+)\Big|_{\by=0}\,.
   \end{align}
As announced, we consider
an on--shell (or `asymptotic') quark which enters the medium coming from far away\footnote{The 
case of an off--shell
quark which is produced by a hard process occurring at some finite time $x_0^+$ can be obtained
be replacing $-\infty\to x_0^+$ in the lower limits of the time integrations in  \eqn{spectrum}.}.
\eqn{spectrum} is a cross--section, so it is obtained by multiplying the  
direct amplitude (DA) times the complex conjugate amplitude (CCA), as illustrated
in Fig.~\ref{fig:gluon1}. The time variables
$y^+$ and $x^+$ are the emission times in the DA and respectively the CCA, so their difference
$\Delta x^+=x^+-y^+$ is indicative of the formation time. We have chosen $x^+ > y^+$ and
multiplied the result by 2 to account for the opposite time ordering. Also, $\bx_0=0$ is the 
transverse position of the quark, which remains unchanged during the process (eikonal approximation)
and is the same in the DA and the CCA (since we do not measure the transverse momentum 
broadening of the quark).  In writing \eqn{spectrum}, we have already set $\bx_0=0$, but
in the subsequent discussion we shall keep the notation $\bx_0$ at intermediate stages,
for more clarity. On the other hand, the  transverse momentum of the emitted gluon {\em is}
measured, so the respective transverse coordinates are different in the DA and the CCA.
Their difference is denoted as $\bx$ in \eqn{spectrum} and it is conjugated to the transverse
momentum $\bk$ via the Fourier transform. 
With these differences in mind, \eqn{spectrum} is quite similar to \eqn{DeltaHeq}
for the emission of a {\em virtual} gluon and can be read by analogy with the latter. 

Once
again, the locality of the medium correlations in $x^+$ has allowed us to factorize the 
process into three stages,  like in \eqn{DeltaH1},  and thus express the cross--section for
gluon radiation in terms of scattering amplitudes for effective dipoles
(see Fig.~\ref{fig:gluon2} for an illustration of this representation) : 

\bigskip
\texttt{(i)} {\em Prior to the first gluon emission, i.e. for time values smaller than $y^+$} : The Wilson lines
describing the color precession of the quark mutually cancel between the DA and the CCA,
by unitarity. Accordingly, there is no imprint of this first stage on the cross--section in
\eqn{spectrum}.

\bigskip
\texttt{(ii)} {\em During the formation time, i.e. for time values between $y^+$ and $x^+$} : 
The partonic system consists in a quark--gluon pair in the DA and the original quark in the CCA.  The
relevant Wilson line correlator reads
 \begin{align}\label{qqg}
 \left\langle U^{\dagger\,ab}_{x^+y^+}[\bu]\, \frac{\rmtr}{N_c}  
   \Big(t^a\, V_{x^+y^+}^{\dagger}(\bx_0) t^b\,V_{x^+y^+}(\bx_0) 
 \Big)\right\rangle\,=\,&\frac{1}{2N_c}  \left\langle \rmTr \,U^{\dagger}_{x^+y^+}[\bu]\,
 U_{x^+y^+}(\bx_0)\right\rangle \nn
 \,=\,& C_F\, {\mcal S}^{^{\rm adj}}_{x^+,y^+}\big([\bu],\bx_0\big)\,,
 \end{align}
where $t^a$ and $t^b$ are color matrices at the emission vertices, 
$\bu(t)$ represents the trajectory of the gluon in the DA for times $y^+\le t \le x^+$, and
$ {\mcal S}^{^{\rm adj}}$ is the average $S$--matrix for a color dipole in the adjoint representation,
\begin{align}\label{Sadj}
 {\mcal S}^{^{\rm adj}}_{x^+,y^+}\big(\bx,\by\big)\,\equiv\,\frac{1}{N_c^2-1}
\left\langle \rmTr \,U^{\dagger}_{x^+y^+}(\bx)\,
 U_{x^+y^+}(\by)\right\rangle \,.\end{align}
The dipole in \eqn{qqg}
is built with one adjoint Wilson line for the emitted gluon and another
one for the precession of the color current of the quark (a color vector in the
adjoint representation). The `dipole propagator' ${\mcal K}(x^+,{\bm x}; y^+,{\bm y}; k^+)$ 
which enters \eqn{spectrum} represents the functional average of the Wilson line 
correlator \eqref{qqg} over the quantum trajectories of the gluon:
 \begin{align}\label{Kdef}
 {\mcal K}(x^+,{\bm x}; y^+,{\bm y}; k^+)
 =\int [\mcal{D}\bm{u}]\ \rme^{i\,\frac{k^+}{2}\int\limits_{y^+}^{x^+}\rmd t\,
 \dot{\bm u}^2(t)}\, {\mcal S}^{^{\rm adj}}_{x^+,y^+}\big([\bu(t)],\bx_0\big)\,,
 \end{align}
with boundary conditions $\bu(y^+)=\by$ and $\bu(x^+)=\bx$. 

\begin{figure}
\begin{center}
\begin{minipage}[b]{0.85\textwidth}
\begin{center}
\includegraphics[width=0.9\textwidth,angle=0]{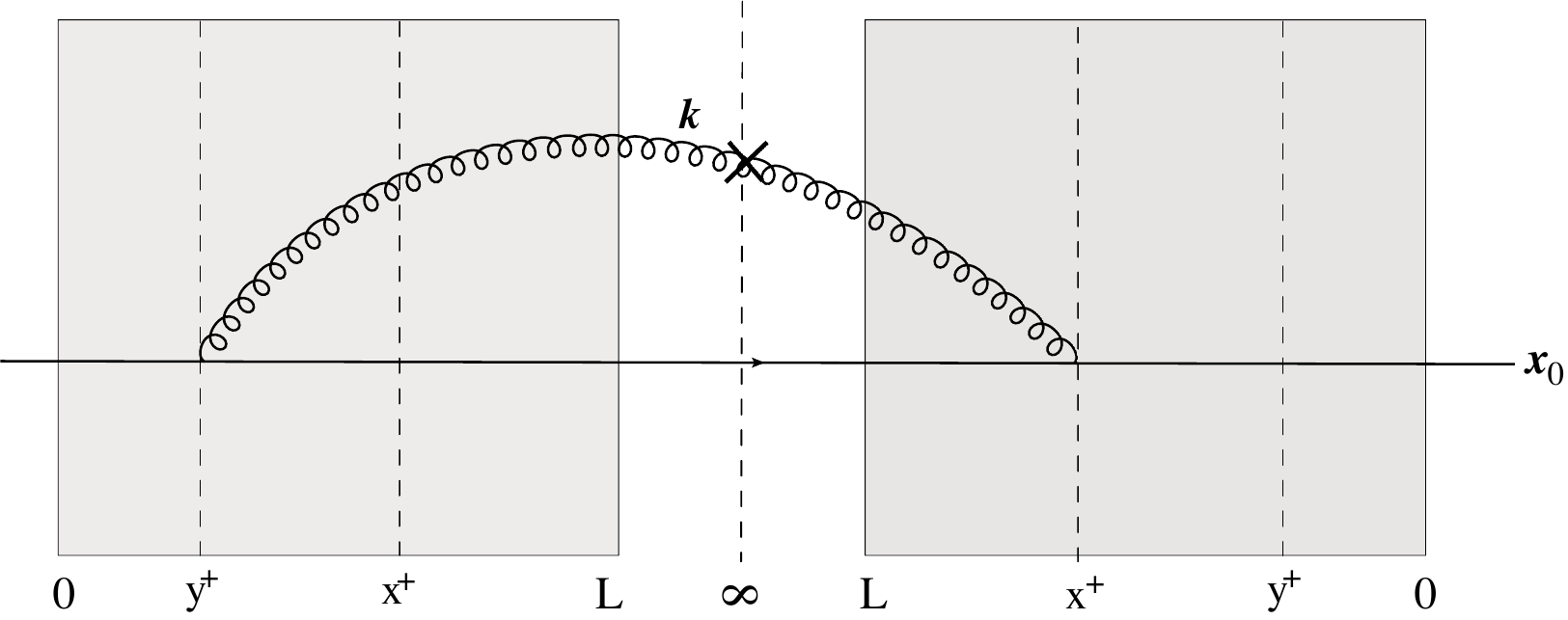}
\end{center}
\end{minipage}
\end{center}
\caption{\label{fig:gluon1} \small\sl Feynman graph contributing to
 the cross--section for producing a gluon with 3--momentum $k=(k^+,\bk)$,
as computed in \eqn{spectrum}. The l.h.s. corresponds to the DA and
the r.h.s. to the CCA. Both emissions times $y^+$ and $x^+$ are chosen 
inside the medium, $0<y^+<x^+<L$, since this is the most interesting configuration
for our present purposes.}
\end{figure}

\bigskip
\texttt{(iii)} {\em After the gluon formation, i.e. for time values larger than $x^+$} : 
In this case too, the quark Wilson lines from the DA and the CCA mutually cancel, 
so we are left with two adjoint Wilson lines, which both refer to the emitted gluon
(one for the DA, the other one for the CCA). 
These Wilson lines combine in the adjoint dipole
$ {\mcal S}^{^{\rm adj}}_{L, x^+}(\bx-\bx_0)$, which describes the transverse momentum broadening 
acquired by the gluon after formation (so long as $x^+ <L$, of course). 
The {\em average} transverse size of this dipole is constant, due to the medium homogeneity
in the transverse plane, and hence it is equal to its original value at time $x^+$, which is
$\bx-\bx_0=\bx$.

If one is not interested in the $k_\perp$--spectrum of
the produced gluon, but only in the energy lost by the quark, then one can integrate
\eqn{spectrum} over $\bk$ and use $ {\mcal S}^{^{\rm adj}}_{L, x^+}({0})=1$,
to deduce
  \begin{align}\label{Espectrum}
 k^+\frac{\rmd N_g}{\rmd k^+}\,=\,\frac{2\alpha_s C_F}{(k^+)^2} \,{\rm Re}
 \int_{-\infty}^{\infty}
  \rmd x^+\! \int_{-\infty}^{x^+} \!  \rmd y^+
 \, \del^i_{\bx}\del^i_{\by}{\mcal K}(x^+,{\bm x}; y^+,{\bm y}; k^+)\Big|_{\bx=\by={0}}\,.
   \end{align}
   
   \begin{figure}
\begin{center}
\begin{minipage}[b]{0.85\textwidth}
\begin{center}
\includegraphics[width=0.8\textwidth,angle=0]{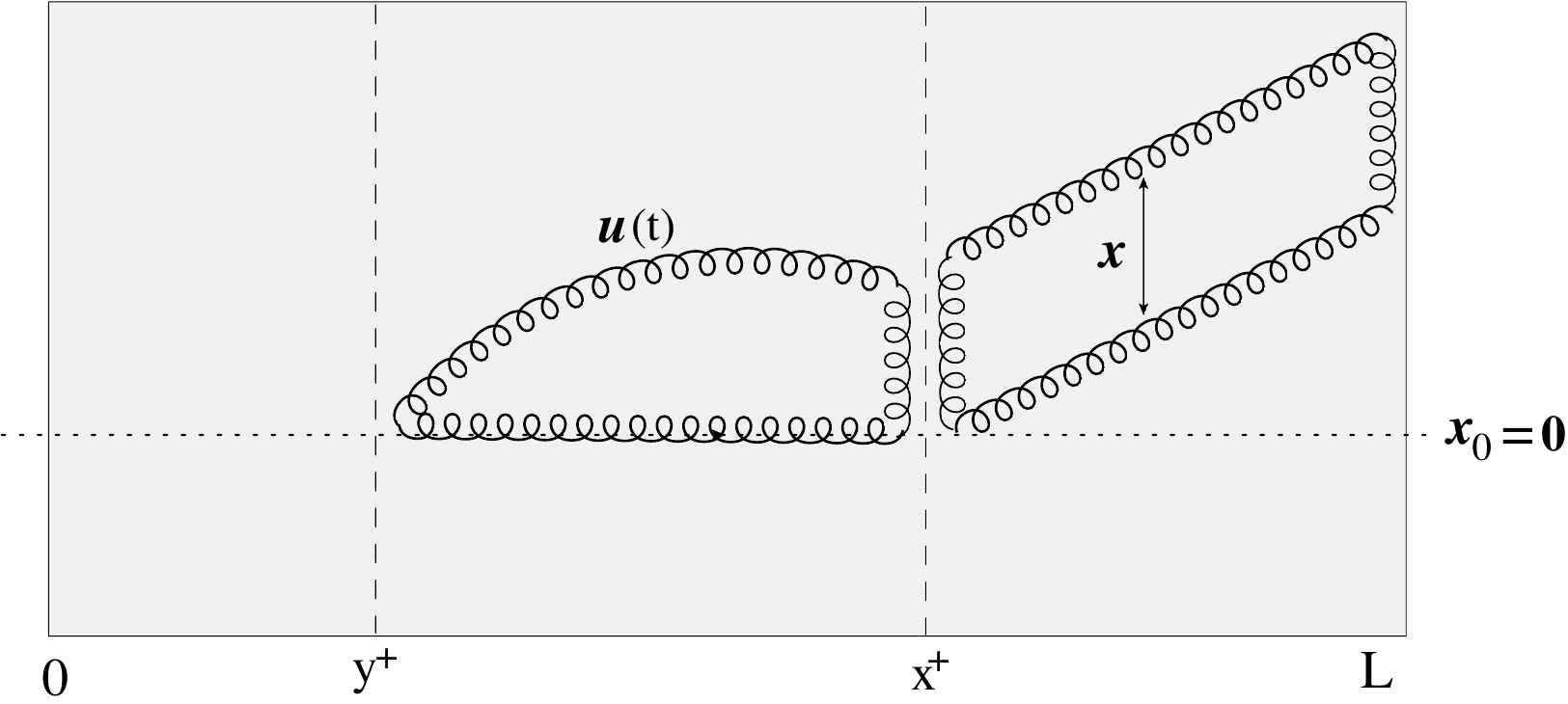}
\end{center}
\end{minipage}
\end{center}
\caption{\label{fig:gluon2} \small\sl Alternative representation for the cross--section in
Fig.~\ref{fig:gluon1}, in terms of dipole amplitudes, which 
 is obtained after performing the
medium average in the Gaussian approximation. The `vertical' lines closing the two dipoles
represent the sum over the color indices.
}
\end{figure}

To be more specific, consider the situation where the emission occurs within the medium
in both the DA and the CCA: $0<y^+<x^+<L$.  This situation yields the dominant contribution
for sufficiently small energies $k^+\ll \omega_c$, but the corresponding result can also be
used when $k^+\sim \omega_c$, at least for parametric estimates.
Then the average dipole $S$--matrix in
\eqn{qqg} can be computed as in Eqs.~\eqref{Sdip0}--\eqref{qhat} and reads
 (we set $\bx_0=0$ from now on) 
 \begin{align}\label{Sadj0}
  {\mcal S}^{^{\rm adj}}_{x^+,y^+}[\bu]\,\simeq
\,\exp\left\{-\frac{1}{4}\,\int_{y^+}^{x^+}\rmd t\, \hat q_{\rm g}(1/u^2)\,\bu^2(t)\right\}\,,\qquad
  \hat q_{\rm g}(Q^2)\,\equiv\,\frac{N_c}{C_F}\,\hat q(Q^2),\end{align}
where the subscript `g' in $\hat q_{\rm g}$ refers to `gluon'. 
(The quantity $\hat q(Q^2)$ without any subscript refers to a quark in the fundamental 
representation and has been introduced in \eqn{qhat}.) 

To compute the path integral in  \eqn{Kdef},
we perform the `harmonic approximation' in \eqn{Sadj0}, that is, we replace
 $\hat q_{\rm g}(1/u^2)$ with the constant quantity $\hat q_{\rm g}(   k_{_{\rm br}}^2)$, where
$   k_{_{\rm br}}^2(k^+)\equiv \sqrt{2k^+ \hat q_{\rm g}}$ is the transverse momentum acquired by the
gluon during formation.  This approximation is appropriate for gluon
emissions triggered by multiple soft scattering in the medium.
Then the path integral yields
  \begin{align}\label{K0}
  \hspace*{-.5cm}
{\mcal K}(x^+,{\bm x}; y^+,{\bm y}; k^+)\,=\,&
   \frac{-\rmi}{2\pi}\,\frac{k^+ \Omega}{\sinh \Omega\Delta \tau}\,
    \exp\left\{\frac{\rmi}{2}\,\frac{k^+ \Omega}{\sinh \Omega\Delta \tau}\,   
  \Big[ (\bx^2 + \by^2)\cosh \Omega\Delta\tau \,-2\,\bx\cdot\by\Big]\right\}\,,
    \end{align}
where $\Delta\tau=x^+-y^+$ and\footnote{The current expressions for 
$ \tau_{_{\rm br}}(k^+)$ and $k_{_{\rm br}}(k^+)$ 
are consistent with their respective definitions in Sect.~\ref{sec:JET} in view
of the relation $\hat q_{\rm g}\simeq2\hat q$ valid at large $N_c$. (Recall
that the discussion in Sect.~\ref{sec:JET} was carried mostly at large $N_c$.)}
 \begin{align}\label{OmegaA}
 \Omega\,=\,\frac{1+\rmi}{\sqrt{2}}
  \frac{1}{  \tau_{_{\rm br}}(k^+)}\,,\qquad   \tau_{_{\rm br}}(k^+)\,=\,\sqrt{\frac
 {2k^+}{\hat q_{\rm g}}}
 \,.\end{align}
It is now straightforward to evaluate the transverse derivatives in \eqn{Espectrum} and
then perform the time integrals within the range $0<y^+<x^+<L$. In this process, one must 
subtract the vacuum piece of \eqn{K0}, this is, its limit when $\hat q_{\rm g}\to 0$ : this
would give a spurious contribution, which is moreover divergent. This procedure yields
the BDMPSZ spectrum for soft energies $k^+\ll  \omega_c$~:
 \begin{align}\label{BDMPS}
 k^+\frac{\rmd N_g}{\rmd k^+}\,\simeq\,\frac{2\alpha_s C_F}{\pi}\,\sqrt{\frac{  \omega_c}
 {2k^+}}\qquad
 \mbox{with} \quad  \omega_c\,=\,\frac{1}{2}\,\hat q_{\rm g} L^2
 \,.\end{align}
This result can be used to check that the integral in
\eqn{Eloss} is indeed dominated by its upper limit. The general result valid for any $k^+$ can be found
in Refs.~\cite{Zakharov:1996fv,Zakharov:1997uu,Baier:1998kq}.

Concerning the $k_\perp$--spectrum, notice that the dominant dependence upon $\bx$ within
the integrand of \eqn{spectrum} is contained in the following product of two Gaussians:
 \begin{align}\label{2Gauss}
 \exp\left\{\frac{\rmi}{2}\,{k^+ \Omega}\,{\coth \Omega\Delta \tau}\, \bx^2 \right\}
 \exp\left\{-\frac{\hat q_{\rm g}}{4}\,(L-x^+) \,\bx^2\right\}\,.
\end{align}
The first factor arises after letting $\by=0$ in \eqn{K0}, while the second one
is the two--gluon dipole $ {\mcal S}^{^{\rm adj}}_{L, x^+}(\bx)$ with $\hat q_{\rm g}$ 
evaluated at a momentum scale $\sim Q_s^2$. 
The transverse momentum spectrum obtained via the Fourier transform of the 
above is clearly Gaussian and peaked at a typical value
\begin{align}\label{form}
\langle k_\perp^2 \rangle\,\simeq\,\sqrt{2k^+ \hat q_{\rm g}}\,+\,{\hat q_{\rm g}}(L-x^+) 
\,\sim\, Q_s^2\equiv \hat q_{\rm g} L\,,\end{align}
where $Q_s^2$ now denotes the {\em gluon} saturation momentum.
In \eqn{form} we recognize the sum of the momentum broadening acquired via collisions
during the formation time and that acquired after the formation. For  $k^+\sim \omega_c$,
both contributions are parametrically of $\order{Q_s^2}$.
More details on the $k_\perp$--spectrum can be found
in Refs.~\cite{Wiedemann:2000za,Wiedemann:2000tf,Blaizot:2012fh}.

Returning to \eqn{K0}, this can be used to read the characteristic 
scales for gluon formation.
The r.h.s. of \eqn{K0} is exponentially suppressed for time separations 
$\Delta\tau\gg  \tau_{_{\rm br}}$ and for transverse separations
$(\bx-\bx_0)^2 \gg 1/   k_{_{\rm br}}^2$ (recall that we set $\bx_0=0$). Accordingly,
the emission of a gluon with energy $k^+$ via the present mechanism takes a
time of order $ \tau_{_{\rm br}}(k^+)$. Also, the maximal transverse
separation between the emitted gluon and its parent parton is of order $1/   k_{_{\rm br}}(k^+)$.
When $k^+\sim \omega_c$, as relevant for the calculation of the energy loss, these scales 
become $ \tau_{_{\rm br}}\sim L$ and $1/   k_{_{\rm br}}\sim 1/Q_s$ --- that is,
they are parametrically similar to those 
underlying the physics of transverse momentum broadening, as discussed in
Sect.~\ref{sec:JET}. Hence, no surprisingly, the respective discussions of the radiative
corrections will be quite similar as well.

\subsection{The dominant radiative corrections}

Without loss of generality, we can restrict our discussion of the evolution to the 
case where the gluon with longitudinal momentum  $k^+\sim \omega_c$ 
(the one which is responsible for the energy loss) is
emitted inside the medium, in both the direct and the complex conjugate amplitudes.
(Indeed this case is the most complicated one, in terms of medium interactions.)
We keep the same conventions as before: the gluon is first emitted in the DA, at time $y^+$,
and then in the CCA, at time $x^+$.

As in Sect.~\ref{sec:JET}, we assume that the high--energy evolution preserves the Gaussian
nature of the medium correlations, cf. \eqn{Gaussian}. It is then quite clear that this
evolution will also preserve the factorization of the cross--section into the three stages
discussed in Sect.~\ref{sec:BDMPS}. This is so because the relevant quantum
fluctuations are short--lived: their coherence time $\tau_{coh}=2\omega/p_\perp^2$ is 
much shorter than the typical duration of any of these three stages. As before, in 
Sect.~\ref{sec:JET}, the variables $\omega$ and $p_\perp$ denote the
`energy' (in the sense of $p^+$) and the transverse momentum of the
evolution gluon, and the most interesting situation is such that $\omega\ll k^+$ and 
$p_\perp\ll Q_s$. In this situation, the
quantum fluctuations which overlap with two different stages (and thus could
break down the factorization) are suppressed by the smallness of their 
longitudinal phase--space. 

Consider e.g. a fluctuation where the soft gluon is emitted by the
quark at some time $t_1<y^+$ and then absorbed by either the quark or the nascent gluon
at some time $t_2$ during the `formation' stage ($y^+ < t_2 < x^+$). This fluctuation has
a lifetime $t_2-t_1\sim \tau_{coh}$, so both $t_1$ and $t_2$ must lie within an interval 
$\sim \tau_{coh}$ around $y^+$. Accordingly, the respective longitudinal phase--space
is of order $\tau_{coh}^2$ and thus is much smaller than that,
of order $(x^+-y^+)\tau_{coh}$, corresponding to fluctuations which fully develop
during the formation time ($y^+ < t_1 < t_2 < x^+$). 
We have indeed $x^+-y^+\sim  \tau_{_{\rm br}}(k^+)\gg \tau_{coh}(\omega)$ for
$\omega\ll k^+$.
This discussion implies that the cross--section for medium--induced radiation can still
be given the factorized structure in \eqn{spectrum}, but with the individual factors
generally modified by radiative corrections. 

\comment{For each of these factors, one can apply the
general arguments in \cite{Mueller:2012bn} to conclude that there is no need to
explicitly distinguish between soft gluon emissions which occur in the DA and respectively 
the CCA. Rather, it is mathematically equivalent and practically simpler to follow the 
evolution of the effective dipole amplitudes which enter the cross--section \eqref{spectrum}.}

With reference to \eqn{spectrum}, it is quite obvious that the evolution has no influence
on the first stage at $t<y^+$, i.e. prior to the emission of the nascent gluon in the DA. During that 
stage, the quarks in the DA and the CCA make up a zero--size `dipole', which does not interact, so
its high--energy evolution cannot be measured. It is furthermore clear that the main effect of the
evolution during the last stage at $t>x^+$ (after gluon formation) is to renormalize the
jet quenching parameter within the two--gluon dipole amplitude $ {\mcal S}^{^{\rm adj}}_{L, x^+}(\bx)$,
in the way explained in Sect.~\ref{sec:DLA} : to double--logarithmic accuracy, the
renormalized adjoint dipole $S$--matrix reads
 \begin{align}\label{SAren}
  {\mcal S}^{^{\rm adj}}_{L, x^+}(\bx)\,\simeq\,\exp\left\{-\frac{1}{4}\,\hat q_{\rm g}(L,Q_s^2)(L-x^+) \,\bx^2\right\}\,,
\end{align}
with $\hat q_{\rm g}(L,Q_s^2)$ the solution to \eqn{DLAint} for $Q_s^2=\hat q_{\rm g} L$.
This differs from the corresponding quark transport coefficient merely by a color factor:
$\hat q_{\rm g}(L,Q_s^2)=(N_c/C_F) \hat q_L(Q_s^2)$. In choosing the scales for $\hat q_{\rm g}$
above, we have 
used the fact that, parametrically, $L-x^+\sim L$ and $x_\perp^2\sim 1/Q_s^2$.

As compared to Sect.~\ref{sec:JET}, the only situation which is
somewhat new is when the fluctuation lives during the formation time of the radiated
gluon. This is new since, unlike in Sect.~\ref{sec:JET}, we do not assume anymore 
the eikonal approximation for the evolving dipole~: the trajectory $\bu(t)$ of the nascent
gluon, which is the same as the size of the effective dipole (since the quark is fixed at 
$\bx_0=0$), is randomly varying via transverse diffusion.  Yet, this transverse motion
looks relatively slow on the typical time scale for quantum fluctuations (since the respective
`transverse mass' $k^+$ is comparatively hard), so we expect some kind of `adiabatic approximation'
to be applicable for the fluctuations. This will be detailed in what follows. 

The respective evolution equation is obtained as in Sect.~\ref{sec:dipole}, that is, by first
acting with the Hamiltonian $\Delta H$ on the (adjoint) dipole scattering operator
$\hat{S}_{x^+,y^+}^{^{\rm adj}}[\bu]$ and then performing the medium average within the Gaussian
approximation \eqref{Gaussian}. This procedure implies
 \begin{align}\label{SAdipEv}
 \mcal {S}_{x^+,y^+}^{^{\rm adj}}([\bu]; \omega)\,=\,\exp\biggl\{-g^2 N_c\int_{y^+}^{x^+}\rmd t\,
 \Gamma_\omega(\bu(t))\biggr\}
 \end{align}
 with the function $\Gamma_\omega(\br)$ now obeying (compare to \eqn{DeltaHGamma})
    \begin{align}
\label{GammaA}
  \hspace*{-.8cm} \int_{y^+}^{x^+}\rmd t\,
 \frac{\del
\Gamma_\omega(\bu(t))}{\del \omega} \,=\,&
  \frac{1}{4\pi \omega^3} 
  \int_{y^+}^{x^+}\rmd t_2 \int_{y^+}^{t_2}\rmd t_1\ 
  \del^i_{\br_1}\del^i_{\br_2}\Bigg\{  \int\big[\mcal{D}\br\big]
  \ \rme^{\rmi \,\frac{\omega}{2}
  \int\limits_{t_1}^{t_2}\rmd t' \,\dot\br^2(t')}\nn
 &  \hspace*{-.8cm} \times 
 \Bigg[\rme^{\! -\frac{g^2N_c}{2}\!  \int\limits_{t_1}^{t_2}\rmd t\,
\big(\Gamma_{\omega} (\bu(t)-\br(t))+
 \Gamma_{\omega} (\br(t))-\Gamma_{\omega} (\bu(t))\big)}
 \,-\, 1\Bigg]
    \Bigg\}\Bigg|^{\br_2=\bu(t_2)}_{\br_2=0}\,
    \Bigg|^{\br_1=\bu(t_1)}_{\br_1=0}\ .
  \end{align}
Within the present approximations, these equations hold for arbitrary $N_c$.
The main difference w.r.t. \eqn{DeltaHGamma} is the fact that the endpoints $\br_1$
and $\br_2$ of the path integral in \eqn{GammaA} (i.e. the transverse positions of the
virtual gluon at the emission points) are time--dependent whenever they refer to emissions
by the nascent gluon.

\begin{figure}
\begin{center}
\begin{minipage}[b]{0.8\textwidth}
\begin{center}
\includegraphics[width=0.7\textwidth,angle=0]{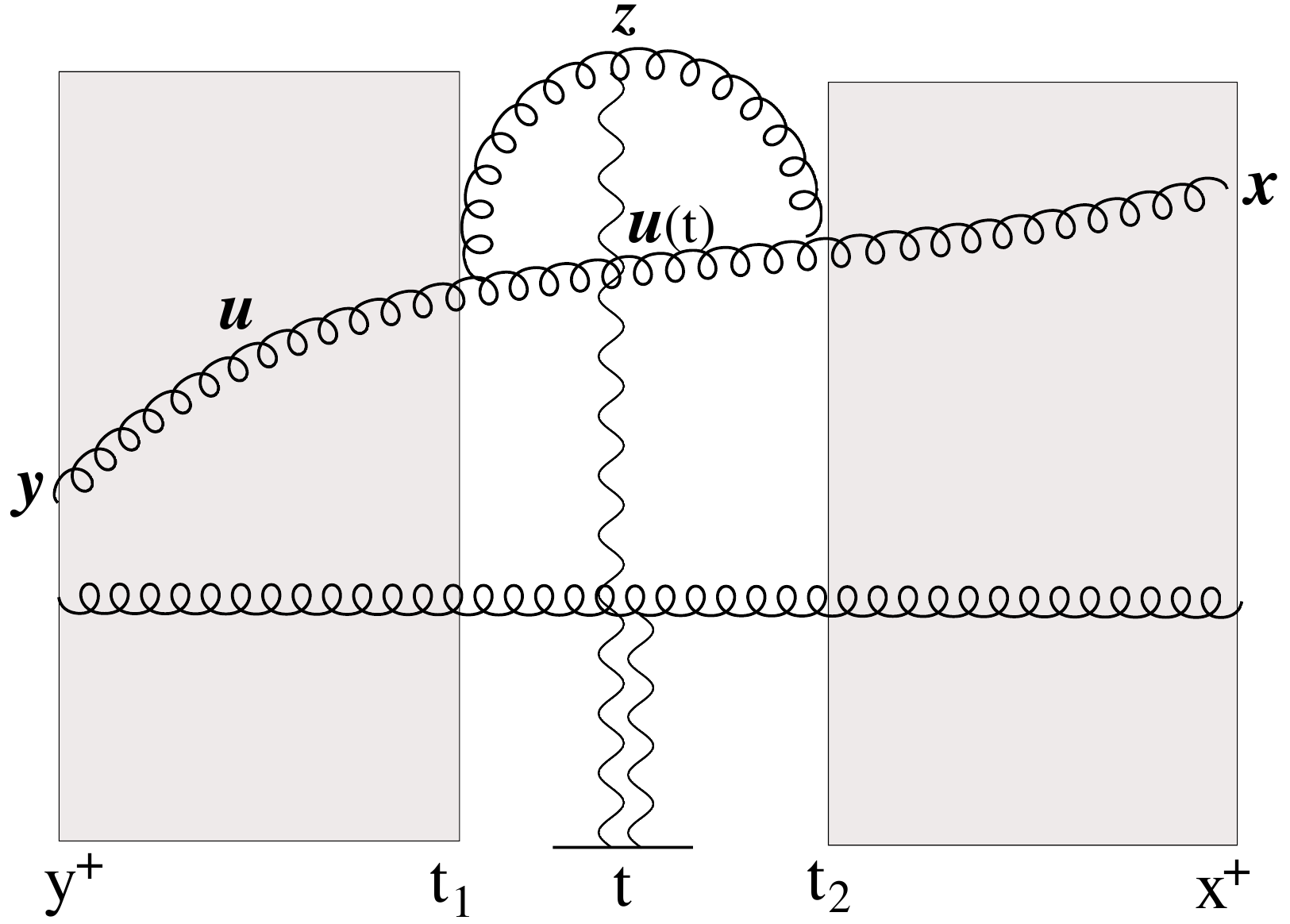}
\end{center}
\end{minipage}
\end{center}
\caption{\label{fig:gluon3} \small\sl A diagram contributing to the evolution of a non--eikonal
dipole, as described by \eqn{EqGamma1A}. The evolving dipole lives fully inside the medium
($0<y^+<x^+<L$). The grey areas prior and after the fluctuation are regions of multiple scattering.
During the lifetime of the fluctuation, between $t_1$ and $t_2$,
the partonic system (effectively made with three gluons) scatters only once, at some
intermediate time $t$.}
\end{figure}

Once again, we are mostly interested in the situation where the partonic
system created by the fluctuation 
scatters only once in the medium. This is illustrated in Fig.~\ref{fig:gluon3} 
and is described by the linearized version of \eqn{GammaA},
as obtained by expanding the exponential. After manipulations similar to those
in Sect.~\ref{sec:SSA}, this can be written as (cf. \eqn{EqGamma1})
 \begin{align}
\label{EqGamma1A}
 \int_{y^+}^{x^+} &\rmd t\, \frac{\del
\Gamma_\omega(\bu(t))}{\del \omega} \,=\,-\frac{\alpha_sN_c}{2}
  \frac{1}{\omega^3}\int_{y^+}^{x^+}\rmd t \int\rmd^2{\bz}\, \Big[\Gamma_{\omega} (\bu(t)-\bz)+
 \Gamma_{\omega} (\bz)-\Gamma_{\omega} (\bu(t))\Big]\nn
 &\times  \int_{y^+}^{t}\rmd t_1  \int_{t}^{x^+}\!\rmd t_2\, 
  \del^i_{\br_1}\del^i_{\br_2}\Bigg\{ 
\mcal{G}_0(t_2-t,\br_2-\bz; \omega)\,\mcal{G}_0(t-t_1,\bz-\br_1; \omega)  
    \Bigg\}\Bigg|^{\br_2=\bu(t_2)}_{\br_2=0}\,
    \Bigg|^{\br_1=\bu(t_1)}_{\br_1=0}\, ,\nn
  \end{align}
where we recall that $\bz$ denotes the position of the virtual gluon at the interaction time $t$,
with $t_1<t<t_2$. For the corresponding equation in Sect.~\ref{sec:SSA}, we have been able
to explicitly perform the integrals over $t_1$ and $t_2$. Here, however, these integrals are
complicated by the time dependence of the endpoints $\br_1$ and $\br_2$. 
To overcome this difficulty, we shall exploit the
separation of time scales between the radiated gluon with energy $k^+$ and the virtual one
with energy $\omega \ll k^+$. During the lifetime $\Delta t\equiv t_2-t_1\lesssim 
\tau_{_{\rm br}}(\omega)$ of the fluctuation, the transverse position of the hard gluon
changes by an amount
 \begin{align} \Delta u_\perp^2 \,\sim\,\frac{2\Delta t}{k^+}\,\lesssim \,
 \frac{2\tau_{_{\rm br}}(\omega)}{k^+}\,,\end{align}
which is small compared to the typical separation $<u_\perp^2>\simeq 4/k_{_{\rm br}}^2(k^+)$
between the hard gluon and the quark~:
 \begin{align}
 \frac{\Delta u_\perp^2}{<u_\perp^2>}\,\sim\,\frac{\tau_{_{\rm br}}(\omega)}
 {\tau_{_{\rm br}}(k^+)}\,\sim\,\sqrt{\frac{\omega}{k^+}}\,\ll\,1\,.
 \end{align}
Hence, when evaluating the endpoints $\br_1$ and $\br_2$ in \eqn{EqGamma1A}, 
one can neglect the small difference 
between $\bu(t_2)$ and $\bu(t_1)$ and approximate  them 
both with the intermediate value  $\bu(t)$ (the transverse
size of the parent dipole at the interaction time). Then the 
integrals over $t_1$ and $t_2$ can be done as in \eqn{int2L}, and one is led to 
  \begin{align}
\label{BFKLint}  \hspace*{-.5cm}
\int_{y^+}^{x^+} \!\rmd t\ \omega\, \frac{\del
\Gamma_\omega(\bu(t))}{\del \omega} \,=\,\frac{\abar}{2\pi}
\int_{y^+}^{x^+}\!\rmd t \int_{\bz}\,\frac{\bm{u}^2(t)}
 {(\bm{z}-\bm{u}(t))^2 \bm{z}^2}\, \Big[\Gamma_{\omega} (\bu(t)-\bz)+
 \Gamma_{\omega} (\bz)-\Gamma_{\omega} (\bu(t))\Big]\,.
  \end{align}
This holds for generic values of the integration limits ${y^+}$ and ${x^+}$
(recall that these variables are themselves integrated over in \eqn{spectrum}), hence it
must hold {\em locally} in $t$~:
  \begin{align}
\label{BFKLt}  \omega\, \frac{\del
\Gamma_\omega(\bu(t))}{\del \omega} \,=\,\frac{\abar}{2\pi}
 \int_{\bz}\,\frac{\bm{u}^2(t)}
 {(\bm{z}-\bm{u}(t))^2 \bm{z}^2}\, \Big[\Gamma_{\omega} (\bu(t)-\bz)+
 \Gamma_{\omega} (\bz)-\Gamma_{\omega} (\bu(t))\Big]\,.
  \end{align}
This is recognized as the BFKL equation for a dipole with time--dependent transverse
size $\bm{u}(t)$. As clear from its above  derivation, this equation  
is valid so long as the relative change in 
$\bm{u}(t)$ remains negligible during the lifetime of the typical fluctuations. 

In particular, the time--dependence of $\bm{u}(t)$ 
is irrelevant for the double--logarithmic approximation that we are primarily
interested in. As explained in Sect.~\ref{sec:DLA}, this is controlled by fluctuations
with relatively large transverse sizes,  which are
only logarithmically sensitive to the parent dipole size. At DLA, \eqn{BFKLt} reduces
to an equation like \eqn{DLAint} which describes the evolution of the jet quenching parameter
$\hat q_{\rm g}(\tau,Q^2)$ with the longitudinal ($\tau$) and transverse ($Q^2$) resolution scales.
For the problem at hand, the relevant scales are
$\tau=\tau_{_{\rm br}}(k^+)$ (the formation time for the radiated gluon) and
$Q^2=k_{_{\rm br}}^2(k^+)=\hat q_{\rm g} \tau_{_{\rm br}}(k^+)$ 
(the transverse momentum squared acquired by
this gluon during formation). Hence, the leading--order radiative correction reads
(compare to \eqn{DLAfin})
 \begin{align}
\label{DLAA}
\delta \hat q_{\rm g}^{(1)}\,=\,\hat q_{\rm g}^{(0)}\,
\frac{\abar}{2}
 \ln^2\frac{\tau_{_{\rm br}}(k^+)}{\lambda} \,=\,\hat q_{\rm g}^{(0)}\,
\frac{\abar}{8}
 \ln^2\frac{k^+}{\omega_0} \,,
  \end{align}
with $\omega_0\equiv\hat q_{\rm g} \lambda^2/2$. For the energy--loss problem, $k^+\sim \omega_c$
and $\tau_{_{\rm br}}(k^+)\sim L$, hence we return to the original version of the logarithm,
as appearing in Eqs.~\eqref{DLAfin} or \eqref{qeff}.

To summarize, the dominant effect of the radiative corrections on the calculation of 
medium--induced gluon radiation consists in the renormalization of the jet quenching 
parameter within the corresponding tree--level calculation. In particular,  the energy loss 
by an energetic quark can be estimated as (cf. Eqs.~\eqref{Eloss} and \eqref{BDMPS})
  \begin{align}\label{ElossDLA}
 \Delta E (L) \,=\,\kappa\,\frac{2\alpha_s C_F}{\pi}\,\omega_c^{ren}
 \,=\,\kappa\,\frac{\alpha_s C_F}{\pi}\,\hat q_{\rm g}(L,Q_s^2)\,L^2\,,
 \end{align}
where $\kappa$ is a number of $\order{1}$ which is fully determined at tree--level.
(\eqn{BDMPS} would predict $\kappa=\sqrt{2}$ but this value changes after using
the correct version of the BDMPSZ spectrum, which remains valid when  
$k^+\sim \omega_c$ \cite{Zakharov:1996fv,Zakharov:1997uu,Baier:1998kq}.)
We thus see that the radiative corrections have the effect to increase the value of the 
energy loss (via the corresponding increase in $\hat q_{\rm g}$) and also to enhance its
dependences upon the medium size $L$ and its temperature $T=1/\lambda$.
In particular, if the medium is sufficiently large, one may approach the asymptotic
scaling $\Delta E (L) \propto L^{2+\gamma_s}$ with $\gamma_s=2\sqrt{\abar}$
the  `saturation exponent'  introduced in
\eqn{Qsasymp}. A similar observation is made in \cite{Blaizot:2014}.
}

\section{Conclusions and perspectives}
\label{sec:conc}

In this paper we have developed the theory for the non--linear evolution of jet quenching and
related phenomena to leading order in perturbative QCD at high energy. This theory can be
viewed as a generalization of the BK--JIMWLK evolution for `dilute--dense' scattering
to the case of a target with an arbitrary longitudinal extent. This generalization is complicated
by the need to go beyond the eikonal approximation in the treatment of multiple scattering
and also to explicitly take into account the non--locality of the quantum fluctuations in time.
Accordingly, the general evolution equations, such as the generalized BK equation 
\eqref{DeltaHeq}, are extremely complicated and the construction of exact solutions appears 
to be prohibitive, except perhaps via numerical methods. 

Fortunately, this theory allows for a drastic simplification
in so far as the dominant radiative corrections are concerned: as originally noticed in 
Ref.~\cite{Liou:2013qya}, these corrections are enhanced by the double logarithm 
$\ln^2(L/\lambda)$ and they can be resumed to all orders by solving 
the relatively simple, linear, equation \eqref{DLAint}, where the non--linear effects enter only via 
the restriction on the transverse phase--space for single scattering. 
This equation, which here emerges via controlled approximations from the generalized BK 
equation alluded to above, has also been obtained by directly computing the relevant Feynman graphs
to DLA accuracy \cite{Liou:2013qya,Blaizot:2014}.  

One of our main results here is to explain the emergence of this remarkable simplification, which
is the DLA, from a physical perspective. As we discuss in Sect.~\ref{sec:gluon}, this is a consequence 
of the special way how gluon saturation occurs in a medium: the saturation momentum
$Q_s^2(x)$ is proportional to the longitudinal wavelength of the gluons and hence it increases very fast
with $1/x$ already in the absence of the quantum evolution. This in turn implies that the
transverse phase--space grows as fast as the longitudinal one when increasing the medium size $L$,
thus favoring a double--logarithmic evolution. 


Within this double--logarithmic approximation, the radiative corrections are sufficiently mild to 
be absorbed into a renormalization of the jet quenching parameter, which then evolves according to \eqn{DLAint}. Here, we have demonstrated this property for two particular observables, the
transverse momentum broadening and the radiative energy loss, which in the approximations of
interest are both related to the scattering amplitude of a color dipole. It would be interesting to understand
whether a similar property remains true in more general situations and for more complicated 
observables, which are also sensitive to other correlations of the Wilson lines, like the quadrupole.
Examples in that sense include the calculation of the medium--induced gluon radiation
beyond the eikonal approximation (for the parent parton) \cite{Zakharov:1997uu,Baier:1998kq,Arnold:2002ja}, the study of color (de)coherence for 
multi--gluon emissions inside a medium \cite{MehtarTani:2010ma,MehtarTani:2011tz,CasalderreySolana:2011rz,Armesto:2011ir}, and the evolution of a jet via successive medium--induced parton 
branchings \cite{Blaizot:2012fh,Blaizot:2013hx,Blaizot:2013vha}. Some generalizations
in that sense, notably to the problem of the
jet evolution, will be presented in  \cite{Blaizot:2014}.

It would be furthermore interesting to have a deeper understanding of the systematics of the
{\em single--logarithmic} corrections, i.e. the terms of order $\abar\ln(L/\lambda)$, or
$\abar\ln(1/x)$, in the evolution equations.
As discussed in Sect.~\ref{sec:multiple}, the situation is quite different in that respect from the
BK--JIMWLK evolution: the individual terms in the multiple scattering series are not separately
enhanced by a large logarithm $\ln(1/x)$, yet they contribute to leading--logarithmic accuracy
in a {\em non--perturbative} way --- their resummation limits the phase--space for the single
scattering approximation. It is not clear to us whether, in this context, it is possible or even useful
to explicitly isolate the terms enhanced by $\ln(1/x)$ within the evolution equations.

An obviously important, open, problem refers to the inclusion of perturbative corrections of
higher loop order, such as the running of the QCD coupling. As noticed in 
Sect.~\ref{sec:gluon}, the leading order correction to the saturation exponent is quite large
for realistic values of $\alpha_s$, a situation which generally signals the importance of higher--order
corrections. A similar problem occurs for the saturation exponent of a 
shockwave and in that case we know that the resummation of higher--order effects
drastically reduces the leading--order estimate (roughly by a factor of 3) 
\cite{Triantafyllopoulos:2002nz,Albacete:2007yr}. The calculation of the NLO corrections
to the BK--JIMWLK equations has just been completed \cite{Balitsky:2013fea,Kovner:2013ona}, 
but the corresponding program for the physics of jet quenching is still awaiting.

Also, it would be important to develop numerical techniques for attacking functional
evolution equations like the generalized BK equation \eqref{DeltaHeq}. The original BK equation 
turned out to be a formidable tool for the phenomenology of  particle production
in $pp$, $pA$, and even $AA$ collisions (especially after
being supplemented with running coupling effects) \cite{Gelis:2010nm,Albacete:2014fwa}, 
and it would be very useful to dispose of a
similar tool for the phenomenology of jet quenching.

Last but not least, it is interesting to notice the convergence between some of the present results at 
weak coupling, e.g. the saturation exponent for the plasma, or the $L$--dependence of the renormalized
$\hat q$ and of the energy loss, and the corresponding results at strong coupling\footnote{More precisely,
we mean here the results concerning the energy loss and momentum broadening of {\em light}
partons, which is the most interesting case
in the high--energy limit. For a more general survey of the related AdS/CFT 
literature, including the important case of a heavy quark, we refer to the review papers
\cite{Gubser:2009md,CasalderreySolana:2011us}.}
\cite{Hatta:2007cs,Hatta:2008tx,Dominguez:2008vd,Gubser:2008as,Chesler:2008uy,Arnold:2010ir}. One may view this as merely a coincidence,
but we do not believe so: as discussed in Refs.~\cite{Hatta:2008tx,Dominguez:2008vd}, the
dominant mechanism for transverse momentum broadening in a strongly coupled plasma is
the recoil associated with medium--induced radiation. (That is, at strong coupling, the same mechanism
is responsible for both energy loss and momentum broadening.) 
The perturbative corrections that we have considered here at weak coupling 
are themselves associated with radiation, so their inclusion naturally
interpolates towards the physical scenario expected at strong coupling. And indeed, the present 
perturbative results, which as we have seen
predict a (global) saturation momentum $Q_s^2(L)=\hat q(L) L\propto L^{1+\gamma_s}$ 
and an energy loss $\Delta E (L) \propto L^{2+\gamma_s}$ with $\gamma_s=2\sqrt{\abar}$, 
suggest a relatively smooth approach towards the 
respective trends at strong coupling, namely $Q_s^2(L)\propto L^2$ \cite{Hatta:2007cs,Dominguez:2008vd}
 and respectively $\Delta E (L) \propto L^{3}$ \cite{Hatta:2008tx,Dominguez:2008vd,Gubser:2008as,Chesler:2008uy,Arnold:2010ir}. It remains to be seen whether such a smooth convergence survives after including
 higher order perturbative corrections.

  \section*{Acknowledgments}
  I am grateful to Al Mueller for many insightful discussions, for repeatedly reading the
  manuscript, and for helping me clarifying some of the arguments.
  During the gestation of this work, I have benefited from many inspiring discussions with
  my colleagues in Saclay, Jean-Paul Blaizot, Fabio Dominguez, Yacine Mehtar-Tani, and  Bin Wu.
This research is supported by the European Research Council under the Advanced Investigator Grant ERC-AD-267258. All figures were made with Jaxodraw \cite{Binosi:2003yf}.

\appendix

 \section{A succinct derivation of the evolution Hamiltonian}
\label{sec:deriv}

In this Appendix, we shall explicitly perform one step in the high energy evolution of the
projectile $S$--matrix and show that the result is indeed equivalent to acting with the
Hamiltonian \eqref{DeltaH} on the original $S$--matrix operator. 
 
The precise nature of the projectile is irrelevant for the present purposes. It suffices to say 
that, in some given frame where the target has rapidity\footnote{Strictly speaking,
the rapidity of the target --- a left--mover --- is negative with our present conventions,
but here we shall reserve the notation $Y_T$ for the positive quantity $|Y_T|$.}
$Y_T$ and the projectile has rapidity 
$Y_P$, with $Y_T+Y_P=Y$ (the total rapidity separation between the projectile and the target), 
the $S$--matrix is represented by some color--singlet operator $\hat{\mcal{O}}_{Y_P}[A^-]$, 
which is built with Wilson lines (one for each partonic constituent of the projectile) and
thus depends upon the target field $A^-$. 
The actual structure of this operator can be quite complicated, as it reflects 
the evolution of the projectile wavefunction over the rapidity interval $Y_P$.
The {\em average} $S$--matrix, which is the physical observable, is obtained by
averaging $\hat{\mcal{O}}_{Y_P}[A^-]$ over all the realizations of the
random field $A^-$, according to
the CGC weight function $W_{Y_T}[A^-]$ (see e.g.
\cite{Iancu:2002xk,Gelis:2010nm}): 
\begin{align}
 \label{oave}
 \langle \hat{\mcal{O}} \rangle_Y = \int [DA^-]\, W_{Y_T}[A^-]\, \hat{\mcal{O}}_{Y_P}[A^-].
 \end{align}
As indicated by the notations above, the physical $S$--matrix depends only upon the
total rapidity separation $Y$, by boost invariance. The CGC weight function $W_{Y_T}[A^-]$ 
is a functional probability density for the target field $A^-$, which encodes the relevant
information about the target wavefunction (including its high--energy evolution up to
rapidity $Y_T$) in the approximations of interest. \eqn{oave} expresses the CGC factorization
of the high--energy evolution between the projectile and the target, for a dense--dilute scattering.
Strictly speaking, this factorization has been established \cite{Gelis:2010nm} 
only for the case where the target
is a large nucleus localized near $x^+=0$ (a `shockwave'), but the results of this paper
demonstrate that it also holds for a target  with a finite, possibly large, 
extent in $x^+$. As a matter of facts, a factorization similar to \eqn{oave}, but
with $W_{Y_T}[A^-]$ taken to be simply a Gaussian in $A^-$, has been used
in several previous studies of jet quenching and energy loss, which however did not
address the problem of the high--energy  evolution of the observables.

To study this high--energy evolution, we shall now increase the rapidity
difference $Y$ by an amount $\Delta Y$, by boosting the projectile:
$Y_P\to Y_P+\Delta Y$. To understand the physical consequences of this boost, let us
first remind some general facts about the high--energy evolution:

\texttt{(i)} The wavefunction of the projectile with rapidity $Y_P$ includes quanta
--- the valence partons and the relatively soft gluons produced via radiation ---
with longitudinal momenta $p^+$ within the strip $\Lambda_0 < p^+ < \Lambda_0
\rme^{Y_P}$. Here, $\Lambda_0$ is the infrared cutoff used to properly define the
wavefunction (the softest longitudinal momentum that can be measured).

\texttt{(ii)} When the projectile is boosted by an amount $\Delta Y$, the already
existing partons act as sources for the emission of additional gluons within the
range $\Lambda_0 < p^+ < \Lambda_0 \rme^{\Delta Y}$. The new gluons are
much softer than their sources (whose typical $p^+$ momenta are very large
as compared to  $ \Lambda_0 \rme^{\Delta Y}$), so their emission can be 
computed in the eikonal approximation.

\texttt{(iii)} Albeit soft relative to their sources, the `evolution' gluons are still 
fast enough as compared to the target, so their scattering off the latter
can be described by Wilson lines. 

\texttt{(iv)} The probability for a soft gluon emission within the range
$\Lambda_0 < p^+ < \Lambda_0 \rme^{\Delta Y}$ is of order
$\abar\Delta Y$, with $\abar\equiv \alpha_sN_c/\pi$. Hence, by keeping
$\Delta Y\ll 1/\abar$, one can ensure that there is only one additional gluon emission,
which can be treated in perturbation theory.

\comment{
These considerations, notably point \texttt{(iv)} above, are strictly true for soft gluon emissions
which proceed via bremsstrahlung {\em in the vacuum}. (This includes the case where the target
is a nuclear shockwave; see Sect.~\ref{sec:JIM} below.) 
Yet, as we shall later discover, they are also useful when the emissions
take place inside a dense, partonic, medium, although in that case the emission rate can,
and generally will, be influenced by the medium.
}

In particular, the above discussion shows that, when $\alpha_s Y_P\ll 1$, the high--energy 
evolution of the projectile is negligible, so the associated wavefunction reduces to the valence partons.  
One then speaks about a `bare' projectile, like a  $q\bar q$ color dipole, whose $S$--matrix
operator is relatively simple (recall \eqn{Sdip}). Conversely, a projectile with $Y_p\gtrsim 1/\abar$
has generally a complicated structure, including a large number of soft gluons, which increases
exponentially with $Y_P$. 

After such general considerations, let us return to the high--energy evolution of \eqn{oave}
under the boost $Y_P\to Y_P+\Delta Y$ of the projectile, with $\abar\Delta Y\ll 1$. This is
obtained by inserting into \eqn{oave} a path integral describing the quantum gluons
with longitudinal momenta within the range $\Lambda_0 < |p^+| < \Lambda_0 \rme^{\Delta Y}$,
together with their eikonal couplings to their sources (the partons already included
in the structure of $\hat{\mcal{O}}_{Y_P}$) and also to the target :
\begin{align}
 \label{oevol}
 \langle \hat{\mcal{O}} \rangle_{Y+\Delta Y} = \int [DA^-]\, W_{Y_T}[A^-]\ {Z_{\Delta Y}^{-1}}
\int_{\Delta Y} [D a^\mu] \,\delta(a^+)\ \rme^{\rmi S_0[a^\mu; A^-]}\
  \hat{\mcal{O}}_{Y_P}[A^-+a^-].
 \end{align}
Here $a^\mu_a(x)$ are the gauge fields describing the soft quantum fluctuations. The path integral
is written in the projectile light--cone gauge $a^+=0$. The effective action $S_0[a^\mu; A^-]$
is obtained by keeping only the terms quadratic in $a^\mu$ in the expansion of the Yang--Mills action 
$S_{YM}[A^\mu + a^\mu]$ around the background field $A^\mu =\delta^{\mu -}A^-$ :
 \begin{align}\label{S2}
 S_0[a^\mu; A^-]\,=\,\frac{1}{2}\int\rmd^4x\Big[a^i\big(-D^2\big)a^i
 \,+\,\big(\del^+a^-+\del_ia^i\big)^2\Big]\,,
  \end{align}
where $D^2=2\del^+D^--\lap$, with $D^-=\del^--igA^-$ (the covariant derivative built with
the background field), and we recall that $\del_i=\del/\del x^i=-\del^i$.
The action \eqref{S2} is formally written as a 4--dimensional integral, but the integrand
is homogeneous in $x^-$ and it is understood that the integral over the corresponding modes
$p^+$ is restricted to the strip  $\Lambda_0 < |p^+| < \Lambda_0 \rme^{\Delta Y}$.
The quadratic action \eqref{S2}  generates, as usual, the propagator $G^{\mu\nu}$ of the  soft
gluons in the background field and in the LC gauge $a^+=0$. Namely, \eqn{S2} is tantamount to
 \beq\label{S2G}
 iS_0[a^\mu; A^-]\,=\,-\frac{1}{2}
\int\limits_{\rm strip}\frac{\rmd p^+}{2\pi}\int\limits_{x^+,y^+}
 \int\limits_{\bx,\by} a^\mu_b(x^+,\bx,p^+) \,G^{-1,bc}_{\mu\nu}(x^+,\bx; y^+,\by; p^+)\,
 a^\nu_c(y^+,\by,p^+) \,,
 \eeq
where $G^{\mu\nu}$ is the background--field gluon propagator in the LC gauge,
to be constructed in Appendix~\ref{sec:G}.
The eikonal approximation for the interactions between the quantum gluons and the target
is enforced by keeping within $S_0$ only the `large' component $A^-$ of the target field.
Also, the fact that the soft emissions are treated in the eikonal approximation is manifest
in the fact that the quantum gluons are coupled to their sources via a simple shift $A^-
\to A^-+a^-$ of the functional argument of the $S$--matrix $\hat{\mcal{O}}_{Y_P}$~;
that is, the relatively fast partons included in $\hat{\mcal{O}}_{Y_P}$ couple to both the target
field and the soft gluons to be emitted via Wilson lines. Finally, the normalization factor
$Z_{\Delta Y}$ is given by a similar path integral, but without the factor $\hat{\mcal{O}}_{Y_P}$.

In writing the action  \eqref{S2} we have ignored the self--interactions of the quantum gluons,
which is indeed correct to leading order in $\alpha_s$. As a bookkeeping, the target field is
strong, $gA^-\sim\order{1}$, and must be treated exactly, whereas the quantum fields are weak,
$ga^\mu \ll 1$, and should be expanded out in perturbation theory. For consistency, one should
also expand the scattering operator $\hat{\mcal{O}}_{Y_P}[A^-+a^-]$ in powers of $a^-$,
up to quadratic order. (This is tantamount to considering a single gluon emission.) The linear
term in this expansion vanishes after performing the path integral, whereas the second
order term, proportional to $G^{--}$, expresses the change in the average $S$--matrix 
due to a soft gluon emission, to the accuracy of interest:
\begin{align}
 \label{ochange}
 \langle \hat{\mcal{O}} \rangle_{Y+\Delta Y} - \langle \hat{\mcal{O}} \rangle_{Y}
 &= \int [DA^-]\, W_{Y_T}[A^-]\ \Delta H \hat{\mcal{O}}_{Y_P}[A^-]\,.
 \end{align}
 We have recognized here the evolution `Hamiltonian' $\Delta H $ according to its definition,
 \eqn{DeltaH}, in which we identify $\Lambda \equiv \Lambda_0 \rme^{\Delta Y}$ and
 $x\equiv\rme^{-\Delta Y}$ (so that $x\Lambda = \Lambda_0$). \eqn{ochange}, which can be 
 rewritten in operator form as 
  \begin{align} \label{DeltaO}
\hat{\mcal{O}}_{Y_P+\Delta Y}[A^-] \,-\, \hat{\mcal{O}}_{Y_P}[A^-]\,=\,
  \Delta H \,\hat{\mcal{O}}_{Y_P}[A^-]\,,
 \end{align}
confirms that $\Delta H $ is indeed the right evolution operator.

Returning to \eqn{ochange}, note that the functional derivatives implicit in $\Delta H$
can be integrated by parts and thus made to act on the CGC weight function $W_{Y_T}[A^-]$.
Accordingly, the change in the average $S$--matrix can be alternatively associated with a
{\em target} evolution, of the form
  \begin{align} \label{DeltaW}
  W_{Y_T+\Delta Y}[A^-]\,-\,W_{Y_T}[A^-] \,=\,
  \Delta H  \,W_{Y_T}[A^-]\,. \end{align}
This is a consequence of boost invariance: one can increase the rapidity difference
$Y$ between the projectile and the target by either boosting the projectile, or by boosting
the target in the opposite direction; both procedures should give the same evolution
for the $S$--matrix.

\section{The background field gluon propagator}
\label{sec:G} 

In this Appendix, we construct the
background field gluon propagator in the light--cone gauge $A^+=0$
and collect  some related formul{\ae} that were used
in the main text. Our presentation will be brief
since similar constructions can be found in the literature.
(See e.g. Sect. 6 in Ref.~\cite{Iancu:2000hn} for a related discussion.)
The propagator is defined as in Appendix~\ref{sec:deriv}, that is,
\begin{align}\label{Gmn}
G^{\mu\nu}_{ab}(x,y)\,\equiv\,&\left\langle {\rm T}\,a^\mu_a(x)\,a^\nu_b(y)
\right\rangle\nn
=& {Z^{-1}}
\int [D a^\mu] \,\delta(a^+)\ a^\mu_a(x)\,a^\nu_b(y)\
 \rme^{\rmi S_0[a^\mu; A^-]}\,,\end{align}
where the symbol T refers to operator ordering in LC time ($x^+$) and  
$x^\mu=(x^+,x^-,\bx)$. It is implicitly
assumed that the longitudinal momentum $p^+$ of the quantum fluctuations is
restricted to the strip \eqref{strip}, whereas the background field 
$A^\mu =\delta^{\mu -}A^-$ carries no $p^+$ momentum (i.e. it is homogeneous
in $x^-$). The action $S_0[a^\mu; A^-]$ is shown in
\eqn{S2} and is quadratic in the quantum fields $a^\mu$. 
It is convenient to bring this action to a diagonal form, by replacing $a^-\to
\tilde a^-$ with
\begin{align} \label{tildea}
\tilde a^-(x)\,\equiv\,a^-(x)\,+\,\frac{\del^i}{\del^+}\,a^i(x)\,.\end{align}
Then the action becomes (we recall that $D^2=2\del^+D^--\lap$
and $D^-=\del^--igA^-$)
 \begin{align}\label{S2tilde}
 S_0[\tilde a^-, a^i; A^-]\,=\,\frac{1}{2}\int\rmd^4x\Big[a^i\big(-D^2\big)a^i
 \,+\,\big(\del^+\tilde a^-\big)^2\Big]\,,
  \end{align}
which implies that the propagator $G^{ij}$ of the transverse fields
is the same as the `scalar' propagator: $G^{ij}=\delta^{ij}G$, with $G_{ab}(x,y)$
obeying \eqn{Geq} (after a Fourier transform $x^--y^-\to p^+$). Also,
\begin{align}
\left\langle {\rm T}\,\tilde a^-_a(x)\,\tilde a^-_b(y)
\right\rangle=\delta_{ab}\int\frac{\rmd^4 p}{(2\pi)^4}\  \rme^{-\rmi p\cdot(x-y)}\,\frac{\rmi}{
(p^+)^2}\,,\qquad \left\langle {\rm T}\,\tilde a^-_a(x)\,a^i_b(y)
\right\rangle\,=\,0\,.\end{align}
After inverting the transformation in \eqn{tildea}, we finally obtain
[in the mixed Fourier representation $(\vec x, p^+)$ with $\vec x\equiv
(x^+,\bx)$ and color indices suppressed]
 \begin{align}\label{GLC}
G^{-i}(\vec x, \vec y; p^+)=\,&\frac{\rmi}{p^+}\,\del^i_{\bx}
G(\vec x, \vec y; p^+)\,,\qquad
G^{i-}(\vec x, \vec y; p^+)=\,-\frac{\rmi}{p^+}\,\del^i_{\by}
G(\vec x, \vec y; p^+)\,,\nn
G^{--}(\vec x, \vec y; p^+)\,=\,&\frac{1}{(p^+)^2}\,\del^i_{\bx}\del^i_{\by}
G(\vec x, \vec y; p^+)\,+\,\frac{\rmi}{(p^+)^2}\,\delta^{(3)}(\vec x-\vec y) \,.
\end{align}
The propagator is to be considered with the  Feynman prescription for the pole at the 
mass-shell. For instance, the free  ($A^-=0$) scalar propagator reads
$G_{0, \,ab}=\delta_{ab}G_{0}$, with 
\begin{align}
G_0(p)\,=\,\frac{\rmi}{2p^+p^--\bp^2+i\epsilon}\,,\end{align}
in momentum space. This implies e.g.
  \begin{align}\label{G--0}
  G^{--}_{0}(p)  \,=\,\frac{\bp^2}{(p^+)^2}\,G_0(p)\,+\,\frac{\rmi}{(p^+)^2}
  \,=\,\frac{2p^-}{p^+}\,\frac{\rmi}{2p^+p^--\bp^2+i\epsilon}\,,
\end{align}
The `axial' pole at $p^+=0$ needs no special prescription, since $p^+$ cannot vanish
within the present context, as manifest on \eqn{strip}. 
The expression of the free propagator
in mixed Fourier representation will also be useful:
 \begin{align}\label{G0m}
 G_0(t, \bx; p^+)\,=\,\frac{1}{2p^+}\,\big[\theta(t)\theta(p^+)-
\theta(-t)\theta(-p^+)\big]\int\frac{\rmd^2\bp}{(2\pi)^2}\ \rme^{\rmi \bp\cdot\bx}
\,\rme^{-\rmi \frac{p_\perp^2}{2p^+}t}\,. \end{align}
Note  that modes with positive (negative) values of $p^+$ propagate forward
(backward) in time.

A formal expression for the `scalar' propagator, which is valid for an arbitrary background 
field but involves a path integral, has been presented in \eqn{Gmedium}.  
Using this formal expression, we shall now to derive a fully explicit
formula for the case where the target is a shockwave localized near $x^+=0$. 
When $x^+$ and $y^+$ are both positive, or both negative (i.e. they are on the same 
side of the shockwave), then the propagator in \eqn{Gmedium}
reduces to the free propagator $G_0$. Consider now the case where the
gluon crosses the shockwave: $y^+<0$ and $x^+>0$. Then for a localized target field 
$A^-(t,\bz)\sim \delta(t)$, one can approximate the Wilson line in  \eqn{Gmedium} as 
  \begin{align}\label{USW}
 U^{\dagger}_{x^+y^+}[\br(t)]\,\simeq\,
 \rmP \, \rme^{\rmi g \int  \dif t \,A^-(t, {\br}(0))}\,=\,\int\rmd^2{\bz}\
 \delta^{(2)}\big(\bz-
 {\br}(0)\big)\ \rmP \,\rme^{\rmi g \int  \dif t \,A^-(t,\bz)}\,,
 \end{align}
and the path integral can be computed as follows:
\begin{align}\label{fact}
\hspace*{-.7cm}
  \int\!\big[\mcal{D}\br(t)\big]
  \exp\bigg\{\rmi \,\frac{p^+}{2}\!
  \int_{y^+}^{x^+}\rmd t \,\dot\br^2(t)\bigg\}\,\delta^{(2)}\big(\bz-
 {\br}(0)\big)=\mcal{G}_0(x^+,\bx-\bz; p^+)\mcal{G}_0(-y^+,\bz-\by; p^+),
 \end{align}
where $\mcal{G}_0=2p^+G_0$ is the free `reduced' propagator. The last remaining 
case, where $y^+>0$ and $x^+<0$, can be deduced by using the symmetry property
\eqref{Gsym}. One finally has
 \begin{align}\label{GSWapp}
G(x^+,\bx; y^+,\by; p^+)\,=\,& G_{0}(x^+-y^+,\bx-\by; p^+)\big[\theta(x^+)\theta(y^+)
+\theta(-x^+)\theta(-y^+)\big]\nn
 &+2 p^+ \int_{\bz} G_{0}(x^+,\bx-\bz; p^+) \,G_{0}(-y^+,\bz-\by; p^+)\nn
 &\qquad\quad
 \times \big[\theta(x^+)\theta(-y^+) U^\dagger_{\bz} - \theta(-x^+)\theta(y^+) U_{\bz}\big] \,.
 \end{align}

\section{A sum rule for the light--cone gauge propagator}
\label{sec:ident} 

In this Appendix, we shall demonstrate the identity \eqref{ident} which 
has played an important role in the construction of the high--energy evolution equations, 
notably in relation with the probability conservation and
the cancellation of ultraviolet divergences.  The careful treatment of the $\epsilon$--dependence
introduced by the adiabatic prescription will be essential for this purpose.
Specifically, we shall show that the double time integral in the l.h.s. of  \eqref{ident}
gives a result of $\order{\epsilon}$ and hence vanishes in the limit $\epsilon\to 0$.

We separately consider the two pieces
in the decomposition  \eqref{G--0} of the free propagator and use the mixed Fourier representation
$G_0(t_2-t_1, \bp; p^+)$, cf. \eqn{G0m}. We focus on the case $p^+>0$ for definiteness. 
Then the `radiation' piece of the propagator is retarded ($\propto\theta(t_2-t_1)$) 
and yields
 \begin{align}\label{identRad}
\hspace*{-.5cm} \int \rmd t_1
 \int \rmd t_2\, G_{0,\,rad}^{--}(t_2-t_1, \bp; p^+)\,\rme^{-\epsilon(|t_1|+|t_2|)}\,=\,
&\frac{p_\perp^2}{2(p^+)^3}\int_{-\infty}^\infty\rmd t_2 \int_{-\infty}^{t_2}\rmd t_1\ 
 \,\rme^{-\rmi p^- (t_2-t_1)}
 \,\rme^{-\epsilon(|t_1|+|t_2|)}\nn
 \,=\,\frac{p_\perp^2}{2(p^+)^3}\ \frac{1}{2\epsilon}\,\left[
  \frac{1}{\rmi p^-+\epsilon} + \frac{1}{\rmi p^--\epsilon}\right]
  \,=\,&-\frac{1}{\epsilon}\,
  \frac{\rmi}{(p^+)^2}\,+\,\order{\epsilon}\,,
 \end{align}
where $p^-\equiv {p_\perp^2}/{2p^+}$.
 Note that, as compared to the previous, related, calculation in \eqn{timeRR},
 the final result here is a purely divergent contribution, without any additional finite term.
 The respective contribution of the Coulomb piece is, clearly,
 \begin{align}\label{identCoul}
 \int \rmd t_1
 \int \rmd t_2\, G_{0,\,Coul}^{--}(t_2-t_1, \bp; p^+)\,\rme^{-\epsilon(|t_1|+|t_2|)}\,=\,
  \frac{\rmi}{(p^+)^2}\int_{-\infty}^\infty\rmd t\ \rme^{-2\epsilon|t|}
  \,=\,&\frac{1}{\epsilon}\,
  \frac{\rmi}{(p^+)^2}\,.
  \end{align}
As anticipated, this precisely cancels the divergent piece of the `radiative' contribution,
thus leaving a net result of $\order{\epsilon}$.
 
 \section{Finite--$N_c$ corrections within the mean field approximation}
\label{sec:Nc}
When constructing the evolution equation for a dipole in a medium,
in Sect.~\ref{sec:dipole}, we have used the large--$N_c$ limit to simplify some  
arguments and the various formul{\ae}. But as also announced there,
all the results obtained within the Gaussian approximation \eqref{Gaussian} for the 
medium correlations can be extended to finite values for $N_c$. In this Appendix,
we present some tools which are useful in that sense. Such tools have been developed
in applications of the CGC formalism and we refer to the original literature for their
derivation and more details \cite{Kovner:2001vi,Blaizot:2004wv,
Kovchegov:2008mk,Dominguez:2011wm,Dumitru:2011vk,Iancu:2011nj}.

As visible e.g. on \eqref{DeltaH1}, the evolution of the dipole $S$--matrix within the
Gaussian approximation involves only two distinct Wilson line correlators: the dipole itself 
and a correlator built with three Wilson lines for the partonic system which exists 
during the fluctuation. For more generality, let us consider a dipole in some generic representation 
$R$ of the SU$(N_c)$ algebra.
Then, the relevant correlators read (within the
mean field approximation, of course)
\begin{align}\label{SdipR} 
 \mcal {S}_R(\bx,\by)\,\equiv\,\frac{1}{d_R}\left\langle
\rmtr_R\,\big[ V^{\dagger}_R({\bx}){V}_R({\by}) \big]\right\rangle\,=\,
\exp\biggl\{-g^2 C_R\int\rmd t\,
 \Gamma(t,\bx,\by)\biggr\}\,,
 \end{align}
and respectively (see e.g. Appendix B in Ref.~\cite{Kovchegov:2008mk} for
a rapid derivation)
  \begin{align}\label{fluct}
\hspace*{-.5cm}
 \left\langle U^{\dagger\,ab}(\bz)\, \frac{\rmtr_R}{d_R}  
   \Big(t^a_R\, V_{R}^{\dagger}(\bx) t^b_R\,V_{R}(\by) 
 \Big)\right\rangle\,=\,C_R
 \,\rme^{ -g^2\!  \int\rmd t\,
\Big[ \frac{N_c}{2}\big(\Gamma(t,\bx,\bz)+
 \Gamma(t,\bz,\by)\big) -\big(\frac{N_c}{2} - C_R\big)
 \Gamma(t,\bx,\by)\Big]}
  \end{align}
In these expressions, $d_R$ is the dimension of the representation ($d_F=N_c$ for the fundamental,
$d_A=N_c^2-1$ for the adjoint, etc.),  $C_R$ is the corresponding second Casimir 
($C_F=(N_c^2-1)/2N_c$, $C_A=N_c$, etc.), and the integrals over $t$ run over some
arbitrary time interval (e.g. the width of the target, or a slice of it). Also, the transverse coordinates
$\bx$, $\by$, and $\bz$ need not be constant during that time interval, that is, \eqn{fluct}
also holds for generic trajectories $\bx(t)$, etc. Finally, the function $\Gamma(t,\bx,\by)$
is related to the function $\bar\Gamma(t,\bx,\by)$ which enters the 2--point function 
\eqref{Gaussian} of the background field via
\beq
\Gamma(t,\bx,\by)\,=\,\frac{1}{2}\big[\bar\Gamma(t,\bx,\bx)+\bar\Gamma(t,\by,\by)\big]
- \bar\Gamma(t,\bx,\by)\,.\eeq
Using these formul{\ae}, it is straightforward to generalize the results in Sect.~\ref{sec:dipole}
to arbitrary $N_c$. For instance, for a dipole in the color representation $R$, the analog
of  \eqn{DeltaHeq} is obtained by replacing 
\begin{align}\hspace*{-.8cm}
 \frac{N_c}{2}
  \Big[\mcal {S}_{t_2,t_1}(\bx,\br) \mcal{S}_{t_2,t_1}(\br,\by)\mcal{S}_{t_2,t_1}^{-1}(\bx,\by)
 \,-\, 1\Big]\,\to\,C_R\,\bigg\{\rme^{ -\frac{g^2N_c}{2}\!  \int\limits_{t_1}^{t_2}\rmd t\,
\big[ \Gamma_{\omega}(t,\bx,\br)+
 \Gamma_{\omega}(t,\br,\by)- \Gamma_{\omega}(t,\bx,\by)\big]}\,-\,1\bigg\}\,
\end{align}
within the r.h.s. of \eqn{DeltaHeq}. Also, the respective l.h.s. should more generally read
 \beq
  - \frac{\del
 \ln\mcal{S}_R (\bx,\by)}{\del \omega} \,=\,C_R \int_0^L\rmd t \
 \frac{\del
 \Gamma_{\omega} (t, \bx,\by)}{\del \omega}\,.
\eeq
After these replacements, the overall factor of $C_R$ cancels out and  \eqn{DeltaHeq}
reduces to 
\eqn{DeltaHGamma} for  $\Gamma_{\omega}(\bx,\by)$ for {\em any} value of $N_c$.
Hence, as already mentioned in the main text, this equation is independent of the color 
representation $R$ of the dipole that we have started with.
 

\begin{thebibliography}{10}

\bibitem{Mehtar-Tani:2013pia}
Y.~Mehtar-Tani, J.~G. Milhano, and K.~Tywoniuk, {\it {Jet physics in heavy-ion
  collisions}},  {\em Int.J.Mod.Phys.} {\bf A28} (2013) 1340013,
  [\href{http://xxx.lanl.gov/abs/1302.2579}{{\tt arXiv:1302.2579}}].

\bibitem{Majumder:2010qh}
A.~Majumder and M.~Van~Leeuwen, {\it {The theory and phenomenology of
  perturbative QCD based jet quenching}},
  \href{http://xxx.lanl.gov/abs/1002.2206}{{\tt arXiv:1002.2206}}.

\bibitem{d'Enterria:2009am}
D.~d'Enterria, {\it {Jet quenching}},
  \href{http://xxx.lanl.gov/abs/0902.2011}{{\tt arXiv:0902.2011}}.

\bibitem{CasalderreySolana:2007zz}
J.~Casalderrey-Solana and C.~A. Salgado, {\it {Introductory lectures on jet
  quenching in heavy ion collisions}},  {\em Acta Phys. Polon.} {\bf B38}
  (2007) 3731--3794, [\href{http://xxx.lanl.gov/abs/0712.3443}{{\tt
  arXiv:0712.3443}}].

\bibitem{Kovner:2003zj}
A.~Kovner and U.~A. Wiedemann, {\it {Gluon radiation and parton energy loss}},
  \href{http://xxx.lanl.gov/abs/hep-ph/0304151}{{\tt hep-ph/0304151}}.

\bibitem{Arnold:2008vd}
P.~B. Arnold and W.~Xiao, {\it {High-energy jet quenching in weakly-coupled
  quark-gluon plasmas}},  {\em Phys.Rev.} {\bf D78} (2008) 125008,
  [\href{http://xxx.lanl.gov/abs/0810.1026}{{\tt arXiv:0810.1026}}].

\bibitem{CaronHuot:2008ni}
S.~Caron-Huot, {\it {O(g) plasma effects in jet quenching}},  {\em Phys.Rev.}
  {\bf D79} (2009) 065039, [\href{http://xxx.lanl.gov/abs/0811.1603}{{\tt
  arXiv:0811.1603}}].

\bibitem{Majumder:2012sh}
A.~Majumder, {\it {Calculating the jet quenching parameter $\hat q$ in lattice
  gauge theory}},  {\em Phys.Rev.} {\bf C87} (2013) 034905,
  [\href{http://xxx.lanl.gov/abs/1202.5295}{{\tt arXiv:1202.5295}}].

\bibitem{Liou:2013qya}
T.~Liou, A.~H. Mueller, and B.~Wu, {\it {Radiative $p_\bot$-broadening of
  high-energy quarks and gluons in QCD matter}},  {\em Nucl.Phys.} {\bf A916}
  (2013) 102--125, [\href{http://xxx.lanl.gov/abs/1304.7677}{{\tt
  arXiv:1304.7677}}].

\bibitem{Laine:2013lia}
M.~Laine and A.~Rothkopf, {\it {Light-cone Wilson loop in classical lattice
  gauge theory}},  {\em JHEP} {\bf 1307} (2013) 082,
  [\href{http://xxx.lanl.gov/abs/1304.4443}{{\tt arXiv:1304.4443}}].

\bibitem{Panero:2013pla}
M.~Panero, K.~Rummukainen, and A.~Schäfer, {\it {A lattice study of the jet
  quenching parameter}},  \href{http://xxx.lanl.gov/abs/1307.5850}{{\tt
  arXiv:1307.5850}}.

\bibitem{Mueller:2001fv}
A.~H. Mueller, {\it {Parton saturation: An Overview}},
  \href{http://xxx.lanl.gov/abs/hep-ph/0111244}{{\tt hep-ph/0111244}}.

\bibitem{Mueller:2012bn}
A.~H. Mueller and S.~Munier, {\it {$p_\perp$-broadening and production
  processes versus dipole/quadrupole amplitudes at next-to-leading order}},
  {\em Nucl.Phys.} {\bf A893} (2012) 43--86,
  [\href{http://xxx.lanl.gov/abs/1206.1333}{{\tt arXiv:1206.1333}}].

\bibitem{Wu:2011kc}
B.~Wu, {\it {On $p_T$--broadening of high energy partons associated with the
  LPM effect in a finite-volume QCD medium}},  {\em JHEP} {\bf 1110} (2011)
  029, [\href{http://xxx.lanl.gov/abs/1102.0388}{{\tt arXiv:1102.0388}}].

\bibitem{Dokshitzer:1991wu}
Y.~L. Dokshitzer, V.~A. Khoze, A.~H. Mueller, and S.~I. Troian, {\it {Basics of
  perturbative QCD}}, . Gif-sur-Yvette, France. Ed. Frontieres (1991) 274 p.

\bibitem{Kovchegov:2012mbw}
Y.~V. Kovchegov and E.~Levin, {\em {Quantum chromodynamics at high energy}}.
\newblock {Cambridge University Press}, 2012.

\bibitem{Balitsky:1995ub}
I.~Balitsky, {\it {Operator expansion for high-energy scattering}},  {\em Nucl.
  Phys.} {\bf B463} (1996) 99--160,
  [\href{http://xxx.lanl.gov/abs/hep-ph/9509348}{{\tt hep-ph/9509348}}].

\bibitem{Kovchegov:1999yj}
Y.~V. Kovchegov, {\it {Small-x F2 structure function of a nucleus including
  multiple pomeron exchanges}},  {\em Phys. Rev.} {\bf D60} (1999) 034008,
  [\href{http://xxx.lanl.gov/abs/hep-ph/9901281}{{\tt hep-ph/9901281}}].

\bibitem{JalilianMarian:1997jx}
J.~Jalilian-Marian, A.~Kovner, A.~Leonidov, and H.~Weigert, {\it {The BFKL
  equation from the Wilson renormalization group}},  {\em Nucl. Phys.} {\bf
  B504} (1997) 415--431, [\href{http://xxx.lanl.gov/abs/hep-ph/9701284}{{\tt
  hep-ph/9701284}}].

\bibitem{JalilianMarian:1997gr}
J.~Jalilian-Marian, A.~Kovner, A.~Leonidov, and H.~Weigert, {\it {The Wilson
  renormalization group for low x physics: Towards the high density regime}},
  {\em Phys.Rev.} {\bf D59} (1998) 014014,
  [\href{http://xxx.lanl.gov/abs/hep-ph/9706377}{{\tt hep-ph/9706377}}].

\bibitem{JalilianMarian:1997dw}
J.~Jalilian-Marian, A.~Kovner, and H.~Weigert, {\it {The Wilson renormalization
  group for low x physics: Gluon evolution at finite parton density}},  {\em
  Phys. Rev.} {\bf D59} (1999) 014015,
  [\href{http://xxx.lanl.gov/abs/hep-ph/9709432}{{\tt hep-ph/9709432}}].

\bibitem{Kovner:2000pt}
A.~Kovner, J.~G. Milhano, and H.~Weigert, {\it {Relating different approaches
  to nonlinear QCD evolution at finite gluon density}},  {\em Phys. Rev.} {\bf
  D62} (2000) 114005, [\href{http://xxx.lanl.gov/abs/hep-ph/0004014}{{\tt
  hep-ph/0004014}}].

\bibitem{Weigert:2000gi}
H.~Weigert, {\it {Unitarity at small Bjorken x}},  {\em Nucl. Phys.} {\bf A703}
  (2002) 823--860, [\href{http://xxx.lanl.gov/abs/hep-ph/0004044}{{\tt
  hep-ph/0004044}}].

\bibitem{Iancu:2000hn}
E.~Iancu, A.~Leonidov, and L.~D. McLerran, {\it {Nonlinear gluon evolution in
  the color glass condensate. I}},  {\em Nucl. Phys.} {\bf A692} (2001)
  583--645, [\href{http://xxx.lanl.gov/abs/hep-ph/0011241}{{\tt
  hep-ph/0011241}}].

\bibitem{Iancu:2001ad}
E.~Iancu, A.~Leonidov, and L.~D. McLerran, {\it {The renormalization group
  equation for the color glass condensate}},  {\em Phys. Lett.} {\bf B510}
  (2001) 133--144, [\href{http://xxx.lanl.gov/abs/hep-ph/0102009}{{\tt
  hep-ph/0102009}}].

\bibitem{Iancu:2001md}
E.~Iancu and L.~D. McLerran, {\it {Saturation and universality in QCD at small
  x}},  {\em Phys. Lett.} {\bf B510} (2001) 145--154,
  [\href{http://xxx.lanl.gov/abs/hep-ph/0103032}{{\tt hep-ph/0103032}}].

\bibitem{Ferreiro:2001qy}
E.~Ferreiro, E.~Iancu, A.~Leonidov, and L.~McLerran, {\it {Nonlinear gluon
  evolution in the color glass condensate. II}},  {\em Nucl. Phys.} {\bf A703}
  (2002) 489--538, [\href{http://xxx.lanl.gov/abs/hep-ph/0109115}{{\tt
  hep-ph/0109115}}].

\bibitem{Iancu:2002tr}
E.~Iancu, K.~Itakura, and L.~McLerran, {\it {Geometric scaling above the
  saturation scale}},  {\em Nucl. Phys.} {\bf A708} (2002) 327--352,
  [\href{http://xxx.lanl.gov/abs/hep-ph/0203137}{{\tt hep-ph/0203137}}].

\bibitem{Mueller:2002zm}
A.~H. Mueller and D.~N. Triantafyllopoulos, {\it {The Energy dependence of the
  saturation momentum}},  {\em Nucl.Phys.} {\bf B640} (2002) 331--350,
  [\href{http://xxx.lanl.gov/abs/hep-ph/0205167}{{\tt hep-ph/0205167}}].

\bibitem{Munier:2003vc}
S.~Munier and R.~B. Peschanski, {\it {Geometric scaling as traveling waves}},
  {\em Phys. Rev. Lett.} {\bf 91} (2003) 232001,
  [\href{http://xxx.lanl.gov/abs/hep-ph/0309177}{{\tt hep-ph/0309177}}].

\bibitem{Kang:2013raa}
Z.-B. Kang, E.~Wang, X.-N. Wang, and H.~Xing, {\it {QCD factorization for
  semi-inclusive deeply inelastic scattering at twist-4 in next-to-leading
  order}},  \href{http://xxx.lanl.gov/abs/1310.6759}{{\tt arXiv:1310.6759}}.

\bibitem{Iancu:2002xk}
E.~Iancu, A.~Leonidov, and L.~McLerran, {\it {The Color glass condensate: An
  Introduction}},  \href{http://xxx.lanl.gov/abs/hep-ph/0202270}{{\tt
  hep-ph/0202270}}.

\bibitem{Gelis:2010nm}
F.~Gelis, E.~Iancu, J.~Jalilian-Marian, and R.~Venugopalan, {\it {The Color
  Glass Condensate}},  {\em Ann.Rev.Nucl.Part.Sci.} {\bf 60} (2010) 463--489,
  [\href{http://xxx.lanl.gov/abs/1002.0333}{{\tt arXiv:1002.0333}}].

\bibitem{Kovner:2001vi}
A.~Kovner and U.~A. Wiedemann, {\it {Eikonal evolution and gluon radiation}},
  {\em Phys. Rev.} {\bf D64} (2001) 114002,
  [\href{http://xxx.lanl.gov/abs/hep-ph/0106240}{{\tt hep-ph/0106240}}].

\bibitem{Iancu:2002aq}
E.~Iancu, K.~Itakura, and L.~McLerran, {\it {A Gaussian effective theory for
  gluon saturation}},  {\em Nucl. Phys.} {\bf A724} (2003) 181--222,
  [\href{http://xxx.lanl.gov/abs/hep-ph/0212123}{{\tt hep-ph/0212123}}].

\bibitem{Blaizot:2004wv}
J.~P. Blaizot, F.~Gelis, and R.~Venugopalan, {\it {High energy p A collisions
  in the color glass condensate approach. II: Quark production}},  {\em Nucl.
  Phys.} {\bf A743} (2004) 57--91,
  [\href{http://xxx.lanl.gov/abs/hep-ph/0402257}{{\tt hep-ph/0402257}}].

\bibitem{Kovchegov:2008mk}
Y.~V. Kovchegov, J.~Kuokkanen, K.~Rummukainen, and H.~Weigert, {\it
  {Subleading--$N_c$ corrections in non-linear small-x evolution}},  {\em Nucl.
  Phys.} {\bf A823} (2009) 47--82,
  [\href{http://xxx.lanl.gov/abs/0812.3238}{{\tt arXiv:0812.3238}}].

\bibitem{Dominguez:2011wm}
F.~Dominguez, C.~Marquet, B.-W. Xiao, and F.~Yuan, {\it {Universality of
  Unintegrated Gluon Distributions at small x}},  {\em Phys. Rev.} {\bf D83}
  (2011) 105005, [\href{http://xxx.lanl.gov/abs/1101.0715}{{\tt
  arXiv:1101.0715}}].

\bibitem{Iancu:2011ns}
E.~Iancu and D.~Triantafyllopoulos, {\it {Higher-point correlations from the
  JIMWLK evolution}},  {\em JHEP} {\bf 1111} (2011) 105,
  [\href{http://xxx.lanl.gov/abs/1109.0302}{{\tt arXiv:1109.0302}}].

\bibitem{Iancu:2011nj}
E.~Iancu and D.~Triantafyllopoulos, {\it {JIMWLK evolution in the Gaussian
  approximation}},  {\em JHEP} {\bf 1204} (2012) 025,
  [\href{http://xxx.lanl.gov/abs/1112.1104}{{\tt arXiv:1112.1104}}].

\bibitem{Dumitru:2011vk}
A.~Dumitru, J.~Jalilian-Marian, T.~Lappi, B.~Schenke, and R.~Venugopalan, {\it
  {Renormalization group evolution of multi-gluon correlators in high energy
  QCD}},  {\em Phys.Lett.} {\bf B706} (2011) 219--224,
  [\href{http://xxx.lanl.gov/abs/1108.4764}{{\tt arXiv:1108.4764}}].

\bibitem{Baier:1996kr}
R.~Baier, Y.~L. Dokshitzer, A.~H. Mueller, S.~Peigne, and D.~Schiff, {\it
  {Radiative Energy Loss of High Energy Quarks and Gluons in a Finite-Volume
  Quark-Gluon Plasma}},  {\em Nucl. Phys.} {\bf B483} (1997) 291--320,
  [\href{http://xxx.lanl.gov/abs/hep-ph/9607355}{{\tt hep-ph/9607355}}].

\bibitem{Baier:1996sk}
R.~Baier, Y.~L. Dokshitzer, A.~H. Mueller, S.~Peigne, and D.~Schiff, {\it
  {Radiative Energy Loss and P(T)-Broadening of High Energy Partons in
  Nuclei}},  {\em Nucl. Phys.} {\bf B484} (1997) 265--282,
  [\href{http://xxx.lanl.gov/abs/hep-ph/9608322}{{\tt hep-ph/9608322}}].

\bibitem{Zakharov:1996fv}
B.~G. Zakharov, {\it {Fully Quantum Treatment of the Landau-Pomeranchuk-Migdal
  Effect in QED and QCD}},  {\em JETP Lett.} {\bf 63} (1996) 952--957,
  [\href{http://xxx.lanl.gov/abs/hep-ph/9607440}{{\tt hep-ph/9607440}}].

\bibitem{Zakharov:1997uu}
B.~G. Zakharov, {\it {Radiative Energy Loss of High Energy Quarks in
  Finite-Size Nuclear Matter and Quark-Gluon Plasma}},  {\em JETP Lett.} {\bf
  65} (1997) 615--620, [\href{http://xxx.lanl.gov/abs/hep-ph/9704255}{{\tt
  hep-ph/9704255}}].

\bibitem{Baier:1998yf}
R.~Baier, Y.~L. Dokshitzer, A.~H. Mueller, and D.~Schiff, {\it {Radiative
  Energy Loss of High Energy Partons Traversing an Expanding {QCD} Plasma}},
  {\em Phys. Rev.} {\bf C58} (1998) 1706--1713,
  [\href{http://xxx.lanl.gov/abs/hep-ph/9803473}{{\tt hep-ph/9803473}}].

\bibitem{Baier:1998kq}
R.~Baier, Y.~L. Dokshitzer, A.~H. Mueller, and D.~Schiff, {\it {Medium-Induced
  Radiative Energy Loss: Equivalence Between the Bdmps and Zakharov
  Formalisms}},  {\em Nucl. Phys.} {\bf B531} (1998) 403--425,
  [\href{http://xxx.lanl.gov/abs/hep-ph/9804212}{{\tt hep-ph/9804212}}].

\bibitem{Wiedemann:2000za}
U.~A. Wiedemann, {\it {Gluon Radiation Off Hard Quarks in a Nuclear
  Environment: Opacity Expansion}},  {\em Nucl. Phys.} {\bf B588} (2000)
  303--344, [\href{http://xxx.lanl.gov/abs/hep-ph/0005129}{{\tt
  hep-ph/0005129}}].

\bibitem{Wiedemann:2000tf}
U.~A. Wiedemann, {\it {Jet Quenching Versus Jet Enhancement: a Quantitative
  Study of the Bdmps-Z Gluon Radiation Spectrum}},  {\em Nucl. Phys.} {\bf
  A690} (2001) 731--751, [\href{http://xxx.lanl.gov/abs/hep-ph/0008241}{{\tt
  hep-ph/0008241}}].

\bibitem{Arnold:2001ba}
P.~B. Arnold, G.~D. Moore, and L.~G. Yaffe, {\it {Photon Emission from
  Ultrarelativistic Plasmas}},  {\em JHEP} {\bf 11} (2001) 057,
  [\href{http://xxx.lanl.gov/abs/hep-ph/0109064}{{\tt hep-ph/0109064}}].

\bibitem{Arnold:2001ms}
P.~B. Arnold, G.~D. Moore, and L.~G. Yaffe, {\it {Photon Emission from Quark
  Gluon Plasma: Complete Leading Order Results}},  {\em JHEP} {\bf 12} (2001)
  009, [\href{http://xxx.lanl.gov/abs/hep-ph/0111107}{{\tt hep-ph/0111107}}].

\bibitem{Arnold:2002ja}
P.~B. Arnold, G.~D. Moore, and L.~G. Yaffe, {\it {Photon and Gluon Emission in
  Relativistic Plasmas}},  {\em JHEP} {\bf 06} (2002) 030,
  [\href{http://xxx.lanl.gov/abs/hep-ph/0204343}{{\tt hep-ph/0204343}}].

\bibitem{Blaizot:2014}
J.-P. Blaizot and Y.~Mehtar-Tani, {\it to appear}, .

\bibitem{Chen:1995pa}
Z.~Chen and A.~H. Mueller, {\it {The Dipole picture of high-energy scattering,
  the BFKL equation and many gluon compound states}},  {\em Nucl.Phys.} {\bf
  B451} (1995) 579--604.

\bibitem{Blaizot:2004wu}
J.~P. Blaizot, F.~Gelis, and R.~Venugopalan, {\it {High-energy pA collisions in
  the color glass condensate approach. 1. Gluon production and the Cronin
  effect}},  {\em Nucl.Phys.} {\bf A743} (2004) 13--56,
  [\href{http://xxx.lanl.gov/abs/hep-ph/0402256}{{\tt hep-ph/0402256}}].

\bibitem{Mueller:2001uk}
A.~H. Mueller, {\it {A Simple derivation of the JIMWLK equation}},  {\em
  Phys.Lett.} {\bf B523} (2001) 243--248,
  [\href{http://xxx.lanl.gov/abs/hep-ph/0110169}{{\tt hep-ph/0110169}}].

\bibitem{Jeon:2013zga}
S.~Jeon, {\it {Color Glass Condensate in Schwinger-Keldysh QCD}},  {\em Annals
  Phys.} {\bf 340} (2014) 119--170,
  [\href{http://xxx.lanl.gov/abs/1308.0263}{{\tt arXiv:1308.0263}}].

\bibitem{Caron-Huot:2013fea}
S.~Caron-Huot, {\it {When does the gluon reggeize?}},
  \href{http://xxx.lanl.gov/abs/1309.6521}{{\tt arXiv:1309.6521}}.

\bibitem{Binosi:2014xua}
D.~Binosi, A.~Quadri, and D.~Triantafyllopoulos, {\it {High-energy QCD
  evolution from BRST symmetry}},
  \href{http://xxx.lanl.gov/abs/1402.4022}{{\tt arXiv:1402.4022}}.

\bibitem{Hatta:2005as}
Y.~Hatta, E.~Iancu, K.~Itakura, and L.~McLerran, {\it {Odderon in the color
  glass condensate}},  {\em Nucl. Phys.} {\bf A760} (2005) 172--207,
  [\href{http://xxx.lanl.gov/abs/hep-ph/0501171}{{\tt hep-ph/0501171}}].

\bibitem{McLerran:1993ni}
L.~D. McLerran and R.~Venugopalan, {\it {Computing quark and gluon distribution
  functions for very large nuclei}},  {\em Phys. Rev.} {\bf D49} (1994)
  2233--2241, [\href{http://xxx.lanl.gov/abs/hep-ph/9309289}{{\tt
  hep-ph/9309289}}].

\bibitem{McLerran:1994vd}
L.~D. McLerran and R.~Venugopalan, {\it {Green's functions in the color field
  of a large nucleus}},  {\em Phys. Rev.} {\bf D50} (1994) 2225--2233,
  [\href{http://xxx.lanl.gov/abs/hep-ph/9402335}{{\tt hep-ph/9402335}}].

\bibitem{Arnold:2008iy}
P.~B. Arnold, {\it {Simple Formula for High-Energy Gluon Bremsstrahlung in a
  Finite, Expanding Medium}},  {\em Phys. Rev.} {\bf D79} (2009) 065025,
  [\href{http://xxx.lanl.gov/abs/0808.2767}{{\tt arXiv:0808.2767}}].

\bibitem{Kovchegov:2001sc}
Y.~V. Kovchegov and K.~Tuchin, {\it {Inclusive gluon production in DIS at high
  parton density}},  {\em Phys.Rev.} {\bf D65} (2002) 074026,
  [\href{http://xxx.lanl.gov/abs/hep-ph/0111362}{{\tt hep-ph/0111362}}].

\bibitem{Lipatov:1976zz}
L.~Lipatov, {\it {Reggeization of the Vector Meson and the Vacuum Singularity
  in Nonabelian Gauge Theories}},  {\em Sov.J.Nucl.Phys.} {\bf 23} (1976)
  338--345.

\bibitem{Kuraev:1977fs}
E.~Kuraev, L.~Lipatov, and V.~S. Fadin, {\it {The Pomeranchuk Singularity in
  Nonabelian Gauge Theories}},  {\em Sov.Phys.JETP} {\bf 45} (1977) 199--204.

\bibitem{Balitsky:1978ic}
I.~Balitsky and L.~Lipatov, {\it {The Pomeranchuk Singularity in Quantum
  Chromodynamics}},  {\em Sov.J.Nucl.Phys.} {\bf 28} (1978) 822--829.

\bibitem{Blaizot:2013vha}
J.-P. Blaizot, F.~Dominguez, E.~Iancu, and Y.~Mehtar-Tani, {\it {Probabilistic
  picture for medium-induced jet evolution}},
  \href{http://xxx.lanl.gov/abs/1311.5823}{{\tt arXiv:1311.5823}}.

\bibitem{Hatta:2007cs}
Y.~Hatta, E.~Iancu, and A.~Mueller, {\it {Deep inelastic scattering off a N=4
  SYM plasma at strong coupling}},  {\em JHEP} {\bf 0801} (2008) 063,
  [\href{http://xxx.lanl.gov/abs/0710.5297}{{\tt arXiv:0710.5297}}].

\bibitem{Hatta:2008tx}
Y.~Hatta, E.~Iancu, and A.~Mueller, {\it {Jet evolution in the N=4 SYM plasma
  at strong coupling}},  {\em JHEP} {\bf 0805} (2008) 037,
  [\href{http://xxx.lanl.gov/abs/0803.2481}{{\tt arXiv:0803.2481}}].

\bibitem{Dominguez:2008vd}
F.~Dominguez, C.~Marquet, A.~Mueller, B.~Wu, and B.-W. Xiao, {\it {Comparing
  energy loss and p-perpendicular - broadening in perturbative QCD with strong
  coupling N = 4 SYM theory}},  {\em Nucl.Phys.} {\bf A811} (2008) 197--222,
  [\href{http://xxx.lanl.gov/abs/0803.3234}{{\tt arXiv:0803.3234}}].

\bibitem{Hatta:2007he}
Y.~Hatta, E.~Iancu, and A.~Mueller, {\it {Deep inelastic scattering at strong
  coupling from gauge/string duality: The Saturation line}},  {\em JHEP} {\bf
  0801} (2008) 026, [\href{http://xxx.lanl.gov/abs/0710.2148}{{\tt
  arXiv:0710.2148}}].

\bibitem{MehtarTani:2006xq}
Y.~Mehtar-Tani, {\it {Relating the Description of Gluon Production in Pa
  Collisions and Parton Energy Loss in Aa Collisions}},  {\em Phys. Rev.} {\bf
  C75} (2007) 034908, [\href{http://xxx.lanl.gov/abs/hep-ph/0606236}{{\tt
  hep-ph/0606236}}].

\bibitem{CasalderreySolana:2011rz}
J.~Casalderrey-Solana and E.~Iancu, {\it {Interference Effects in
  Medium-Induced Gluon Radiation}},  {\em JHEP} {\bf 08} (2011) 015,
  [\href{http://xxx.lanl.gov/abs/1105.1760}{{\tt arXiv:1105.1760}}].

\bibitem{Blaizot:2012fh}
J.-P. Blaizot, F.~Dominguez, E.~Iancu, and Y.~Mehtar-Tani, {\it {Medium-induced
  gluon branching}},  {\em JHEP} {\bf 1301} (2013) 143,
  [\href{http://xxx.lanl.gov/abs/1209.4585}{{\tt arXiv:1209.4585}}].

\bibitem{MehtarTani:2010ma}
Y.~Mehtar-Tani, C.~A. Salgado, and K.~Tywoniuk, {\it {Antiangular Ordering of
  Gluon Radiation in QCD Media}},  {\em Phys. Rev. Lett.} {\bf 106} (2011)
  122002, [\href{http://xxx.lanl.gov/abs/1009.2965}{{\tt arXiv:1009.2965}}].

\bibitem{MehtarTani:2011tz}
Y.~Mehtar-Tani, C.~A. Salgado, and K.~Tywoniuk, {\it {Jets in QCD Media: from
  Color Coherence to Decoherence}},  {\em Phys. Lett.} {\bf B707} (2012)
  156--159, [\href{http://xxx.lanl.gov/abs/1102.4317}{{\tt arXiv:1102.4317}}].

\bibitem{Armesto:2011ir}
N.~Armesto, H.~Ma, Y.~Mehtar-Tani, C.~A. Salgado, and K.~Tywoniuk, {\it
  {Coherence Effects and Broadening in Medium-Induced QCD Radiation Off a
  Massive Q ${\Bar Q}$ Antenna}},  {\em JHEP} {\bf 01} (2012) 109,
  [\href{http://xxx.lanl.gov/abs/1110.4343}{{\tt arXiv:1110.4343}}].

\bibitem{Blaizot:2013hx}
J.-P. Blaizot, E.~Iancu, and Y.~Mehtar-Tani, {\it {Medium-induced QCD cascade:
  democratic branching and wave turbulence}},  {\em Phys.Rev.Lett.} {\bf 111}
  (2013) 052001, [\href{http://xxx.lanl.gov/abs/1301.6102}{{\tt
  arXiv:1301.6102}}].

\bibitem{Triantafyllopoulos:2002nz}
D.~Triantafyllopoulos, {\it {The Energy dependence of the saturation momentum
  from RG improved BFKL evolution}},  {\em Nucl.Phys.} {\bf B648} (2003)
  293--316, [\href{http://xxx.lanl.gov/abs/hep-ph/0209121}{{\tt
  hep-ph/0209121}}].

\bibitem{Albacete:2007yr}
J.~L. Albacete and Y.~V. Kovchegov, {\it {Solving high energy evolution
  equation including running coupling corrections}},  {\em Phys.Rev.} {\bf D75}
  (2007) 125021, [\href{http://xxx.lanl.gov/abs/0704.0612}{{\tt
  arXiv:0704.0612}}].

\bibitem{Balitsky:2013fea}
I.~Balitsky and G.~A. Chirilli, {\it {Rapidity evolution of Wilson lines at the
  next-to-leading order}},  {\em Phys.Rev.} {\bf D88} (2013) 111501,
  [\href{http://xxx.lanl.gov/abs/1309.7644}{{\tt arXiv:1309.7644}}].

\bibitem{Kovner:2013ona}
A.~Kovner, M.~Lublinsky, and Y.~Mulian, {\it {Complete JIMWLK Evolution at
  NLO}},  \href{http://xxx.lanl.gov/abs/1310.0378}{{\tt arXiv:1310.0378}}.

\bibitem{Albacete:2014fwa}
J.~L. Albacete and C.~Marquet, {\it {Gluon saturation and initial conditions
  for relativistic heavy ion collisions}},
  \href{http://xxx.lanl.gov/abs/1401.4866}{{\tt arXiv:1401.4866}}.

\bibitem{Gubser:2009md}
S.~S. Gubser and A.~Karch, {\it {From gauge-string duality to strong
  interactions: A Pedestrian's Guide}},  {\em Ann.Rev.Nucl.Part.Sci.} {\bf 59}
  (2009) 145--168, [\href{http://xxx.lanl.gov/abs/0901.0935}{{\tt
  arXiv:0901.0935}}].

\bibitem{CasalderreySolana:2011us}
J.~Casalderrey-Solana, H.~Liu, D.~Mateos, K.~Rajagopal, and U.~A. Wiedemann,
  {\it {Gauge/String Duality, Hot QCD and Heavy Ion Collisions}},
  \href{http://xxx.lanl.gov/abs/1101.0618}{{\tt arXiv:1101.0618}}.

\bibitem{Gubser:2008as}
S.~S. Gubser, D.~R. Gulotta, S.~S. Pufu, and F.~D. Rocha, {\it {Gluon energy
  loss in the gauge-string duality}},  {\em JHEP} {\bf 0810} (2008) 052,
  [\href{http://xxx.lanl.gov/abs/0803.1470}{{\tt arXiv:0803.1470}}].

\bibitem{Chesler:2008uy}
P.~M. Chesler, K.~Jensen, A.~Karch, and L.~G. Yaffe, {\it {Light quark energy
  loss in strongly-coupled N = 4 supersymmetric Yang-Mills plasma}},  {\em
  Phys.Rev.} {\bf D79} (2009) 125015,
  [\href{http://xxx.lanl.gov/abs/0810.1985}{{\tt arXiv:0810.1985}}].

\bibitem{Arnold:2010ir}
P.~Arnold and D.~Vaman, {\it {Jet quenching in hot strongly coupled gauge
  theories revisited: 3-point correlators with gauge-gravity duality}},  {\em
  JHEP} {\bf 10} (2010) 099, [\href{http://xxx.lanl.gov/abs/1008.4023}{{\tt
  arXiv:1008.4023}}].

\bibitem{Binosi:2003yf}
D.~Binosi and L.~Theussl, {\it {JaxoDraw: A Graphical user interface for
  drawing Feynman diagrams}},  {\em Comput.Phys.Commun.} {\bf 161} (2004)
  76--86, [\href{http://xxx.lanl.gov/abs/hep-ph/0309015}{{\tt
  hep-ph/0309015}}].

\end{thebibliography}

\providecommand{\href}[2]{#2}\begingroup\raggedright\endgroup

\end{document}